\documentclass[final,5p,times,twocolumn]{elsarticle}

\usepackage{amssymb}
\usepackage{amsmath}

\usepackage[T2A]{fontenc}
\usepackage[ukrainian,english]{babel}

\usepackage{xurl}

\begin{document}

\begin{frontmatter}



\title{Modeling the contact area and intersurface forces between Pb nanoparticles and suspended graphene}


\author{Mykola Prodanov\corref{cor1}} 
\ead{prodanov.my@gmail.com}
\cortext[cor1]{Corresponding author}
\author{Anton Samilyk}
\ead{samilykanton@gmail.com}
\author{Oleksii Khomenko}
\ead{o.khomenko@mss.sumdu.edu.ua}
\author{Antonina Biesiedina}
\ead{a.besedina@med.sumdu.edu.ua}

\affiliation{organization={Sumy State University},
            addressline={116 Kharkivska Str.},
            city={Sumy},
            postcode={UA-40007},
            country={Ukraine}}

\begin{abstract}
We perform large-scale atomistic simulations to explore the contact mechanics of Pb nanoparticles (NPs) physisorbed on a suspended graphene sheet at zero external load. The study considers the kinetics of the NPs' relative contact area and the distributions of intersurface atomic force magnitudes acting on the NP during self-assembly from the liquid phase, and the size scaling of these quantities for the solidified NPs. The functional forms of both the repulsive and attractive parts of the normal intersurface force distributions for solidified NPs are size-independent for nanoislands with a linear lateral size larger than about 10 nm (or a surface area-to-volume ratio less than about 1~nm$^{-1}$). To first order, the repulsive tail decays exponentially, while the cutoff power law fits more precisely, especially for the high-probability repulsive forces. In both cases, the decay exponent is around -5~nN$^{-1}$. The exponentially modified Gaussian is the best fit for the attractive part for most NP sizes. The time-averaged relative contact area calculated directly in the simulations shows considerable fluctuations for NPs a few nm in diameter, while converging to approximately 0.44 for larger NPs. Based on fitting the normal-force distributions, we obtain analytical expressions for the relative contact area. Ensemble-averaged relative contact areas for NPs larger than around 10~nm, calculated using the analytical expressions, match the time-averaged MD values within statistical error. Our model is the first step towards developing analytical approximations for quantitative estimates of the relative contact area of Pb NPs adsorbed on suspended graphene.
\end{abstract}

\begin{keyword}
Surface forces \sep Contact mechanics \sep Adhesion \sep Nanoparticle \sep Graphene \sep Molecular dynamics



\end{keyword}

\end{frontmatter}



\section{Introduction}
\label{sec::intro}

Forces acting between surfaces in contact---intersurface or interfacial forces---determine a broad range of phenomena like adhesion, friction, powder clumping, fluid leakage, and wetting~\cite{israelachvili2011intermolecular,Persson2005jpcm,Persson2006suscrep}. Knowledge of the local normal interfacial forces $F_{\mathrm{n}}$, e.g., as a distribution $P(F_{\mathrm{n}})$ of their magnitudes, allows calculating the real contact area $A_{\mathrm{real}}$, which is often measured in units of the nominal (or apparent) contact area $A_0$ and is referred to as the relative contact area $A_{\mathrm{r}} = A_{\mathrm{real}}/A_0$. It enables estimation of the friction coefficient, contact electroconductivity, and contact percolation~\cite{Persson2005jpcm,Persson2006suscrep,Yang2008euphyslett,Cheng2010triblett,Pastewka2014pnas,Mo2009nature,Mo2010prb,Persson2012,Jacobs2022mrsbul}. Quantitative prediction of intersurface forces and $A_{\mathrm{r}}$ is especially relevant for nanosystems involving graphene~\cite{Novoselov2004sc,Szroeder2019om,Sahalianov2020el,Radchenko2021} and nanoparticles (NPs)~\cite{Min2008,Khomenko_NAP2017, Zakharov_NAP2018,Khomenko2022ppm,Badan2022nm,Eker2024mol,Khomenko2025ppm,Suhas2026}, where interfacial forces play a much more profound role because of the high surface area-to-volume ratio $S_{\mathrm{V}}$ compared to macroscopic objects~\cite{Suhas2026,Pozzi2024}.
Emerging practical applications of these nanosystems require detailed knowledge of their contact mechanics (CM). In particular, suspended graphene (SGr)~\cite{Gass2008} interacting with various surfaces is increasingly used in micro-/nanoelectromechanical systems and sensing devices, such as, nano switches and accelerometers~\cite{Huang2021,Fan2024,Zhang2025nanomat,Ding2025}. Graphene sensors are especially promising for real-time detection of nanoclusters and NPs of toxic heavy metals (HM), such as lead (Pb)~\cite{Shtepliuk2017scirep,Shtepliuk2018pccp}, due to graphene's large surface area and unique electronic properties~\cite{Novoselov2004sc,Szroeder2019om,Sahalianov2020el,Radchenko2021,Gass2008,Shtepliuk2017scirep,Shtepliuk2018pccp}. Quantifying interfacial forces and $A_{\mathrm{r}}$ for Pb NPs adsorbed on SGr allows prediction of the NP's contact resistance and the sensor's efficiency. Interactions between Pb NPs and SGr have also attracted attention in the context of the pressing issue of water pollution~\cite{Kommu2017jpca,Kuganathan2021,Chandran2025chemeng,Singhal2025apa}. SGr is a candidate material for membrane separation filters in water purification, removing toxic HM from water via adhesion~\cite{Kommu2017jpca,Kuganathan2021,Chandran2025chemeng,Singhal2025apa}.

Most experiments in the literature have considered Pb interactions with epitaxial, but not suspended/freestanding, graphene~\cite{Shtepliuk2018pccp,Liu2013jpcc,Gong2014dalton,Koch2025jpcm}. Additionally, these works focused on the mobility and interaction energies of single Pb atoms or Pb nanoclusters, without detailed measurements of interfacial Pb--graphene forces for NPs of different sizes. To the best of our knowledge, the literature contains no quantitative experimental results for detailed Pb NPs--SGr interactions.

In general, modern atomic force microscopy (AFM) techniques can measure adhesion and pull-off forces of individual NPs interacting with each other~\cite{Min2008,Endres2021jas} or adsorbed on different substrates~\cite{You2014langmuir,Salameh2017langmur}. Moreover, Khomami et al.~\cite{Khomami2026asusc} recently used sharp silicon AFM tips with a nominal radius of 2~nm to probe the distribution of surface adhesion forces of TiO$_2$ NPs with at least 20~nm in linear size.
However, the tip-based approach has inherent limitations regarding probing contact forces at a NP--substrate interface. First, measuring these forces requires access to the interface, which is currently limited or impossible~\cite{Rovatti2011,Rodriguez2025triblett,Chen2026ns}. Second, forces measured with a sharp tip a few nm in diameter may not reflect intersurface forces in the larger contact because the tip cannot resolve roughness wavelengths longer than its size, nor can it capture the effects of long-range surface deformations~\cite{Persson2005jpcm,Persson2006suscrep,Pastewka2014pnas}.

Currently, numerical simulations and analytical theories are the only techniques that allow estimation of detailed interfacial interactions for nanometer-sized surfaces. However, the literature lacks quantitative results on Pb NP interactions with SGr. For example, although the multiscale Persson theory of CM can predict trends in interfacial stress distribution with reasonable accuracy and has been experimentally and numerically verified for macroscopic surfaces~\cite{Persson2005jpcm,Almqvist2026pre,Prodanov2014triblett,Campana2006prb,Carbone2009epje}, its basic assumptions do not apply to the Pb NPs--SGr system. In particular, it relies on continuum mechanics for bulk, infinite surfaces and assumes infinitely large asperities, which does not hold for finite-size NPs. Instead, nanosystems require an atomistic approach~\cite{Rodriguez2025triblett}.

Several numerical models considering interfacial interactions between various surfaces appear in the literature~\cite{Yang2008euphyslett,Cheng2010triblett,Mo2009nature,Mo2010prb,Campana2006prb,Carbone2009epje,Yang2008prl,Yang2008jpcm}.
However, at least one contacting surface in these models was designed to reflect bulk 3D behavior, and the results may not hold for finite-size NPs interacting with an atomically thin 2D material.

None of the mentioned models or theories directly apply to a 2D membrane in adhesive contact with finite-size Pb NPs. In this work, we employ a numerical atomistic model to study Pb NPs interacting with SGr. For the first time, we use large-scale classical MD simulations to comprehensively investigate the local forces and contact area at the adhesive interface between a Pb NP and graphene. The main goals of our study are:

\begin{itemize}
\item To systematically investigate the behavior of the intersurface forces and $A_{\mathrm{r}}$ for Pb NPs physisorbed on a 2D material and compare it to the Persson theory and other simulations~\cite{Yang2008euphyslett,Cheng2010triblett,Mo2009nature,Mo2010prb,Campana2006prb,Carbone2009epje,
Yang2008prl,Yang2008jpcm}.

\item To develop analytical expressions for $P(F_{\mathrm{n}})$, allowing computation of $A_{\mathrm{r}}$ for Pb NPs without time-consuming simulations.

\item To consider the size scaling of these quantities and identify potential size effects. CM quantities, e.g., interfacial stiffness for randomly rough surfaces~\cite{Pastewka2013} and mean interface separation for metal NPs~\cite{Prodanov2026pccp} can exhibit finite-size effects. We want to check whether $P(F_{\mathrm{n}})$ and $A_{\mathrm{r}}$ of physisorbed Pb NPs exhibit finite-size behavior. To do this, we consider NPs over a broad size range, from 288 to 1124864 Pb atoms per NP, corresponding to a linear lateral size of 2.8--56~nm.

\item To study the kinetics of intersurface forces and $A_{\mathrm{r}}$ during NPs formation as they self-assemble from the liquid phase, mimicking thermal dewetting of thin metal films~\cite{Badan2022nm,Neek-Amal2009,Harsha2024,Kim2023}. Understanding how these quantities evolve during NP formation may help control the dewetting process. Additionally, this would shed light on how NP surface structure, inferred from spectral analysis of the force power spectrum, affects intersurface forces.
\end{itemize}

The paper is organized as follows. Section~\ref{sec::review} introduces the main definitions and details the available theoretical and computational results~\cite{Yang2008euphyslett,Cheng2010triblett,Mo2009nature,Mo2010prb,Campana2006prb,Carbone2009epje,Yang2008prl,
Yang2008jpcm}. Section~\ref{sec::methodology} describes the model and the data analysis procedures. Section~\ref{sec::results} presents the kinetics and scaling results along with the spectral analysis, fitting, and analytical results. The conclusions wrap up the work.

\section{Review and definitions}
\label{sec::review}

Analytical theories often operate with $\sigma(\mathbf{r})$---the perpendicular (normal) stress at the contact interface at lateral position $\mathbf{r}=(x,y)$ and its distribution $P(\sigma)$ rather than utilize $F_{i, z}$---a normal force acting on atom $i$ from the atoms residing on the opposing contacting surface (we choose the normal direction to be along the $z$-axis pointing from the bottom to the top surface, and $xy$ is the lateral plane, so that $F_{\mathrm{n}} \equiv F_{i, z}$ or simply $F_{z}$). In general, the atomic normal stress $\sigma = \sigma_{zz}^{(i)}$ for atom $i$ is given by the Lagrangian virial stress~\cite{Zhou2003procrsl,Liu2009jctn}, meaning the normal stress is determined entirely by the normal interfacial forces.

For a simple case of the maximum surface resolution down to the atomic scale and for the contact area projected onto the $xy$ plane, the normal stress $\sigma_i$ of atom $i$ is proportional to $F_{i, z}$, $\sigma_{i} = F_{i, z}/A_i$,
where $A_i$ is the in-plane area associated with surface atom $i$ and
\begin{equation}
F_{i, z} = \sum_{j \text{ } \in \text{ opposing } \text{atoms}} F_{ij, z}.
\label{eq:atomic_force_def}
\end{equation}
This approach is often used to compare theory results with corresponding atomistic simulations~\cite{Cheng2010triblett,Yang2008prl} and is employed in the current work. Therefore, we use $P(\sigma)$ and $P(F_z)$ interchangeably.

The knowledge of $P(\sigma)$ allows calculating the relative contact area $A_{\mathrm{r}}$~\cite{Persson2005jpcm,Persson2006suscrep,Yang2008prl}:

\begin{equation}
A_{\mathrm{r}} = \int_{0^+}^\infty
P(\sigma)\,d\sigma.
\label{eq:a_r}
\end{equation}
Here, $\sigma = 0$ is deliberately excluded because it is the boundary-condition value that encompasses all non-contact area. Continuum theories use a simple binary criterion for determining contact: 0 local stress corresponds to no contact, while non-zero stress is contact. Integrating $P(\sigma)$ over the full range of $\sigma$, including 0, yields unity, as expected for a probability distribution~\cite{Yang2008jpcm}.

Persson theory for elastic randomly rough surfaces in repulsive contact predicts  that $P(\sigma)$ has the form of a double Gaussian:
\begin{equation}
P(\sigma,\zeta) =
\frac{1}{2\sqrt{\pi G}}
\left(
e^{
-(\sigma-p)^2/4G
}
-
e^{
-(\sigma+p)^2/4G
}
\right),
\label{eq:persson_double_gaussian}
\end{equation}
where $p$ is the nominal squeezing pressure, $G$ is an adjustable parameter that depends on the current ``magnification'' $\zeta = \lambda_0/\lambda$, where $\lambda$ is the shortest wavelength under consideration relative to a reference wavelength $\lambda_0$ (e.g., the roll-off wavelength in the surface roughness power spectrum). Persson theory predicts that the stress distribution broadens as $\zeta$ increases, i.e., as shorter surface-roughness wavelengths are resolved. Substituting Eq.~(\ref{eq:persson_double_gaussian}) into Eq.~(\ref{eq:a_r}) gives $A_r = \mathrm{erf}(p/2G^{1/2})$ for repulsive contacts.

Adhesive contacts change the tensile-stress boundary in Persson theory: contact can sustain \(\sigma<0\) down to a scale-dependent detachment stress \(-\sigma_a(\zeta)\), rather than terminating at \(\sigma=0\)~\citep{Carbone2009epje}. In general, for adhesive contacts $P(\sigma)$ does not follow the double Gaussian functional form~\cite{Yang2008euphyslett} even though some models, e.g., a 1D anisotropic adhesive numerical model for randomly rough surfaces developed in~\cite{Carbone2009epje} reported it as the best fit for $P(\sigma)$ over the whole force range including the negative adhesive tail.

Several numerical simulations have been conducted to verify Persson and other CM theories~\cite{Yang2008euphyslett,Cheng2010triblett,Mo2009nature,Mo2010prb,Campana2006prb,Carbone2009epje,Yang2008prl,
Yang2008jpcm}. One problematic question for numerical, especially atomistic, simulations is how to determine which atoms are in contact, because at the atomic level the binary criterion of zero forces/stress as non-contact does not hold~\cite{Yang2008euphyslett,Cheng2010triblett,Mo2009nature,Mo2010prb,Carbone2009epje,Yang2008prl,
Yang2008jpcm}. One common approach for adhesive surfaces is to count as contact those atoms that experience repulsive forces from the opposite surface~\cite{Cheng2010triblett,Pastewka2014pnas}. Using this approach, Cheng and Robbins~\cite{Cheng2010triblett} conducted comprehensive MD simulations of interfacial forces for simplified atomic surface geometries. Rigid flat and spherical surfaces contacted an elastic substrate, with and without adhesion. The results showed a nearly universal exponential distribution $P(F_z)$ of the magnitude of instantaneous repulsive forces between surfaces, in contrast to Persson theory's prediction for repulsive randomly rough surfaces. Another MD study~\cite{Mo2010prb} simulated an AFM tip made of H-terminated diamond-like carbon interacting with an H-terminated diamond sample. They found that $P(F_z)$ is closer to the double Gaussian predicted by Persson theory. However, their distribution included only forces from atoms that spent more than 30\% of the time in contact, which might have affected the result~\cite{Cheng2010triblett,Mo2010prb}. Smart-block~\cite{Yang2008prl,Yang2008jpcm} and Green's function~\cite{Campana2006prb} MD simulations of bulk randomly rough surfaces confirmed the double-Gaussian predictions of Persson theory for repulsive intersurface interactions~\cite{Campana2006prb,Yang2008prl,Yang2008jpcm}, while for the adhesive case no fitting to the obtained numerical data was done, so no general analytical form for $P(F_z)$ was obtained~\cite{Yang2008euphyslett}.

\begin{figure*}[!htb]
\centering
  \includegraphics[height=2.05cm]{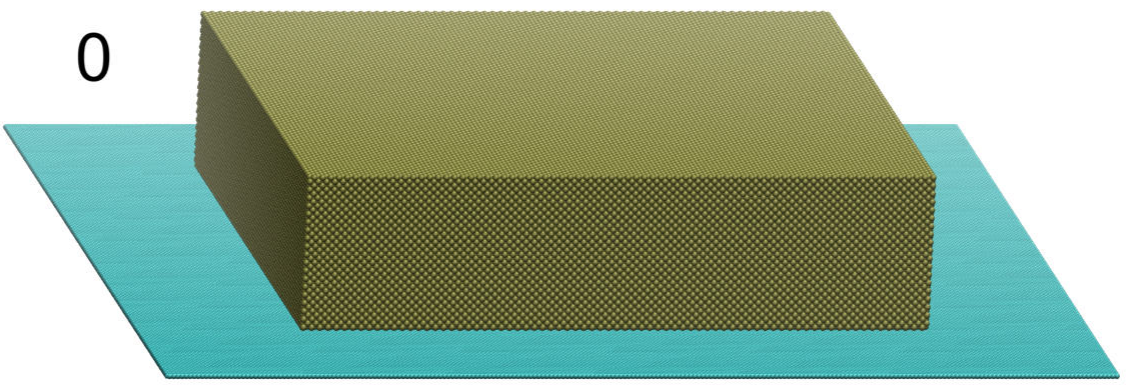}
  \includegraphics[height=2.05cm]{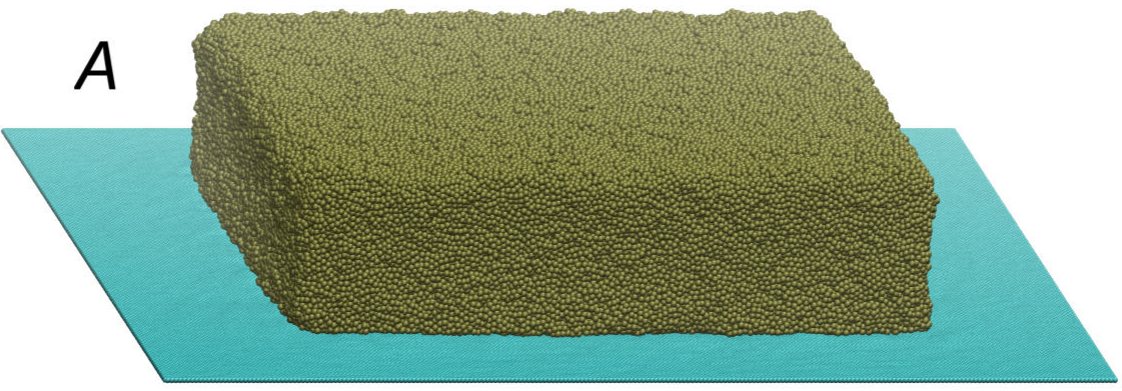}
  \includegraphics[height=2.05cm]{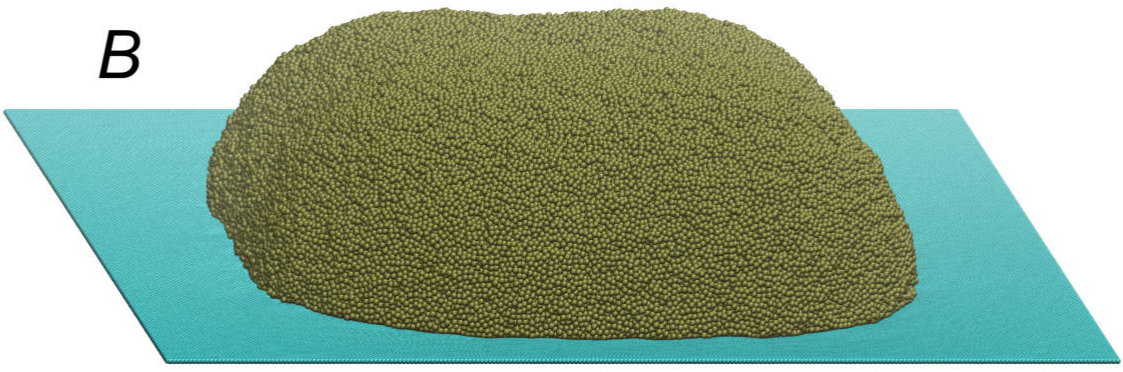}
  \includegraphics[height=2.05cm]{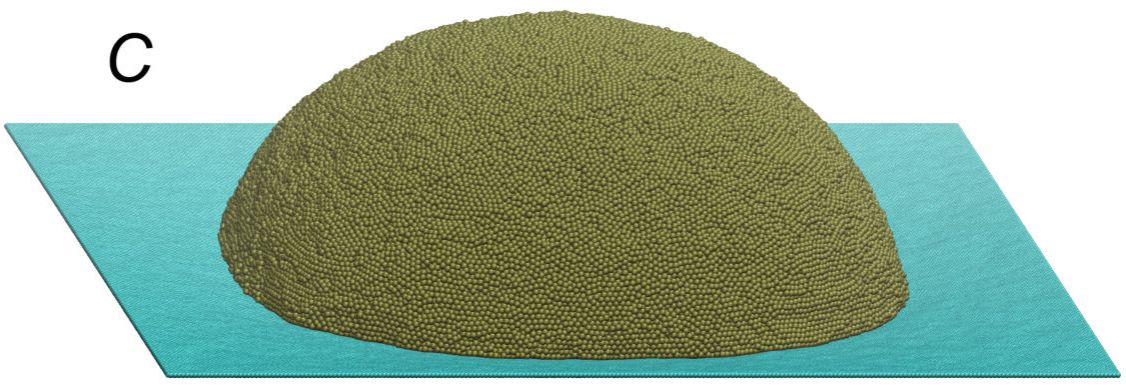}
  \includegraphics[height=2.05cm]{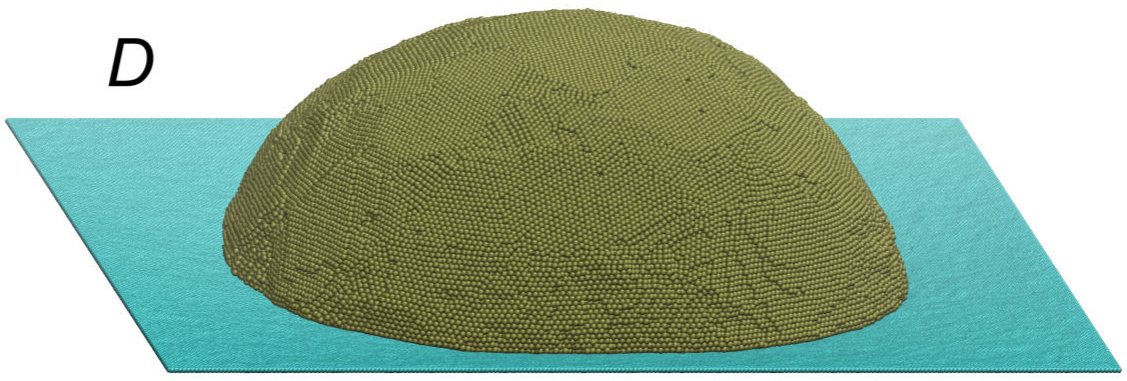}
  \caption{Snapshots of a NP containing 438976 atoms at the simulation start (0 ps), and at time intervals $\mathit{A}$ (10 ps), $\mathit{B}$ (70 ps), $\mathit{C}$ (140 ps), $\mathit{D}$ (600 ps) from Fig.~\ref{fig:time_deps}.}
  \label{fig:snapshots_vs_t}
\end{figure*}

The outlined atomistic models used surface configurations prepared in advance. Additionally, most focused on capturing bulk behavior to avoid size effects and to compare with continuum models of bulk surfaces. Therefore, their results may not apply to finite-size Pb NPs adsorbed on 2D surfaces. We describe our model, which eliminates these inconsistencies, below.

\section{Methodology}
\label{sec::methodology}

\subsection{Model}
\label{subsec::model}

At the start of the simulation, we place Pb atoms in the bulk FCC lattice in a cuboid slab above the SGr substrate, oriented parallel to the $xy$-plane, with the $x$- and $y$-axes along the zigzag and armchair edges of the SGr sheet, cf. Fig.~\ref{fig:snapshots_vs_t} (the snapshots were generated with VMD software~\cite{Humphrey1996}). The height of the cuboid is approximately 1/3 of its width/length. The (110) atomic plane of the FCC lattice is initially parallel to SGr. The perimeter carbon atoms of the substrate are fixed. We use periodic boundary conditions in all directions. Ultrahigh vacuum conditions are maintained. No external load is applied to the metal or carbon atoms.

$x-$ and $y$-dimensions of the carbon substrate are 50\% larger than those of the slab. The total number of atoms $N_{0}$ ranges from 1248 to 1354912, and the number of Pb atoms $N$ in a NP varies from 288 to 1124864. The initial arrangement of metal atoms into a cuboid may affect the NP's final shape. Nevertheless, we kept the initial shape fixed to focus on consistent conditions for NP formation and reduce the number of influencing factors. The Berendsen thermostat~\cite{Griebel2007} controls the system's temperature $T$. We integrate the equations of motion using the leapfrog method~\cite{Rapaport2004} with a $\Delta t = 0.2$~fs time step.

The embedded-atom method (EAM) potential describes the interatomic energy of metal atoms~\cite{Zhou2001}. Carbon atoms inside the graphene layer interact via a spring potential~\cite{Sasaki1996}. The model also accounts for the fact that Pb atoms weakly interact with perfect graphene (i.e., physisorb on graphene), as evidenced by both experimental and numerical results~\cite{Kuganathan2021,Gong2014dalton,Ma2011newjp,Li2016scirep}, and incorporates this into the Lennard-Jones (LJ) potential describing carbon--lead interactions. We use a 6--12 LJ potential to approximate the weak metal--carbon interactions. The LJ energy $\epsilon=0.01751$~eV and distance $\sigma=3.288$~\AA~were used~\cite{Li2016scirep}. Our model does not expect chemical bonding between metal atoms and graphene, even at high temperatures, because the spring potential is not designed to simulate dangling-bond formation. Nevertheless, this limitation should not lead to unrealistic behavior, as experiments do not indicate the formation of chemical Pb--C bonds in defect-free graphene sheets at the 300--600~K temperatures used in our simulations~\cite{Gong2014dalton}.

Our in-house MD code is based on the highly parallel graphics processing unit (GPU) computing platform NVIDIA CUDA~\cite{lindholm2008nvidiacuda,nvidia2020cuda}, which allows treating systems of millions of atoms on a single GPU in a reasonable time. Computations ran on NVIDIA GeForce RTX 5090, NVIDIA RTX 6000 Ada Generation, and NVIDIA GeForce RTX 4070 Laptop GPUs.

\subsection{Data analysis}
\label{subsec::analysis}

\begin{figure}[!htb]
\centering
  \includegraphics[height=3.75cm]{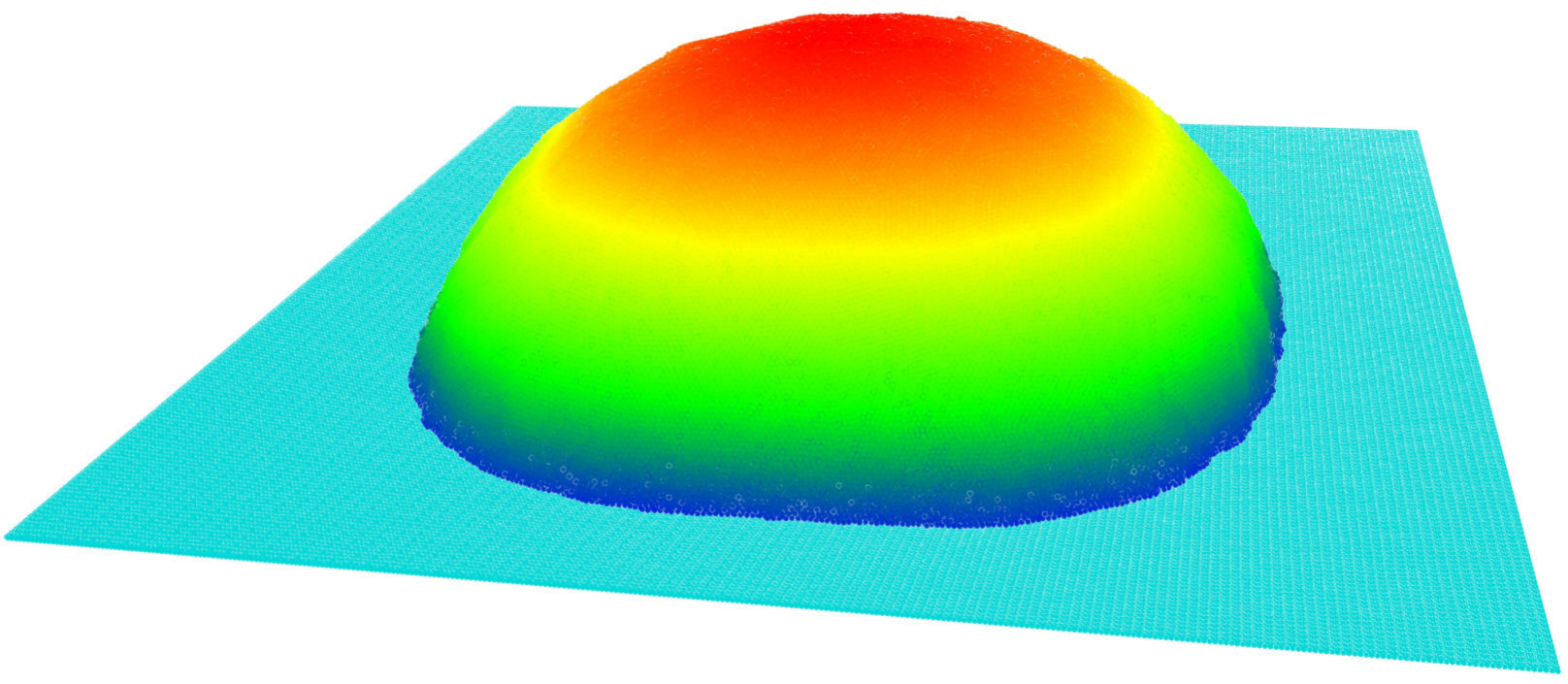}
  \includegraphics[height=3.75cm]{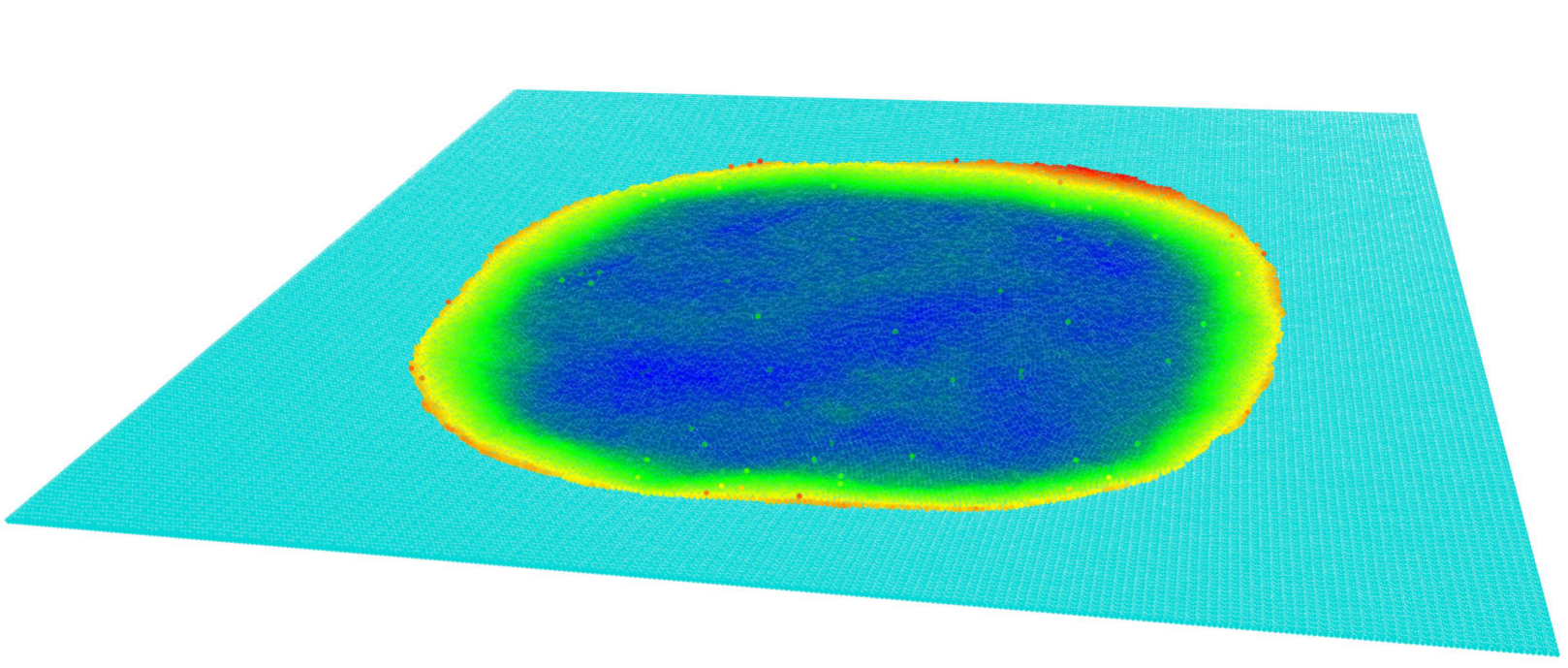}
  \caption{Point clouds (PCDs) of the largest solidified NP (top) with 1124864 atoms and its contact surface atomic layer (bottom). The NP heatmap shows the atomic $z$-coordinate, where the blue--red color interval depicts the smallest--largest coordinates, respectively. PCD figures are produced with CloudCompare~\cite{CloudCompare}.}
  \label{fig:np_pcd}
\end{figure}

In addition to measuring standard MD quantities, such as energies and forces, we performed further data analysis. We consider distributions $P_{\mathrm{NP}}(F_z)$ and histograms $H_{\mathrm{NP}}(F_z)$ of magnitudes of the intersurface normal force $F_z$ acting on each atom of the NP, as well as acting only on the atoms located in the contact layer (denoted as $P(F_z)$ and $H(F_z)$ in this case). We defined the NP contact atomic layer as follows: assuming a smooth NP contact surface, we split the simulation box into 3D cells with a lateral size equal to the Pb nearest-neighbor distance, $a_{nn} = 3.5$~\AA~\cite{Zhou2001}. We set the $z$ threshold to $0.9a_{nn}$ to capture variations in the $z$ coordinate of atoms in neighboring cells. The bottom layer includes all Pb atoms in the lowest cells with the smallest $z$ value, as well as those within the $z$ threshold relative to these atoms. Fig.~\ref{fig:np_pcd} displays the largest NP and its contact atomic layer.

NP total surface area $S$ and volume $V$ were calculated by reconstructing closed triangular surfaces from the atomic coordinates. We employed two approaches: voxel-based surface reconstruction~\cite{Lorensen1987acm,Lewiner2003jgt} and the three-dimensional alpha-shape method~\cite{Edelsbrunner1983}. Both approaches gave very similar results, within a few percent, so we present the voxel-based results. The voxel method achieved an RMS error of less than 6\% for the surface area-to-volume ratio $S_{\mathrm{V}} = S/V$.

One important question raised in~\cite{Cheng2010triblett} is the stability of the contact area under thermal fluctuations of atomic positions, which cause instantaneous changes in the contact area. Continuum theories typically utilize the time-averaged forces and displacements. However, atomistic models often consider both time-averaged and instantaneous values. To be consistent with~\cite{Cheng2010triblett}, we consider instantaneous histograms in the kinetics plots. For scaling plots, we ensemble-average the data over 3--7 runs for each NP size to improve statistics.

Spectral analysis utilized the 2D $C(q_{x}, q_{y})$ and 1D $C(q)$ power spectral densities (PSDs) of the normal intersurface forces $F_z(x, y)$ acting on the contact NP atomic layer~\cite{Jacobs2017stmp}:
\begin{equation}
  C(q_{x}, q_{y}) = (L_{x} L_{y})^{-1} \bigl\lvert \tilde{F_z}(x, y) \bigr\rvert^{2},
  \label{eq:psd_h_x_y}
\end{equation}
where $q_{x}$, $q_{y}$ are the spatial frequencies and $L_{x}$, $L_{y}$ are the linear sizes of the region of interest in the $x$- and $y$-directions, respectively, and $\tilde{F_z}(x, y)$ is the Fourier transform of the normal force field $F_z(x, y)$:
\begin{equation}
  \tilde{F_z}(x, y) = \frac{L_{x} L_{y}}{N_{x} N_{y}} \sum_{x,y} F_z(x, y) e^{-i(q_{x} x + q_{y} y)},
  \label{eq:ft_h_x_y}
\end{equation}
where $N_{x}$, $N_{y}$ are the numbers of grid points in the $x$- and $y$-direction, respectively. In our simulations, $L_{x}$ is the lateral size of the NP in the $x$-direction, which is nearly identical to its size in the $y$-direction because the NP perimeter is almost round, so we refer to $L_{x}$ as the NP diameter. Quantities $q_{x} = 1/\lambda_{x}$ and $q_{y} = 1/\lambda_{y}$ are the spatial frequencies. The factor $2 \pi$ is omitted for convenience in transforming them into the corresponding wavelengths $\lambda_{x}$ and $\lambda_{y}$. The frequency unit of PSDs $1.76 \cdot 10^{-3}$~\AA$^{-1}$ was chosen so that the frequency of the graphene covalent bond length of 1.42~\AA~corresponds to 400~units. 1D $C(q)$ is an analog of the isotropic PSD~\cite{Persson2005jpcm,Jacobs2017stmp}. It is a histogram of $C(q_{x}, q_{y})$ of the radial frequency $q = \sqrt{q_{x}^{2} + q_{y}^{2}}$, so it has the same units of measurements as $C(q_{x}, q_{y})$.

The fitting quality of the plots was assessed using the following coefficients.

\begin{itemize}
\item Root mean square error (RMS), with smaller values indicating better fits; 0 indicates a perfect fit. RMS is useful for the high-probability region of the distribution.

\item Logarithmic root mean square error (log RMSE), where smaller values correspond to better fits with an ideal value of 0. It is useful for distributions spanning several orders of magnitude, where precise matching of the low-probability tails is important.

\item Logarithmic coefficient of determination (\(R^2_{\log}\)). Values closer to 1 are better, with \(R^2_{\log}=1\) being the perfect fit.

\item Logarithmic Akaike information criterion (log AIC). Smaller is better, including negative values. Differences in AIC are more informative than its absolute magnitude.
\end{itemize}

\begin{figure}[!htb]
\includegraphics[height=4.7cm]{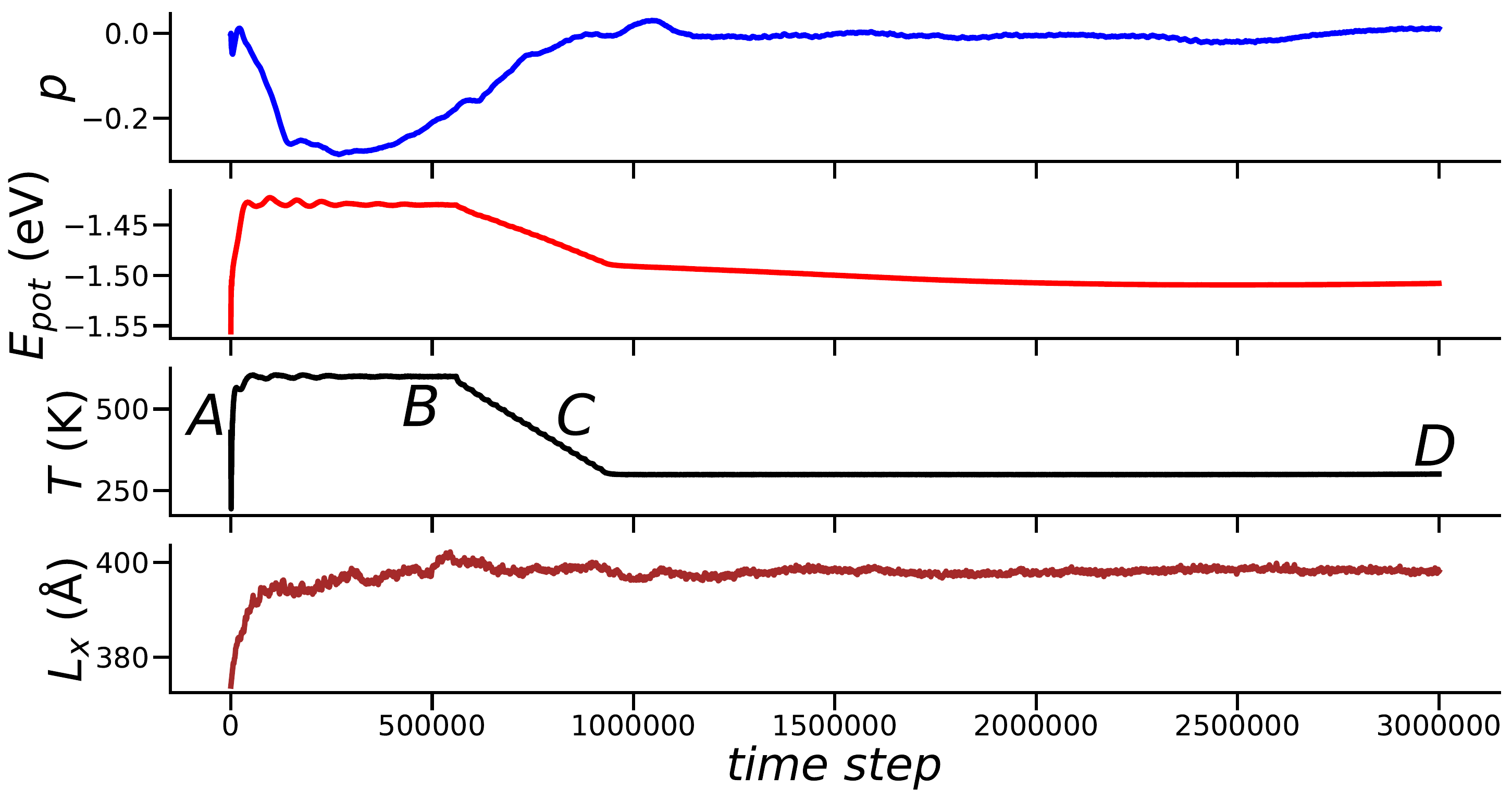}
\caption{Time dependence of the impulse $p$, potential energy $E_{pot}$, system temperature $T$, and the linear NP size $L_x$ of a NP containing 438976 atoms. Labels depict time intervals of particular temperature modes: $\mathit{A}$---fast heating with quickly increasing $T$, $\mathit{B}$---heating with constant $T$, $\mathit{C}$---cooling, $\mathit{D}$---equilibration.}
\label{fig:time_deps}
\end{figure}

\begin{figure}[!htb]
\centering
  \includegraphics[height=7cm]{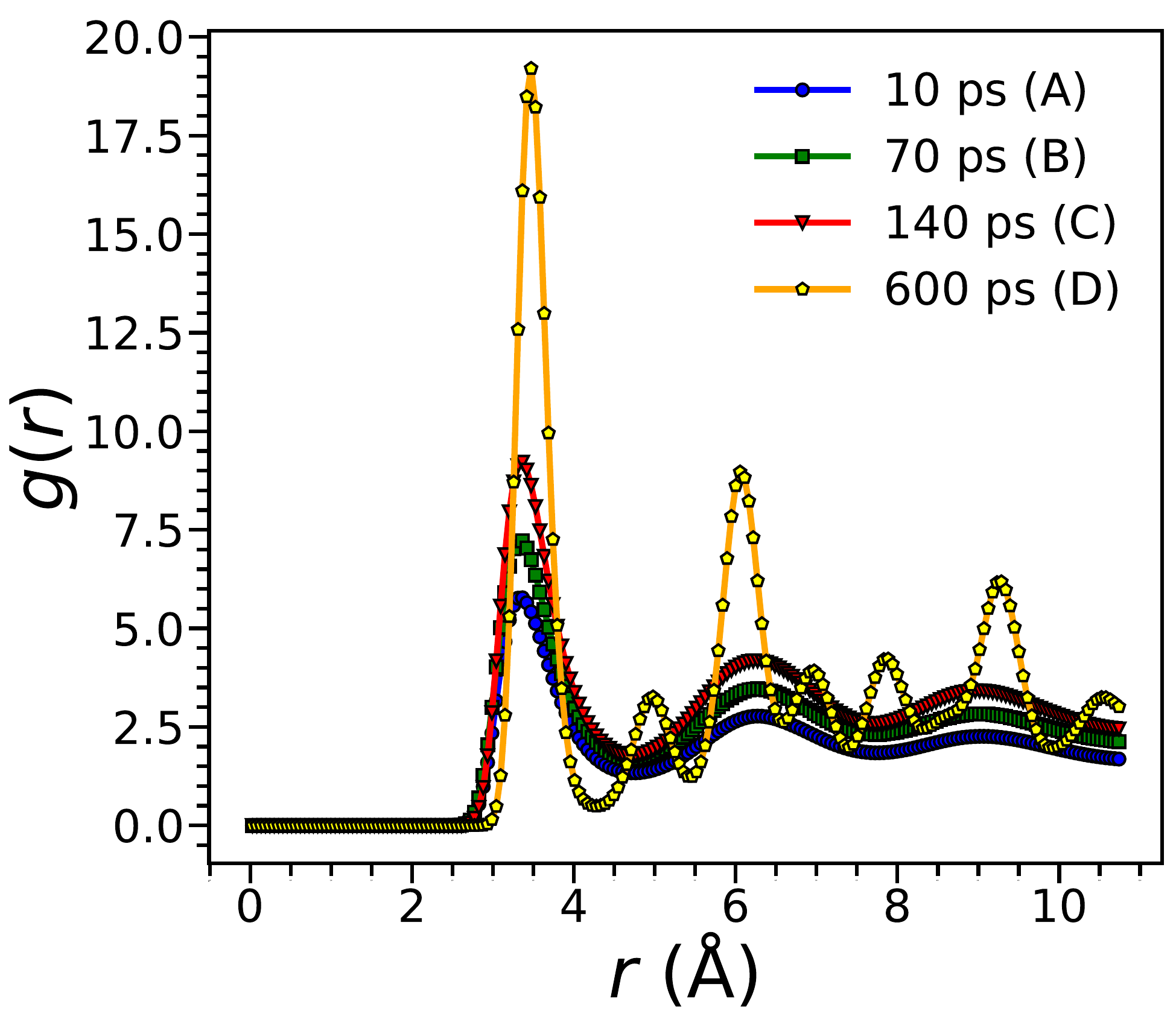}
  \caption{RDFs at time intervals $\mathit{A}$, $\mathit{B}$, $\mathit{C}$, and $\mathit{D}$ in Fig.~\ref{fig:time_deps} of a Pb NP containing 438976 atoms.}
  \label{fig:rdfs_16_05}
\end{figure}

\section{Results and discussion}
\label{sec::results}

\subsection{NP self-assembly}
\label{subsec:self_assembly}
Bulk~\cite{Yang2008euphyslett,Cheng2010triblett,Mo2009nature,Mo2010prb,Campana2006prb,
Yang2008prl,Yang2008jpcm} and nano-sized~\cite{Gao2022fic,Cihan2023prl,Oo2024} CM models usually predefine the shape and atomic arrangements of the contacting surfaces. This limits the model's realism, because, in practice, NP surfaces, especially larger ones, may not have predefined ideal atomic configurations and can exhibit atomic-scale roughness~\cite{Khomami2026asusc,Prodanov2026pccp}. Alternatively, one can emulate NP formation on a graphene surface using established approaches, such as photo-induced electrodeposition~\cite{Xia2020nanoscale} or chemical vapor deposition~\cite{Shtepliuk2017scirep,Shtepliuk2018pccp}. In this work, we obtain NPs using a technique that simulates the final phase of thermal dewetting of thin metal films, where the thin film separates into distinct islands that self-assemble into a liquid droplet and eventually solidify during cooling~\cite{Badan2022nm,Neek-Amal2009,Harsha2024,Kim2023}.

We mimicked the thermal dewetting procedure by adjusting the system temperature $T$ via the Berendsen thermostat. Fig.~\ref{fig:time_deps} shows an example of the time evolution of $T$ and other system parameters for a NP containing 438976 atoms. Initially, $T$ rises quickly from its initial value to 600~$K$, which roughly corresponds to the bulk melting $T$ of Pb; this is the time interval marked with $\mathit{A}$. The target maximum $T$ is the model input parameter. During period $\mathit{B}$, $T$ is constant until the NP's height reaches 30-60\% of the initial metal slab, depending on the system size. Cooling then continues during period $\mathit{C}$ until $T$ reaches 300~K. Finally, equilibration occurs over the time interval $\mathit{D}$. The shape and linear size of the NP depend on when cooling is applied and how fast it occurs. We heat and cool with gradual $T$ changes, and the temperature-time profile matches the experimental plots~\cite{Badan2022nm}.

Fig.~\ref{fig:snapshots_vs_t} shows snapshots of the time evolution of a NP containing 438976 atoms at time instances from each of the mentioned intervals, while Fig.~\ref{fig:rdfs_16_05} shows the corresponding radial distribution functions (RDFs) of the NP. The RDFs suggest that the NP is molten during intervals $\mathit{A}$, $\mathit{B}$, and $\mathit{C}$. During equilibration at $\mathit{D}$, it eventually solidifies into a polycrystalline atomic configuration. Equilibration lasted for more than 1M time steps at constant temperature. The total run varied from 1.5 to 5 million time steps, corresponding to 300~ps and 1~ns, respectively.

MD reproduces the full dynamics of the non-equilibrium process, allowing measurement of the kinetics of interfacial forces throughout the simulation in both the liquid and solid states of the NP. We consider this next.

\subsection{Kinetics}
\label{subsec:kinetics}

\subsubsection{Intersurface forces}
\label{subsubsec::kinetics_forces_np}

\begin{figure}[!ht]
\centering
  \includegraphics[height=3.8cm]{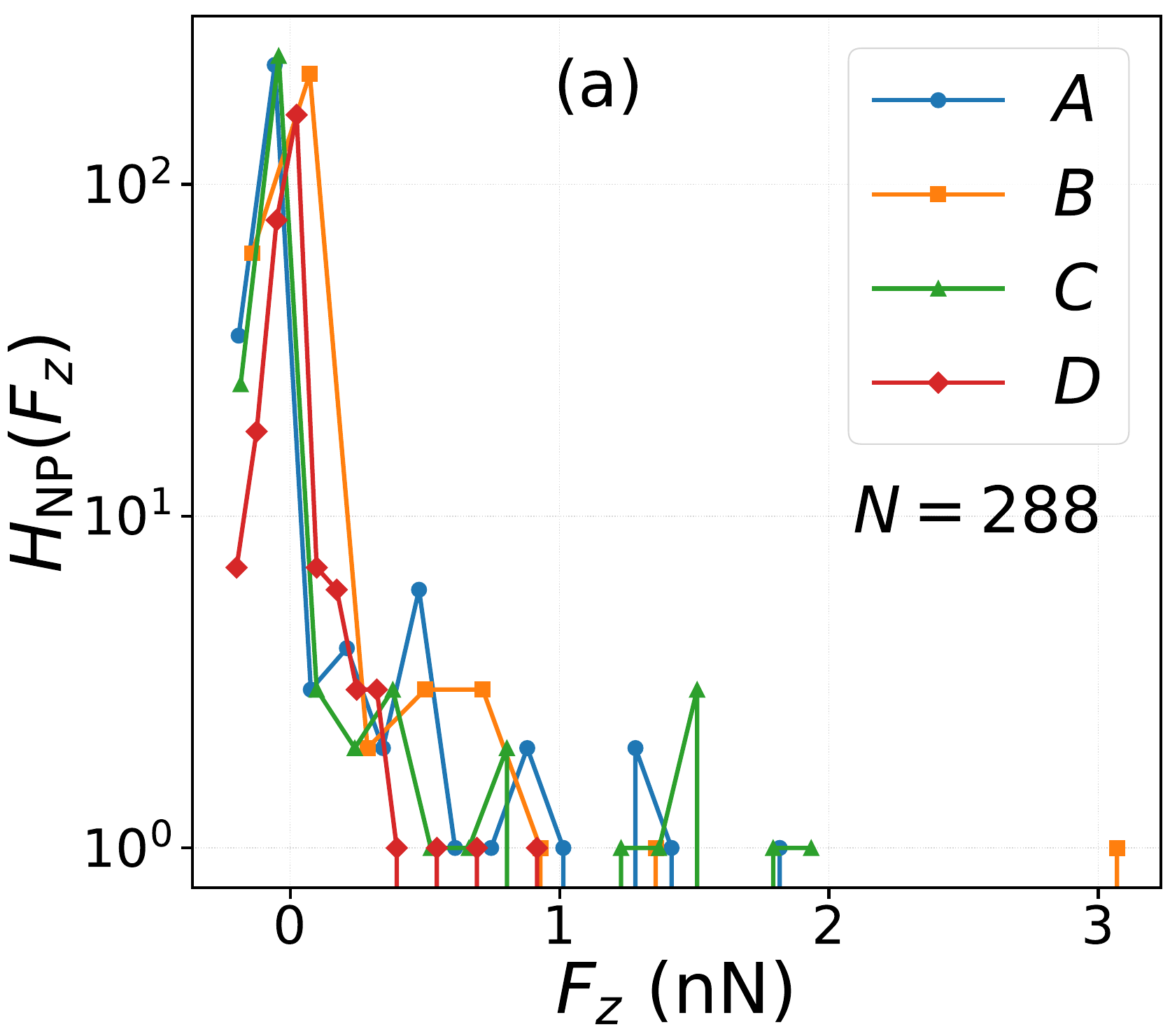}
  \includegraphics[height=3.8cm]{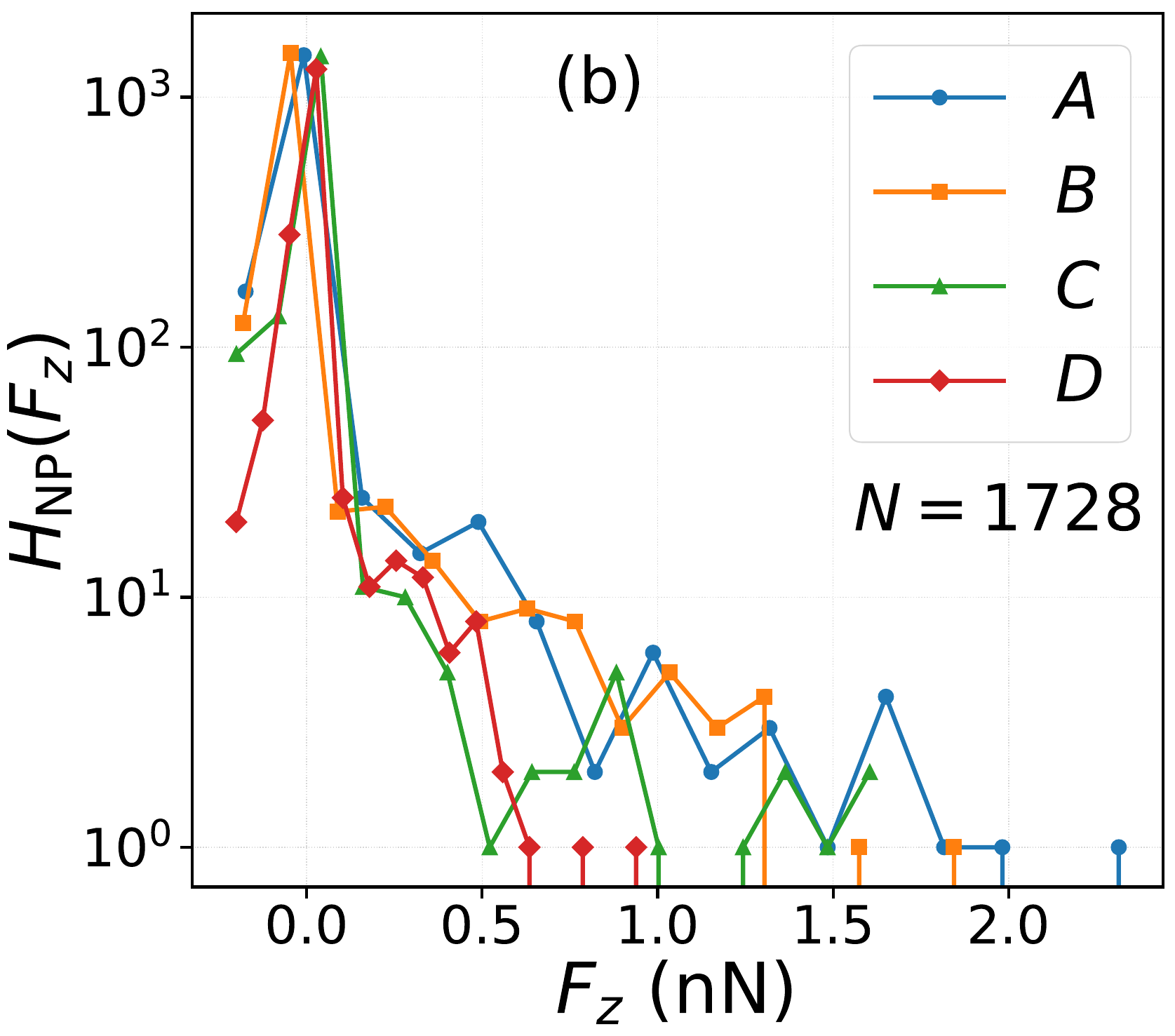}
  \includegraphics[height=3.8cm]{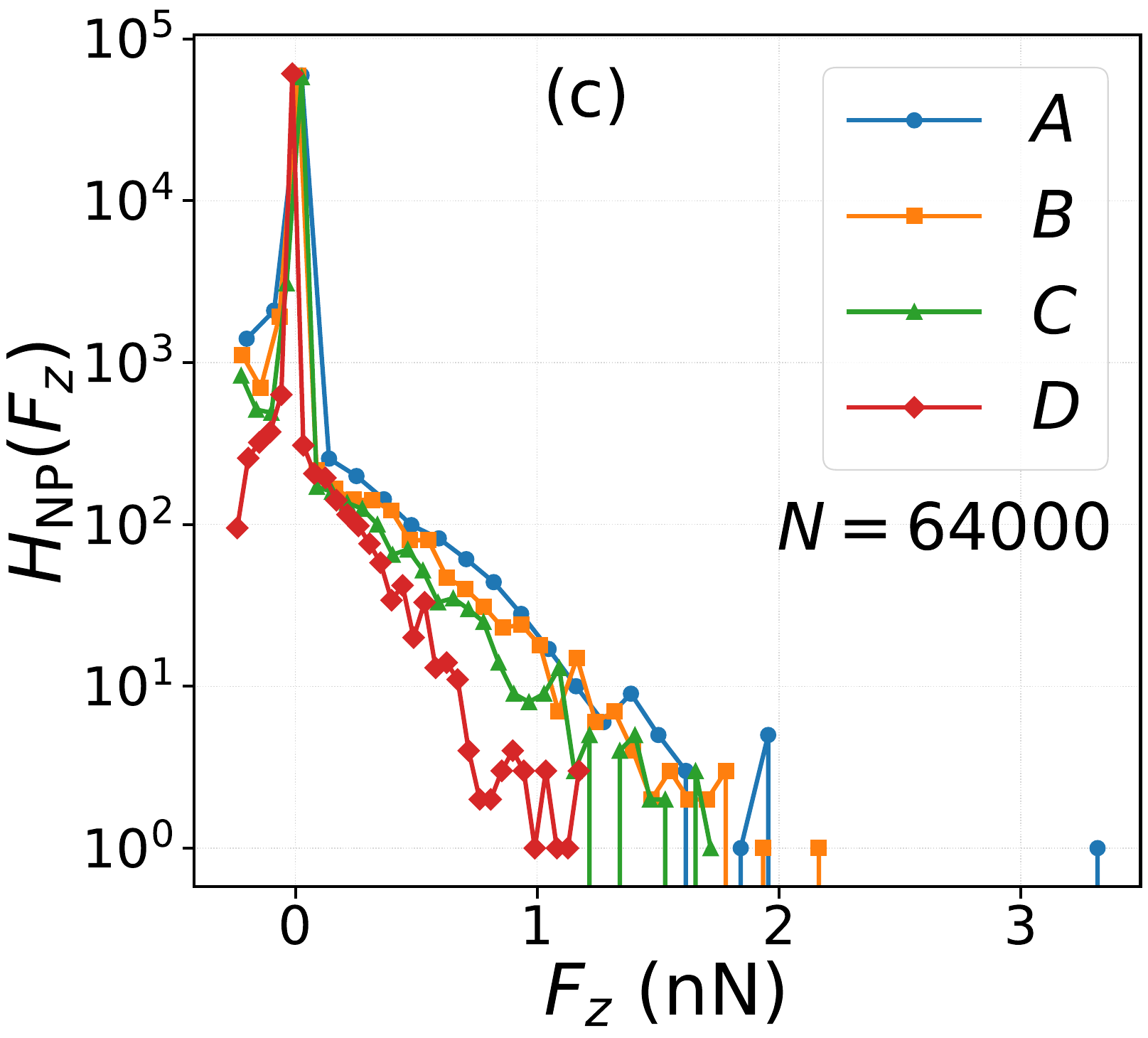}
  \includegraphics[height=3.8cm]{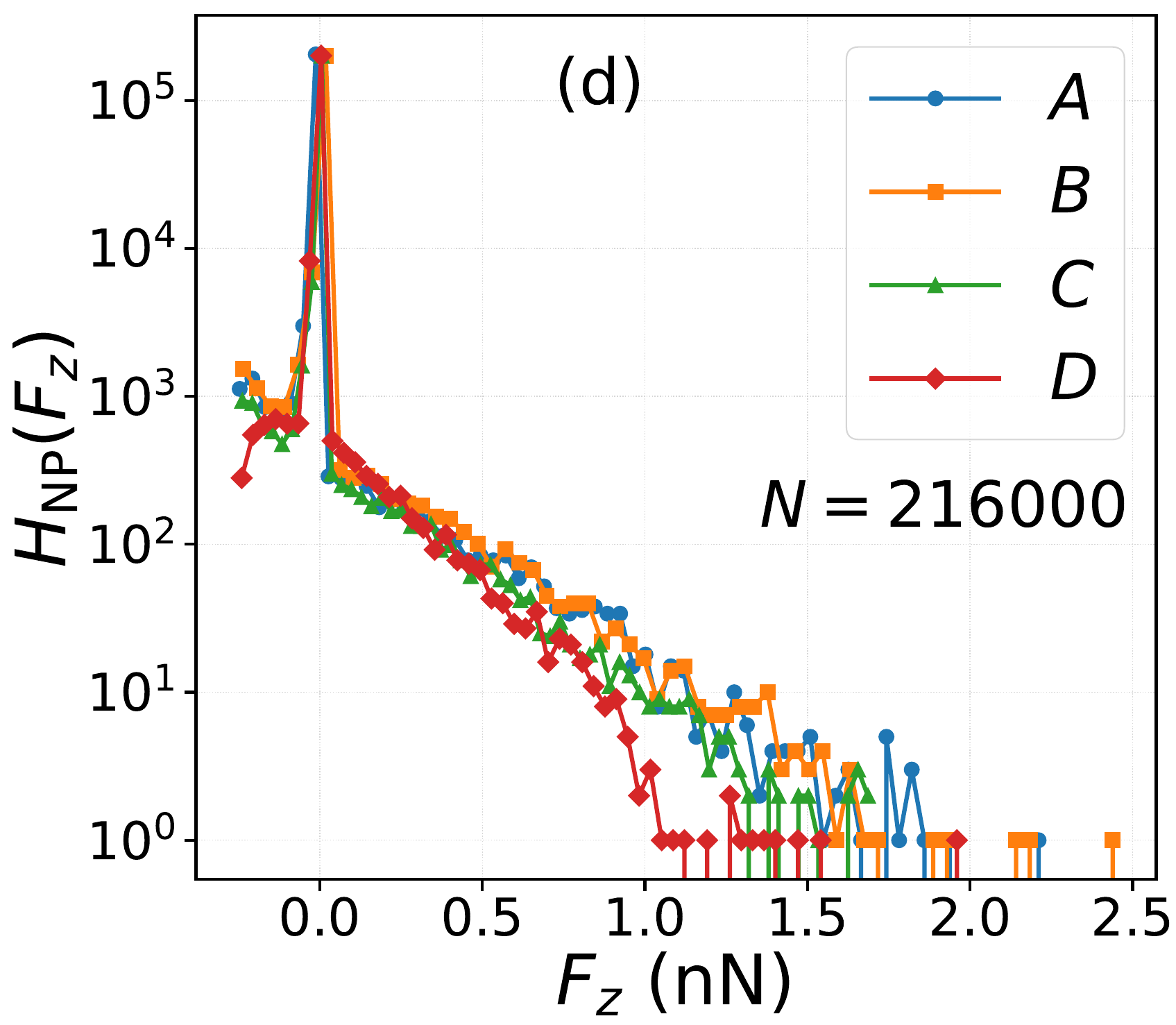}
  \includegraphics[height=3.8cm]{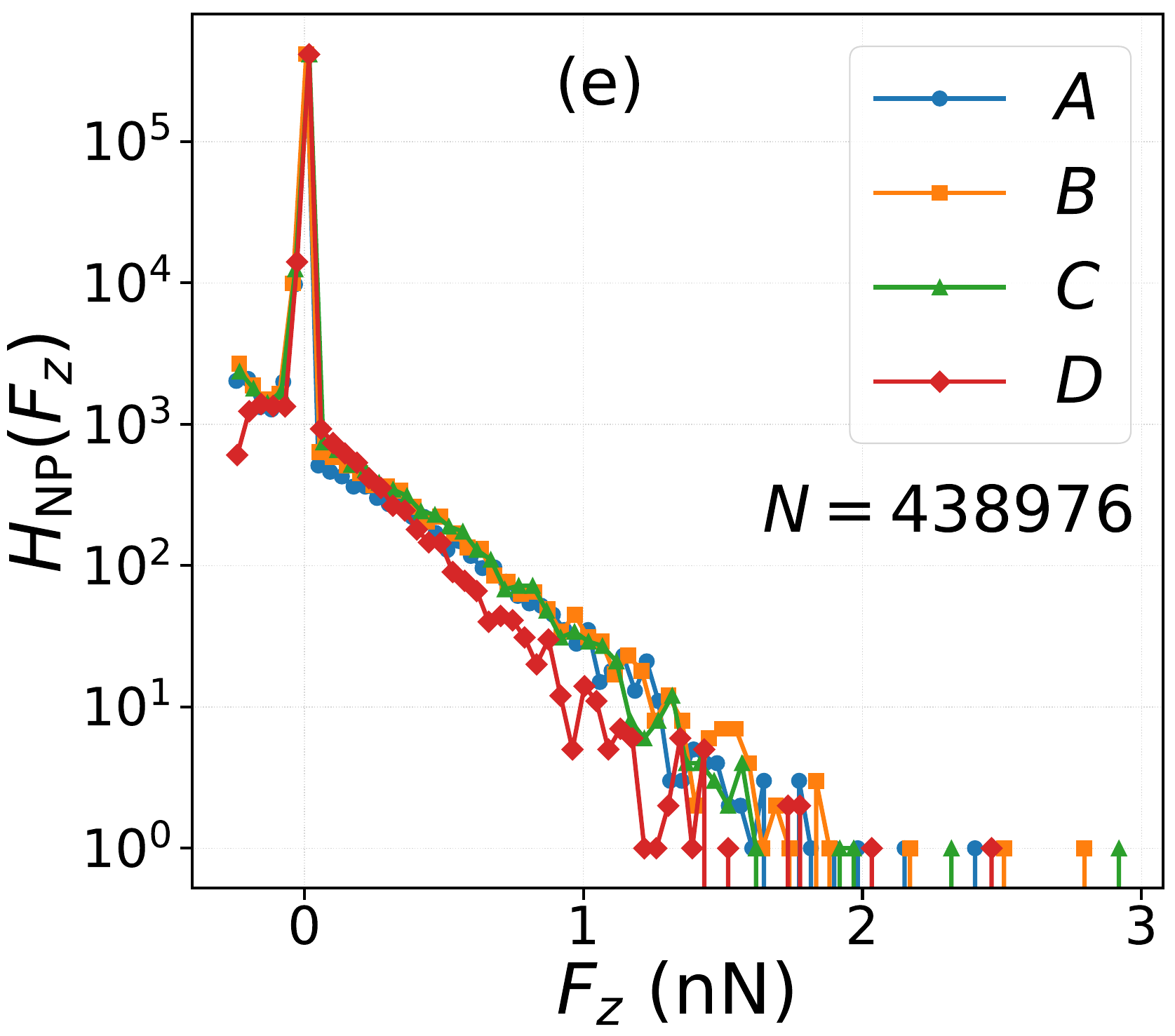}
  \includegraphics[height=3.8cm]{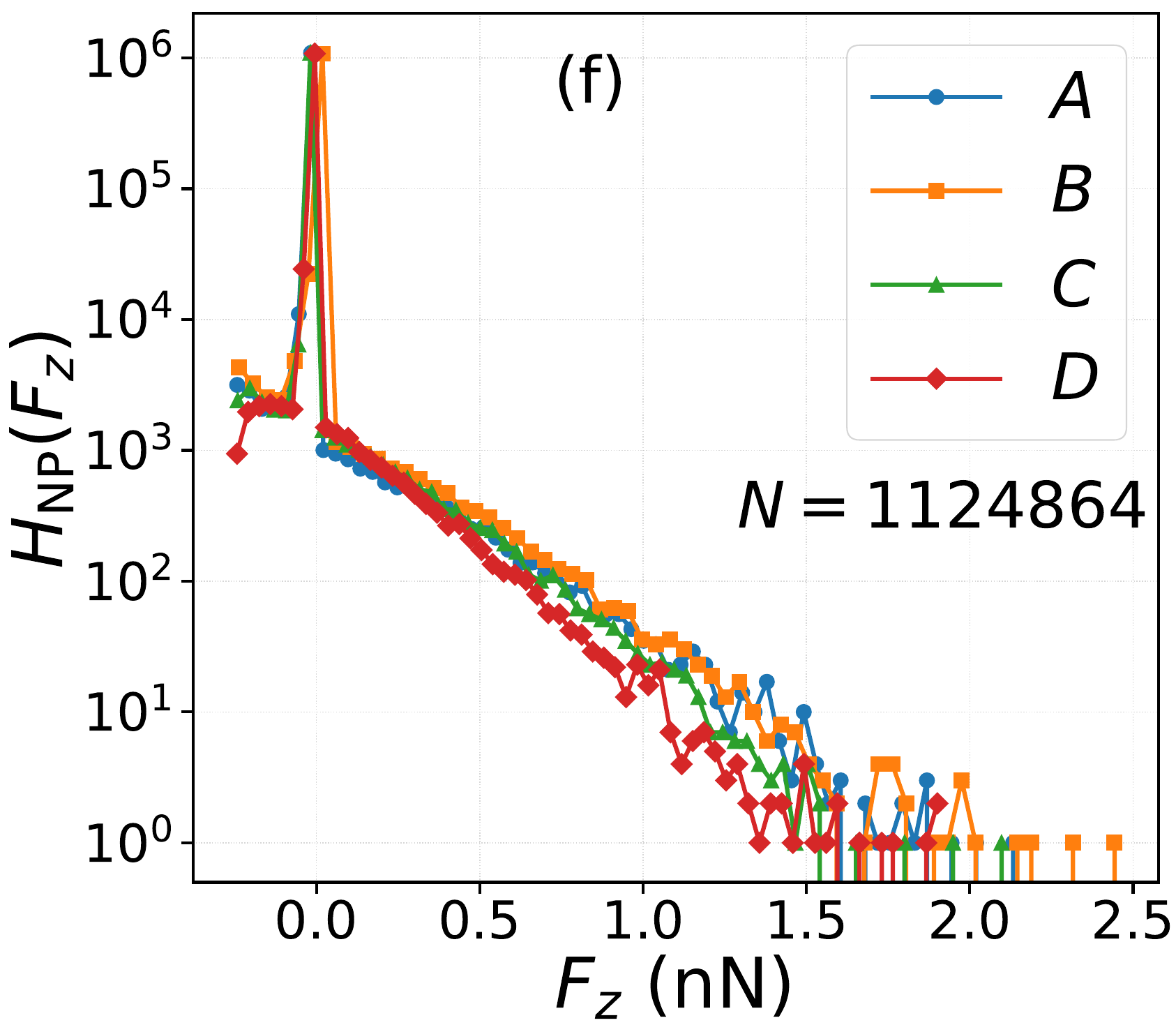}
  \caption{Histograms $H_{\mathrm{NP}}(F_{z})$ of the normal force $F_z$ acting from the carbon atoms onto each Pb atom at time intervals $\mathit{A}$, $\mathit{B}$, $\mathit{C}$, and $\mathit{D}$ for NPs containing 288 (a), 1728 (b), 64000 (c), 216000 (d), 438976 (e), and 1124864 (f) atoms.}
  \label{fig:np_pcfz_vs_t}
\end{figure}

\begin{figure}[!ht]
\centering
  \includegraphics[height=3.8cm]{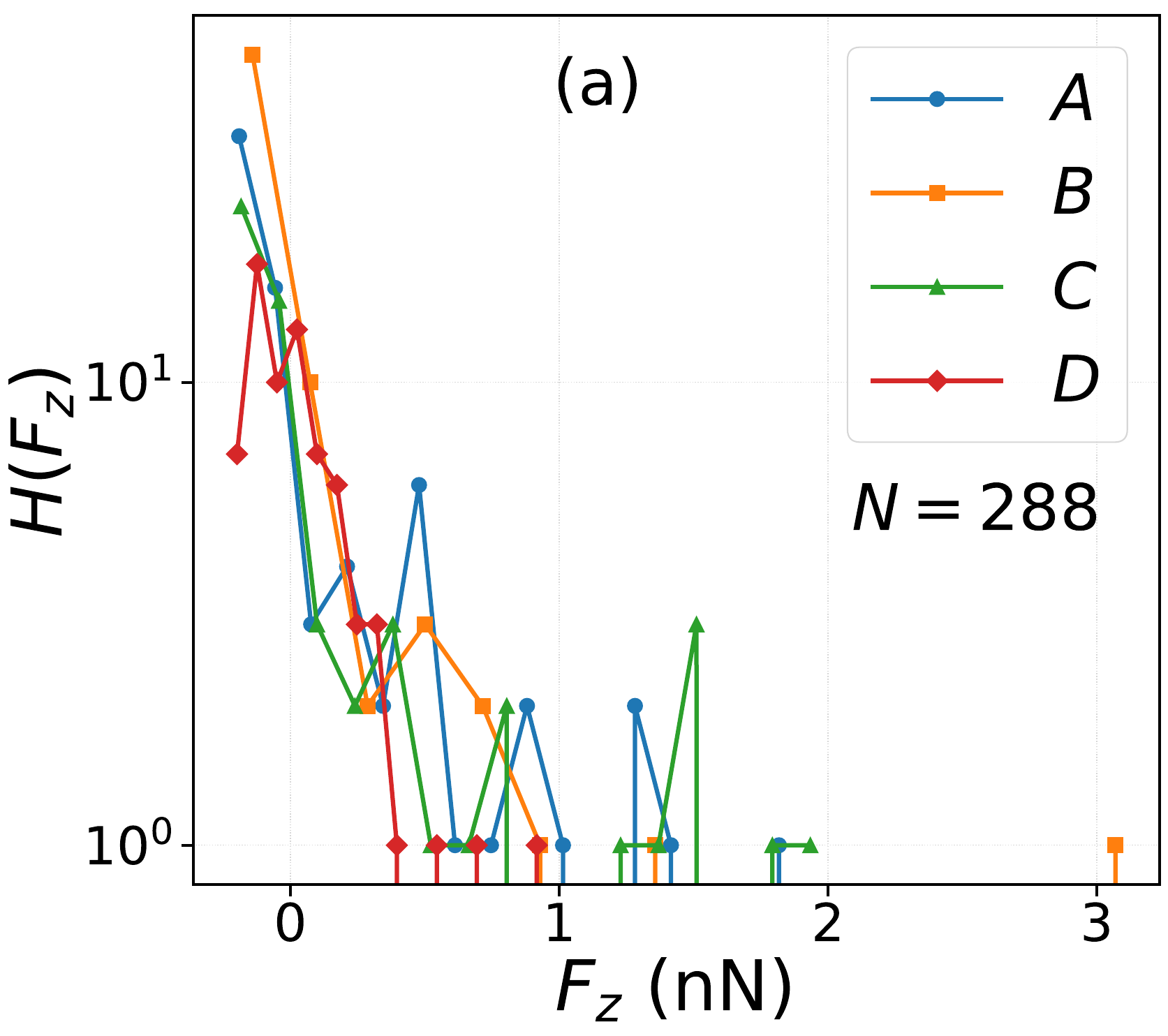}
  \includegraphics[height=3.8cm]{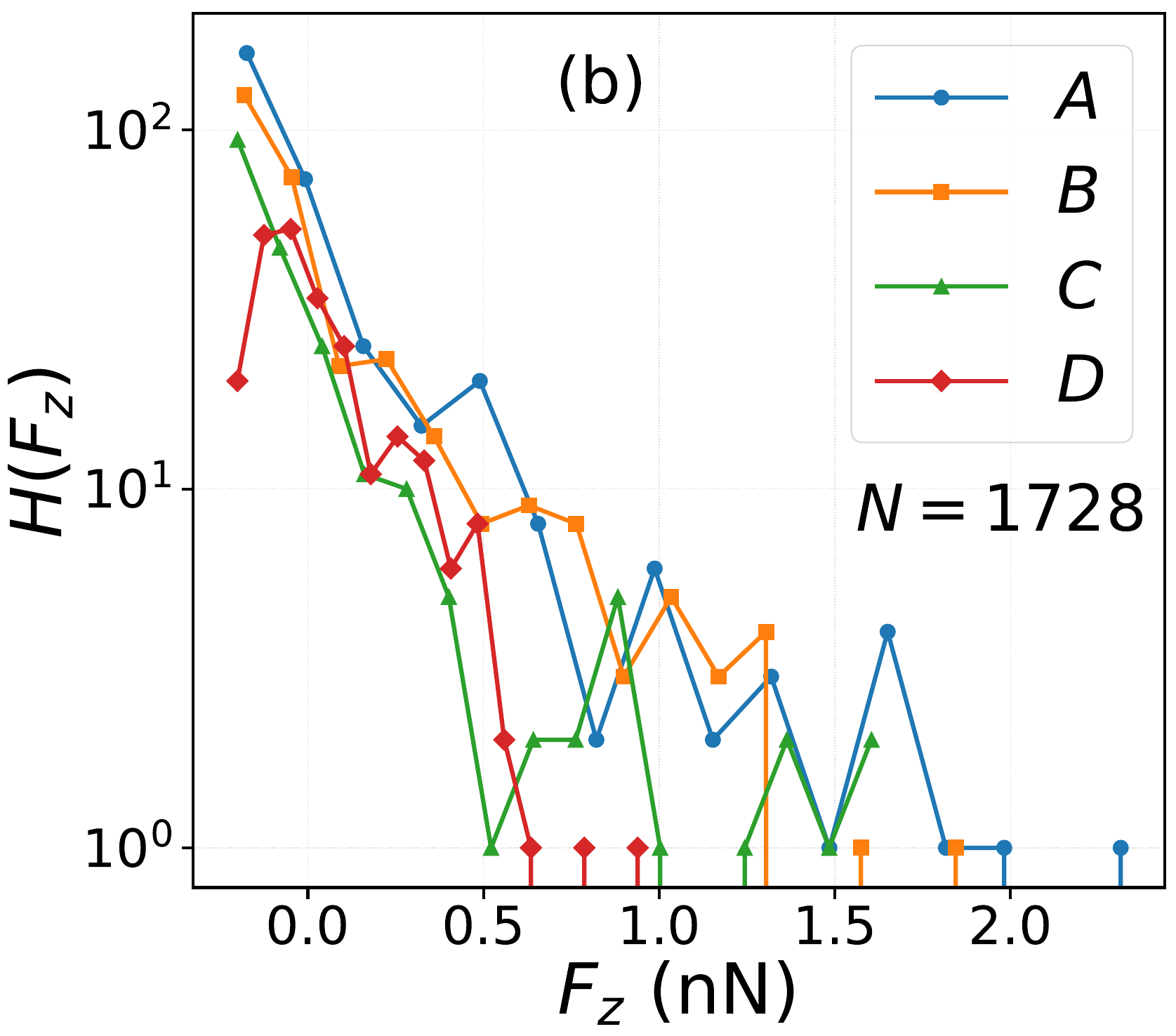}
  \includegraphics[height=3.8cm]{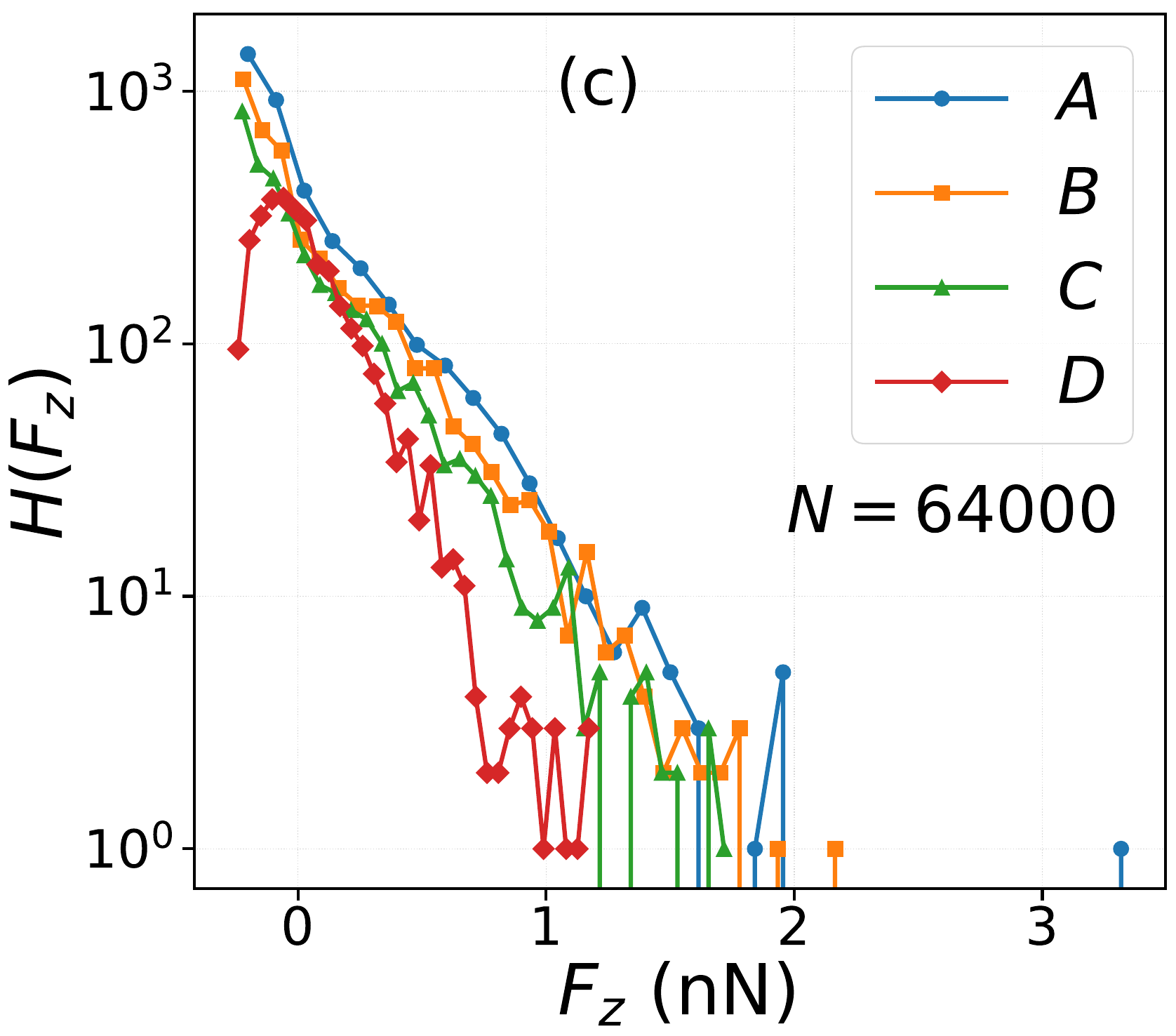}
  \includegraphics[height=3.8cm]{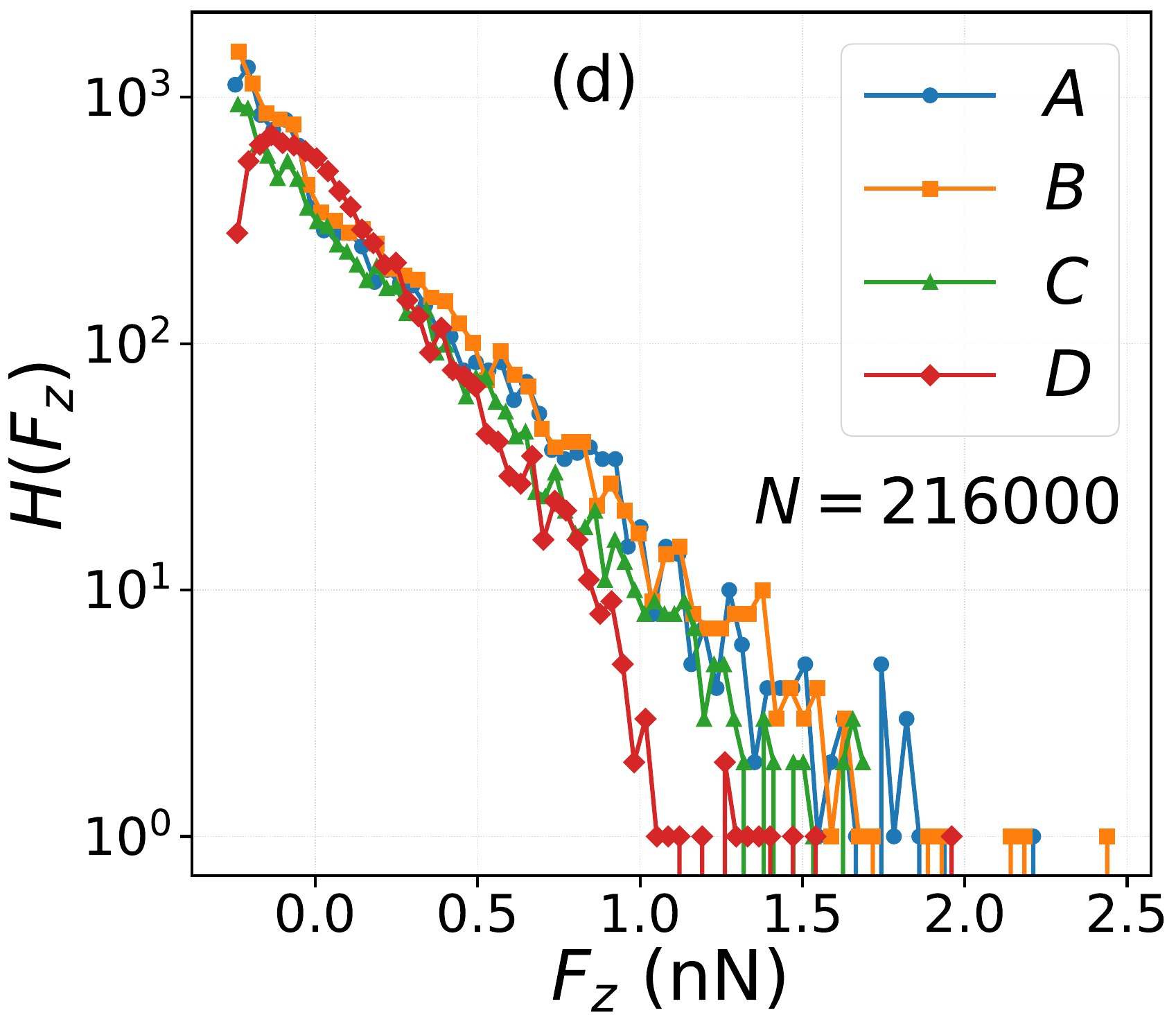}
  \includegraphics[height=3.8cm]{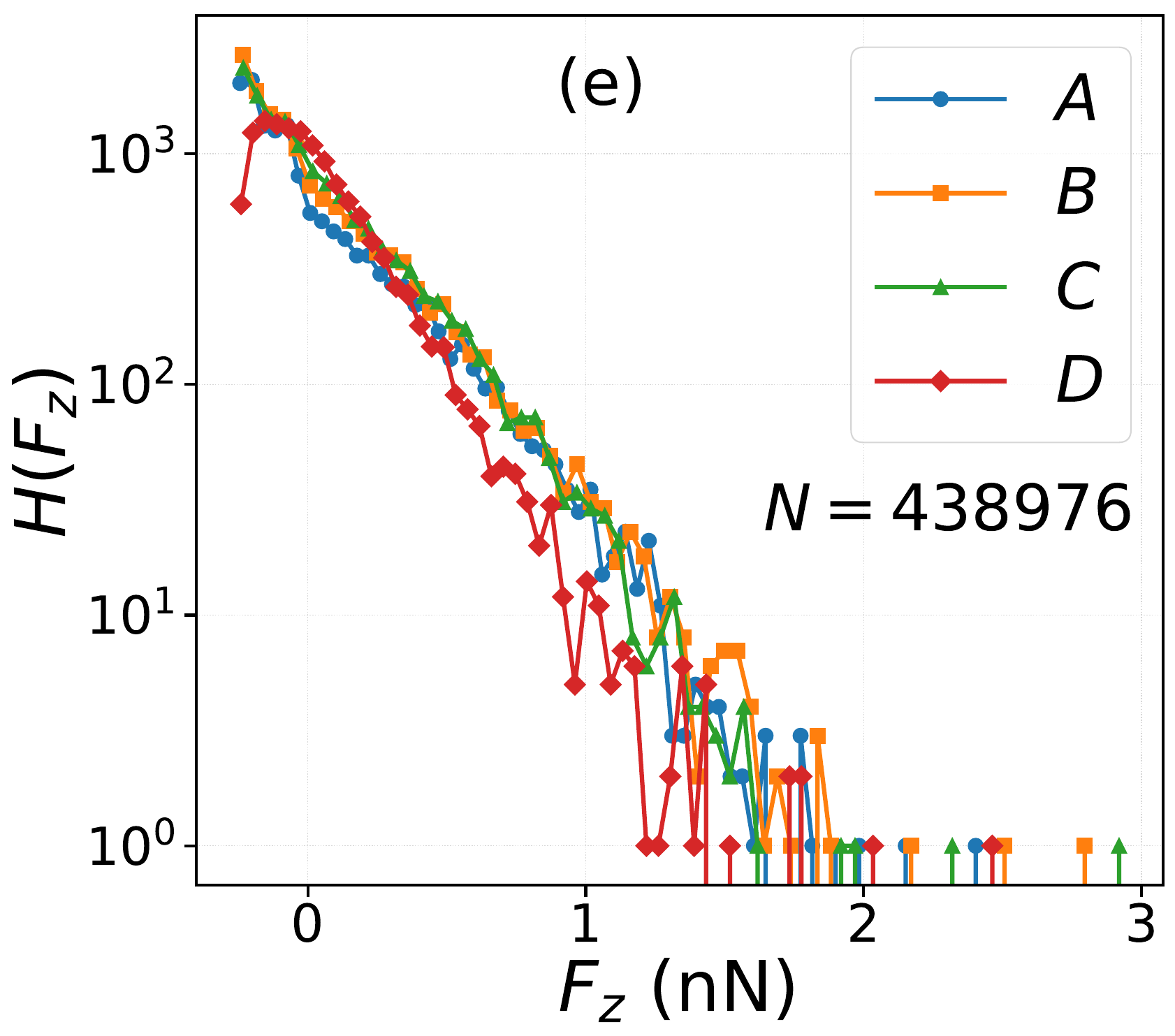}
  \includegraphics[height=3.8cm]{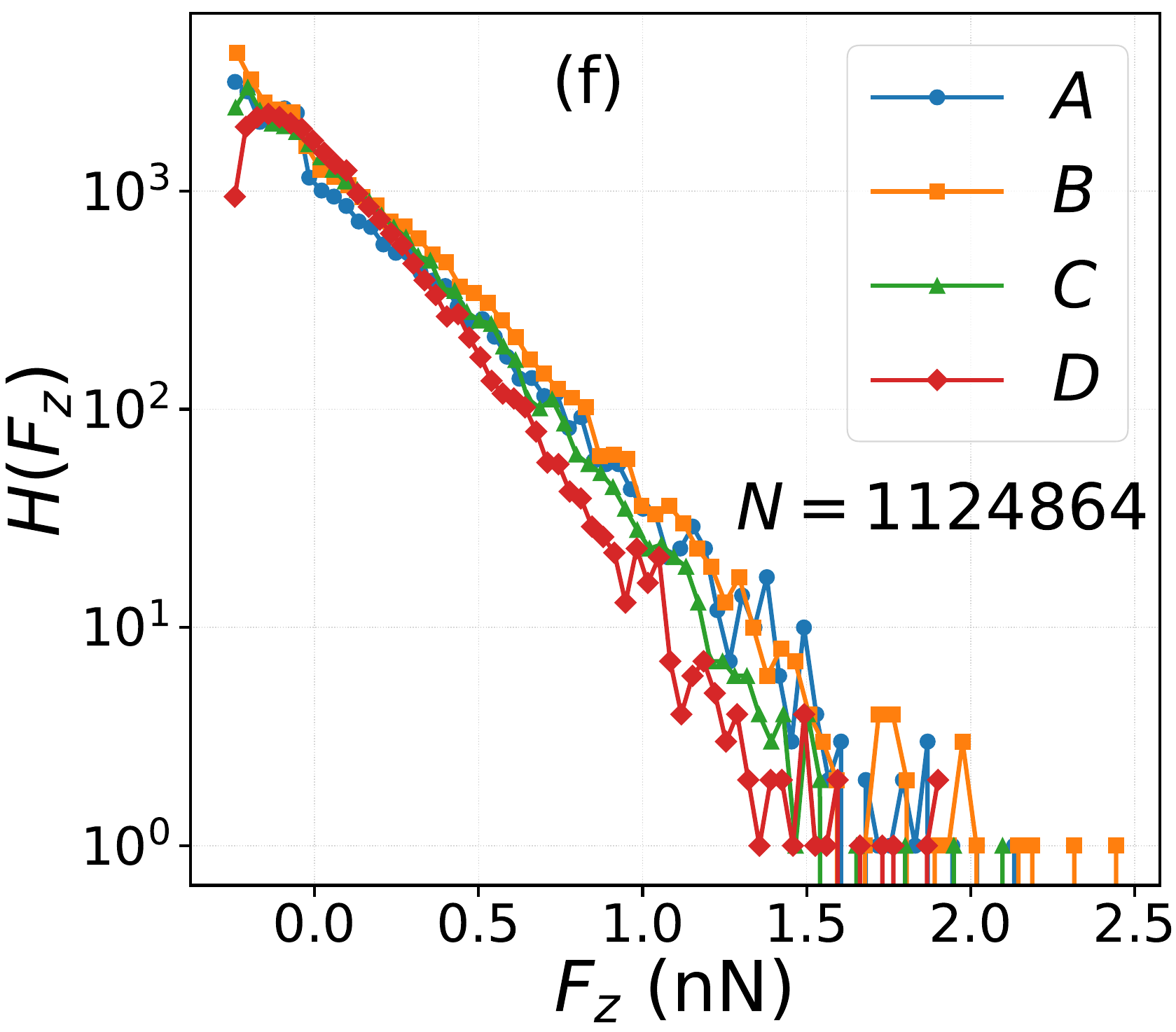}
  \caption{$H{(F_{z})}$ at time intervals $\mathit{A}$, $\mathit{B}$, $\mathit{C}$, and $\mathit{D}$ for NPs containing 288 (a), 1728 (b), 64000 (c), 216000 (d), 438976 (e), and 1124864 (f) atoms.}
  \label{fig:bot_pcfz_vs_t}
\end{figure}

\begin{figure}[!ht]
\centering
  \includegraphics[height=3.7cm]{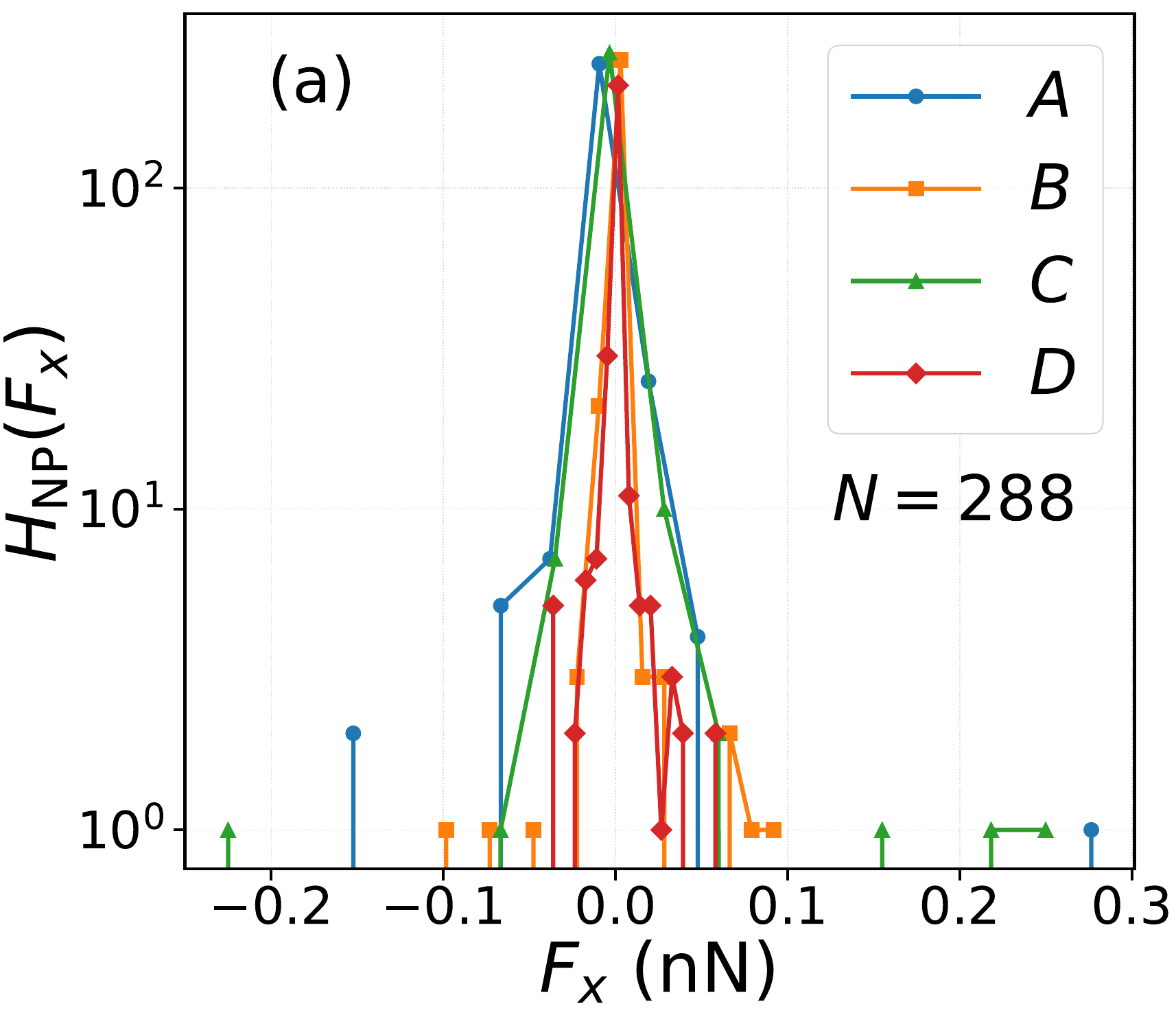}
  \includegraphics[height=3.7cm]{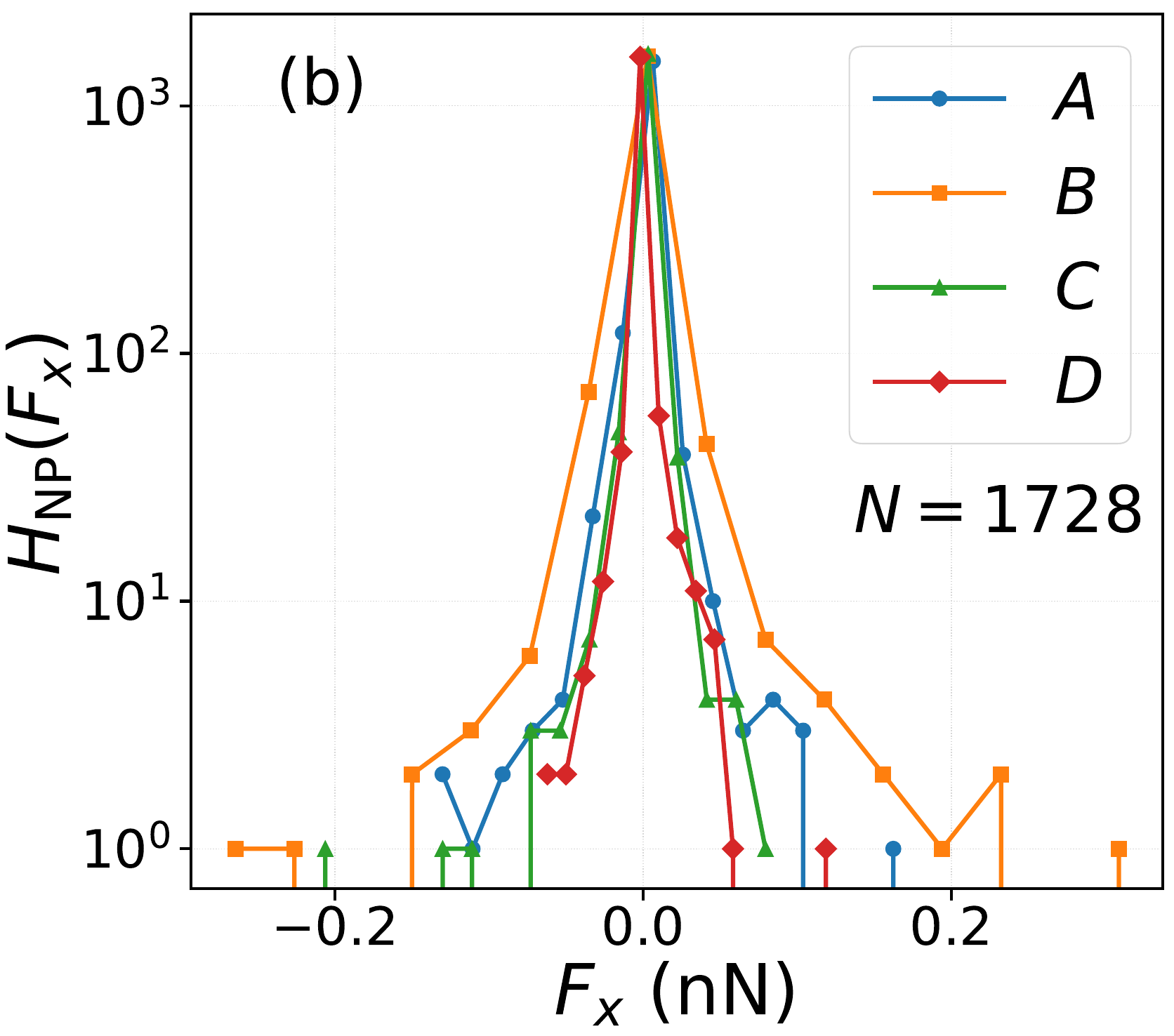}
  \includegraphics[height=3.7cm]{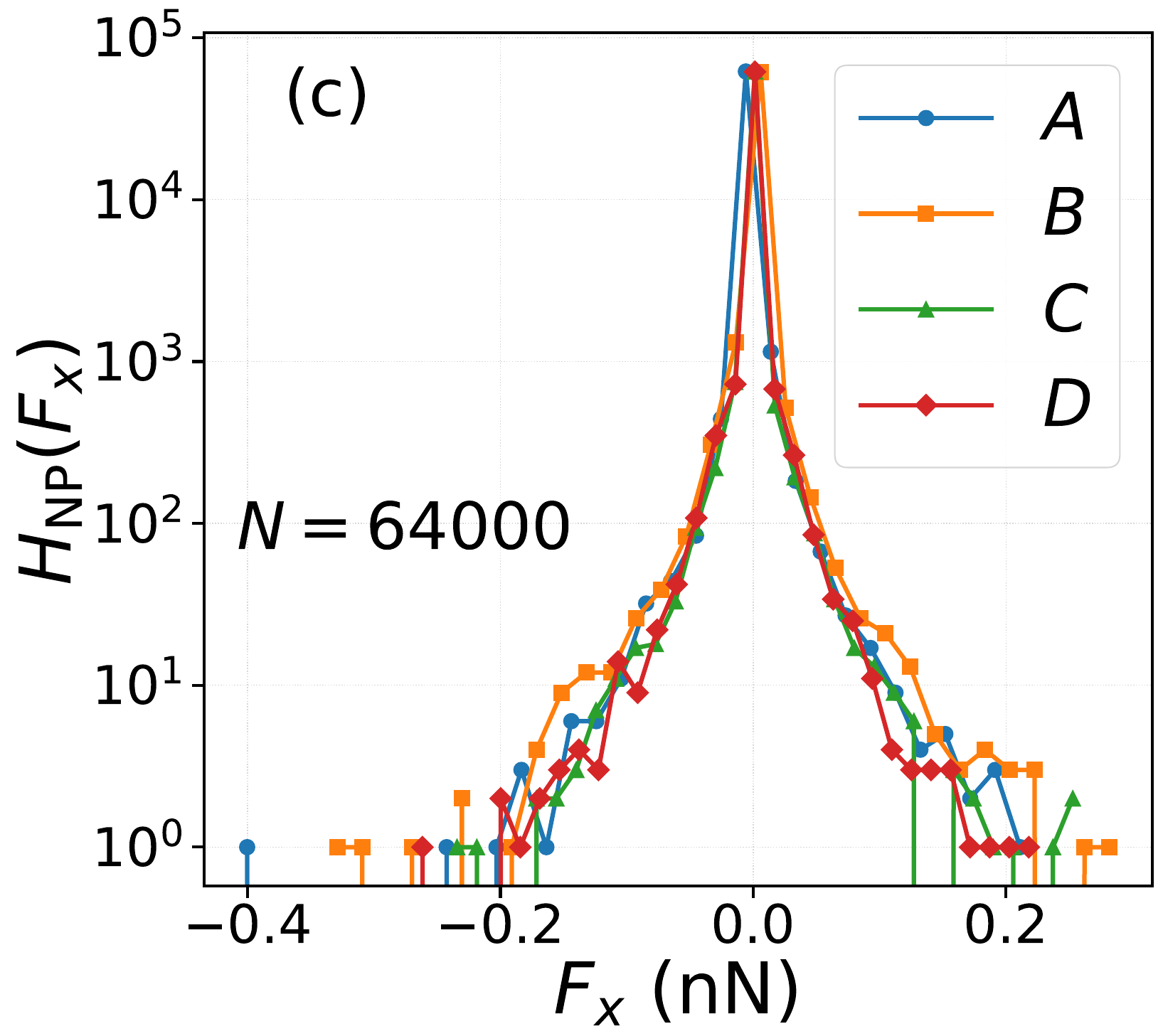}
  \includegraphics[height=3.7cm]{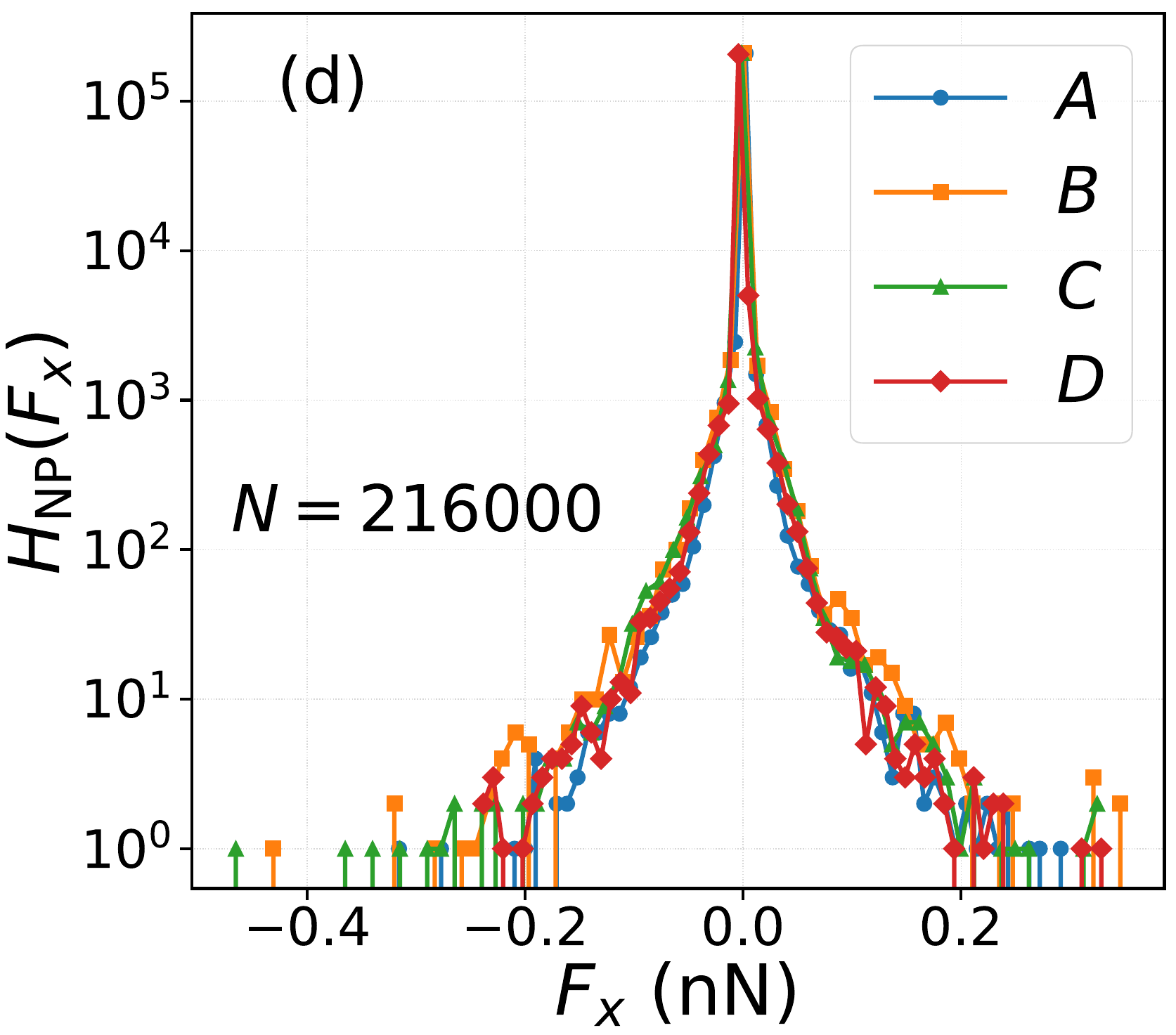}
  \includegraphics[height=3.7cm]{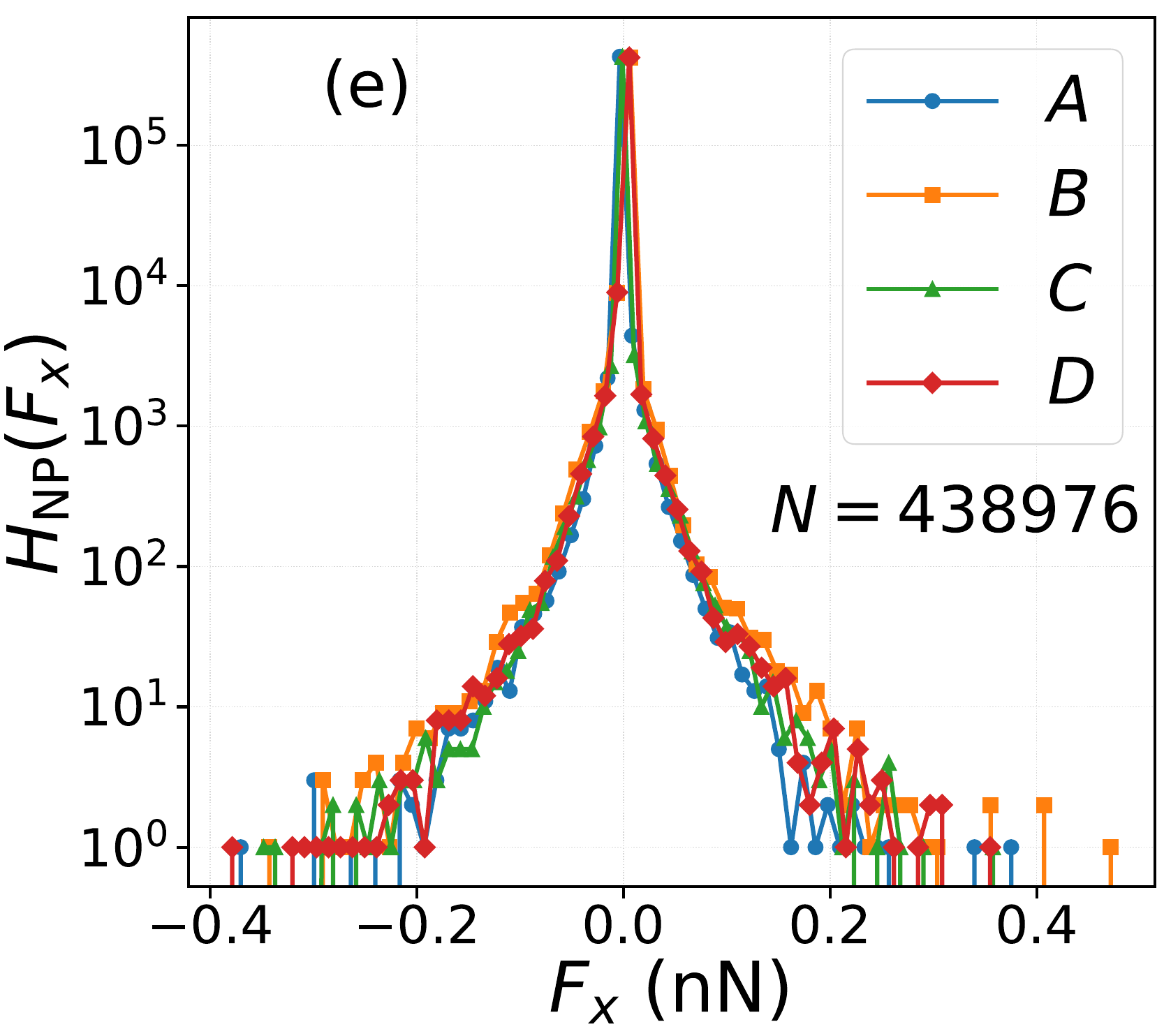}
  \includegraphics[height=3.7cm]{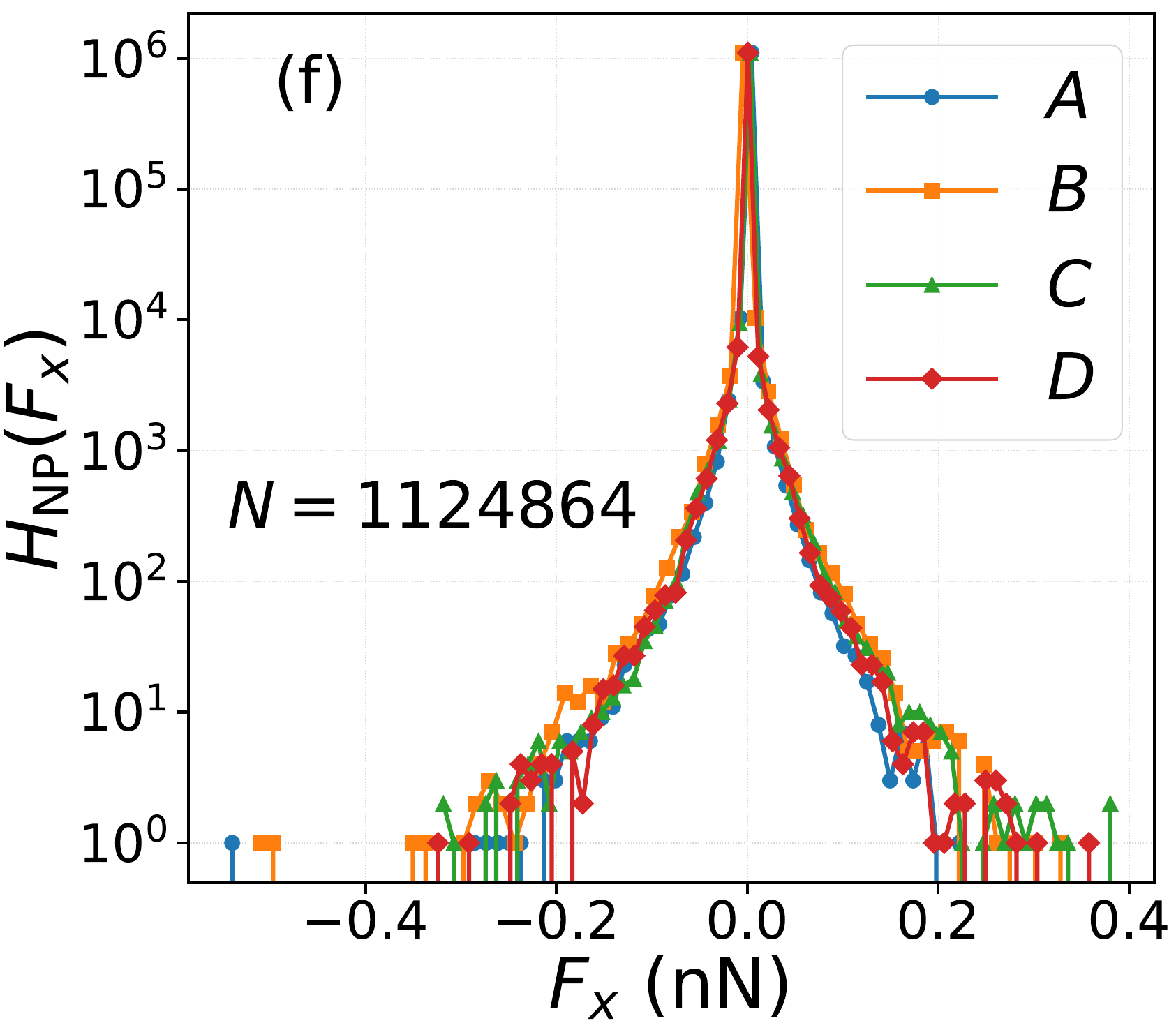}
  \caption{$H(F_{x})$, where $F_{x}$ is the force acting from the C atoms onto each Pb atom along the $x$-direction at time intervals $\mathit{A}$, $\mathit{B}$, $\mathit{C}$, and $\mathit{D}$ for NPs containing 288, 1728, 64000, 216000, 438976, and 1124864 atoms.}
  \label{fig:np_pcfx_vs_t}
\end{figure}

We consider the time evolution of the instantaneous forces acting on NP atoms from SGr carbon atoms at time intervals $\mathit{A}$, $\mathit{B}$, $\mathit{C}$, and $\mathit{D}$ (cf. Fig.~\ref{fig:time_deps}). Normal force $F_z$ was calculated using Eq.~(\ref{eq:atomic_force_def}). The same approach was used for the lateral forces: Eq.~(\ref{eq:atomic_force_def}) was applied to the $x$ and $y$ force components. $H_{\mathrm{NP}}(F_z)$ denotes the histograms of forces acting on each NP atom, whereas $H(F_z)$ is the histogram of forces acting only on the atomic layer of the NP contact surface (see Fig.~\ref{fig:np_pcd} for a visual example of such a layer). The number of bins in the histograms depends on size: 16 bins for the smallest NPs, increasing to 64 for the largest. The latter provides enough data points for attraction (negative force values) and relatively little noise from positive outliers (larger positive $F_z$).

Fig.~\ref{fig:np_pcfz_vs_t} shows the kinetics of $H_{\mathrm{NP}}(F_z)$ for NPs containing 288, 1728, 64000, 216000, 438976, and 1124864 atoms. The histograms are strongly asymmetric: they have a relatively compact negative region, corresponding to attractive forces from graphene (downwards, opposite to the $z$-direction), and a wider positive repulsive tail. As in other MD simulations~\cite{Yang2008euphyslett}, the distributions show a tall spike around $F_z = 0$, indicating that most Pb atoms do not interact with graphene. The spike is wider for negative, attractive forces, reflecting bulk Pb atoms located farther from SGr that still experience slight attraction. Starting already from about 1728 atoms or $L_x = 5$~nm, $S_{\mathrm{V}} = 1.6$~nm$^{-1}$, one can recognize the functional form of $H_{\mathrm{NP}}(F_z)$, which is noisy for small NP and which basically does not change with increasing the NP size. The attractive part in the liquid state ($\mathit{A}$, $\mathit{B}$, $\mathit{C}$) shows an additional maximum at larger attractive forces. At the same time, the repulsive tail roughly coincides across the 3 time moments for all NP sizes. In the solid state ($\mathit{D}$), the attractive part decays as the attractive force increases. Compared to the liquid state, the repulsive tail is higher near $ F_z=0$ and decays faster, covering a narrower interval of $F_z$. The shape of $H_{\mathrm{NP}}(F_z)$ suggests, in the liquid state, more atoms experience stronger attraction, have a wider range of repulsive forces, and fewer atoms feel close to 0 force. Solidification bends the tail and narrows the range of repulsive forces, meaning fewer atoms in the solid state experience larger repulsive forces because atomic thermal fluctuations are less intense. The histograms confirm previous simulations showing that interfacial forces are unevenly distributed among atoms, with a small fraction experiencing much larger forces than the rest~\citep{Cheng2010triblett}.

In contrast to $H_{\mathrm{NP}}(F_z)$, the histograms $H(F_z)$ of the normal intersurface force acting on the contact layer in Fig.~\ref{fig:bot_pcfz_vs_t} lack the near-0 spike, suggesting that most contact surface atoms experience non-zero forces. Nevertheless, $H(F_z)$ exhibits the same trends as $H_{\mathrm{NP}}(F_z)$ above: larger attractive forces, wider repulsive forces spread in the liquid state, and a larger number of atoms with forces close to 0, along with a faster decay of the repulsive tail in the solid state.

The histograms $H_{\mathrm{NP}}(F_x)$ of the lateral force $ F_x$ acting on each metal atom from the carbon atoms in Fig.~\ref{fig:np_pcfx_vs_t} are symmetric around 0, reflecting force balance in the lateral directions. Most atoms have near-zero forces, and the curves differ little between the liquid and solid states, especially for the largest systems. The $y$-component behavior is similar, so we do not show it here.

\begin{figure*}[!htb]
\centering
  \includegraphics[height=5.2cm]{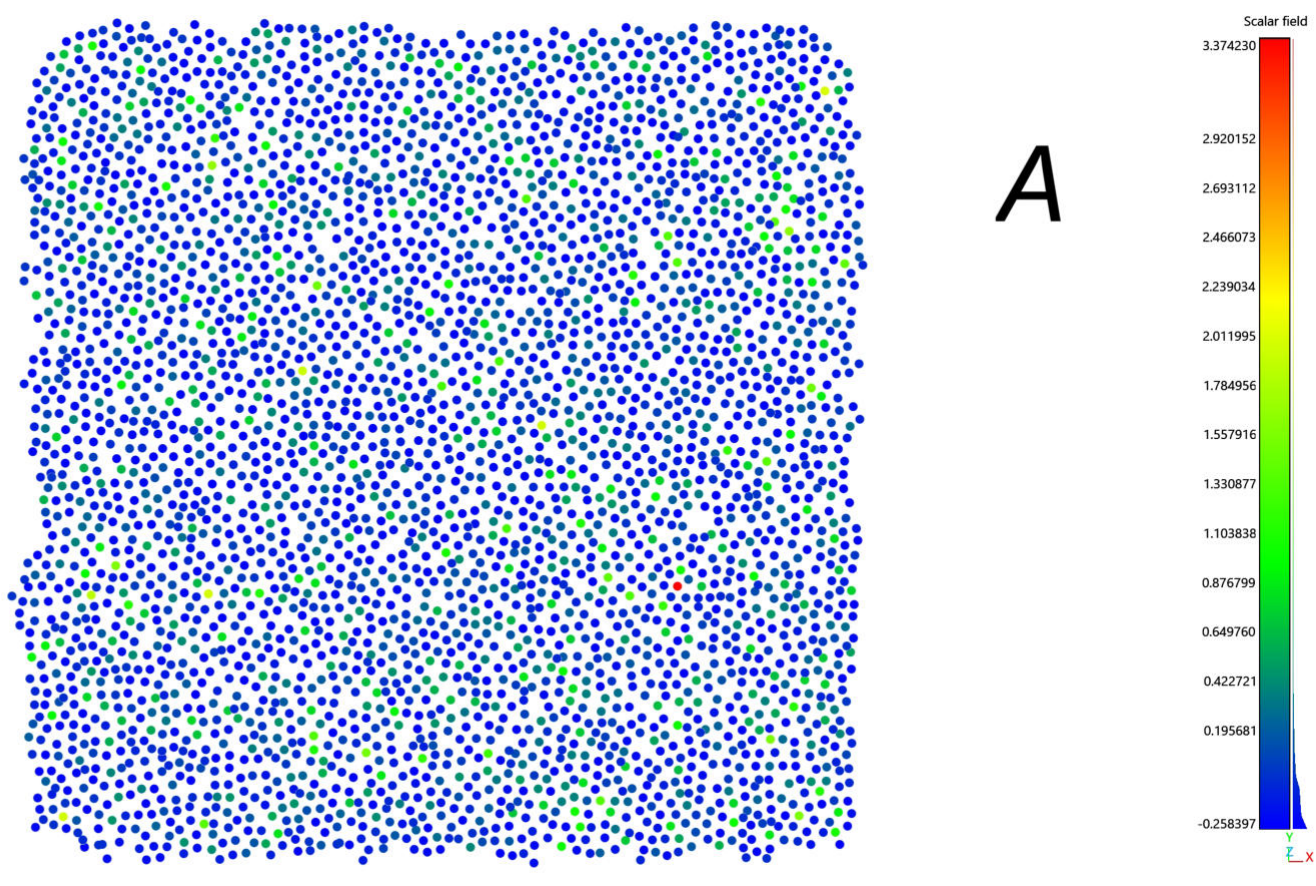}
  \includegraphics[height=5.2cm]{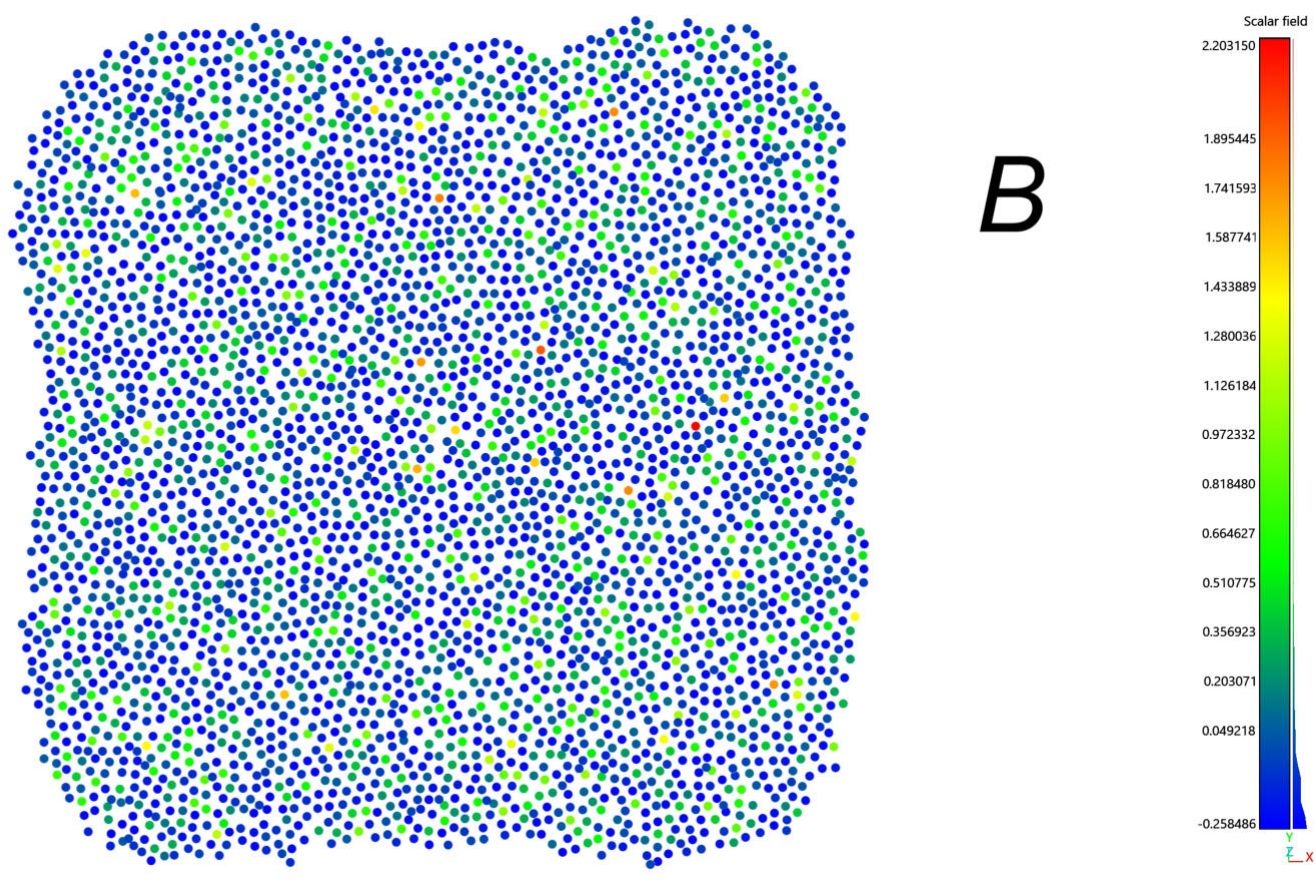}
  \includegraphics[height=5.2cm]{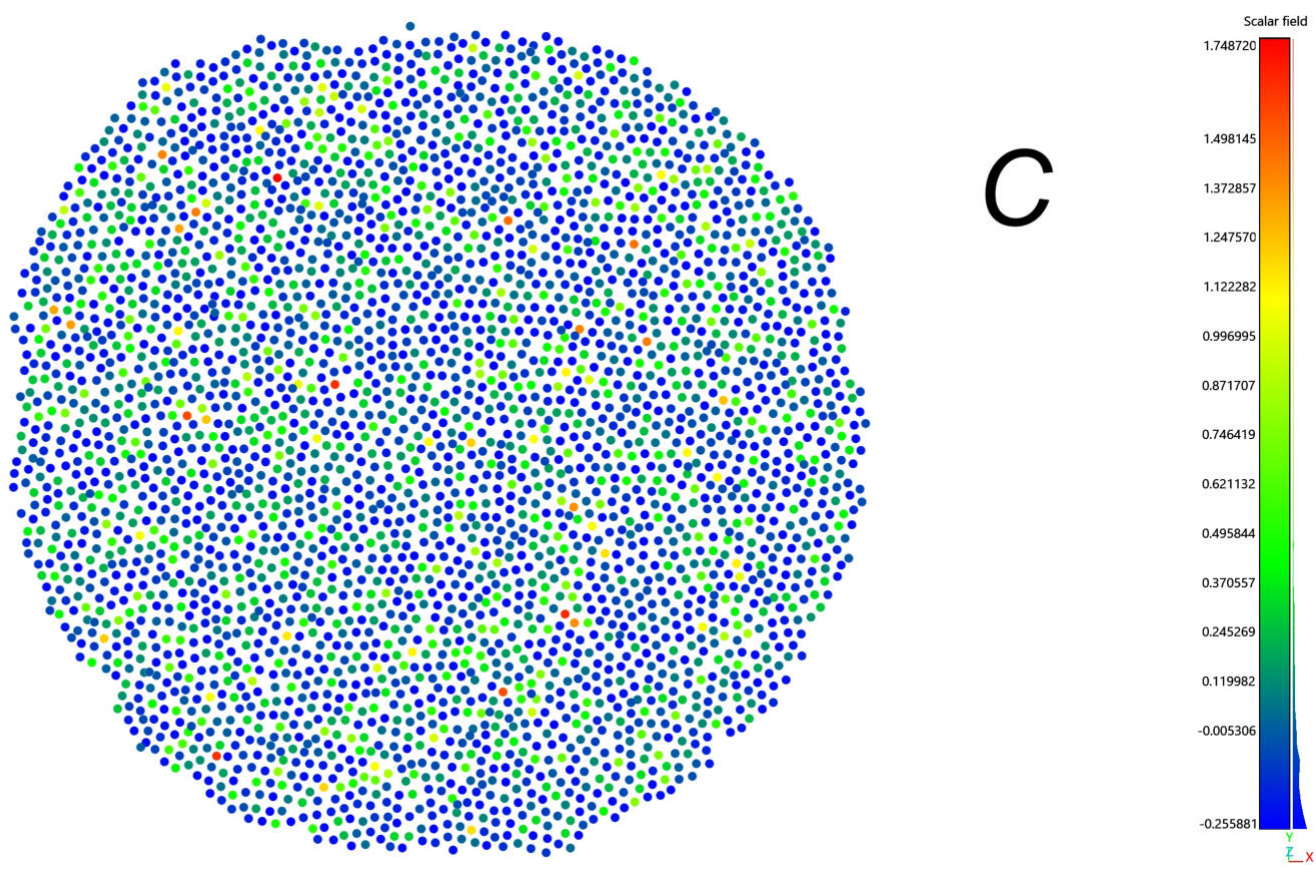}
  \includegraphics[height=5.2cm]{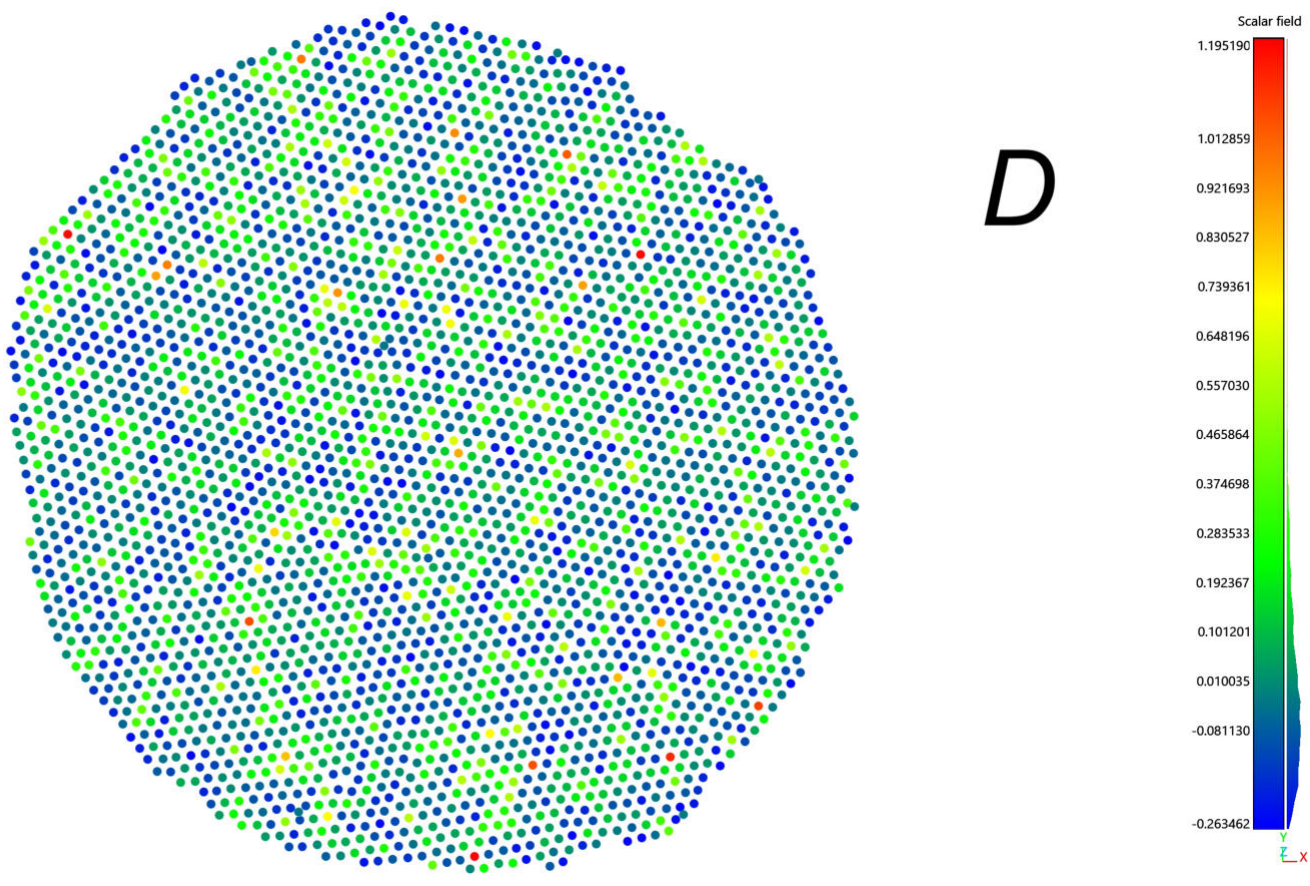}
  \caption{$F_{z}$ heatmaps of the contact atomic layer of a NP containing 64000 atoms at time intervals $\mathit{A}$, $\mathit{B}$, $\mathit{C}$, and $\mathit{D}$ (from Fig.~\ref{fig:time_deps}). Blue color---the smallest values, red---the largest. Produced with CloudCompare~\cite{CloudCompare}.}
  \label{fig:heatmaps_fcz_64k}
\end{figure*}

\begin{figure*}[!htb]
\centering
  \includegraphics[height=5.2cm]{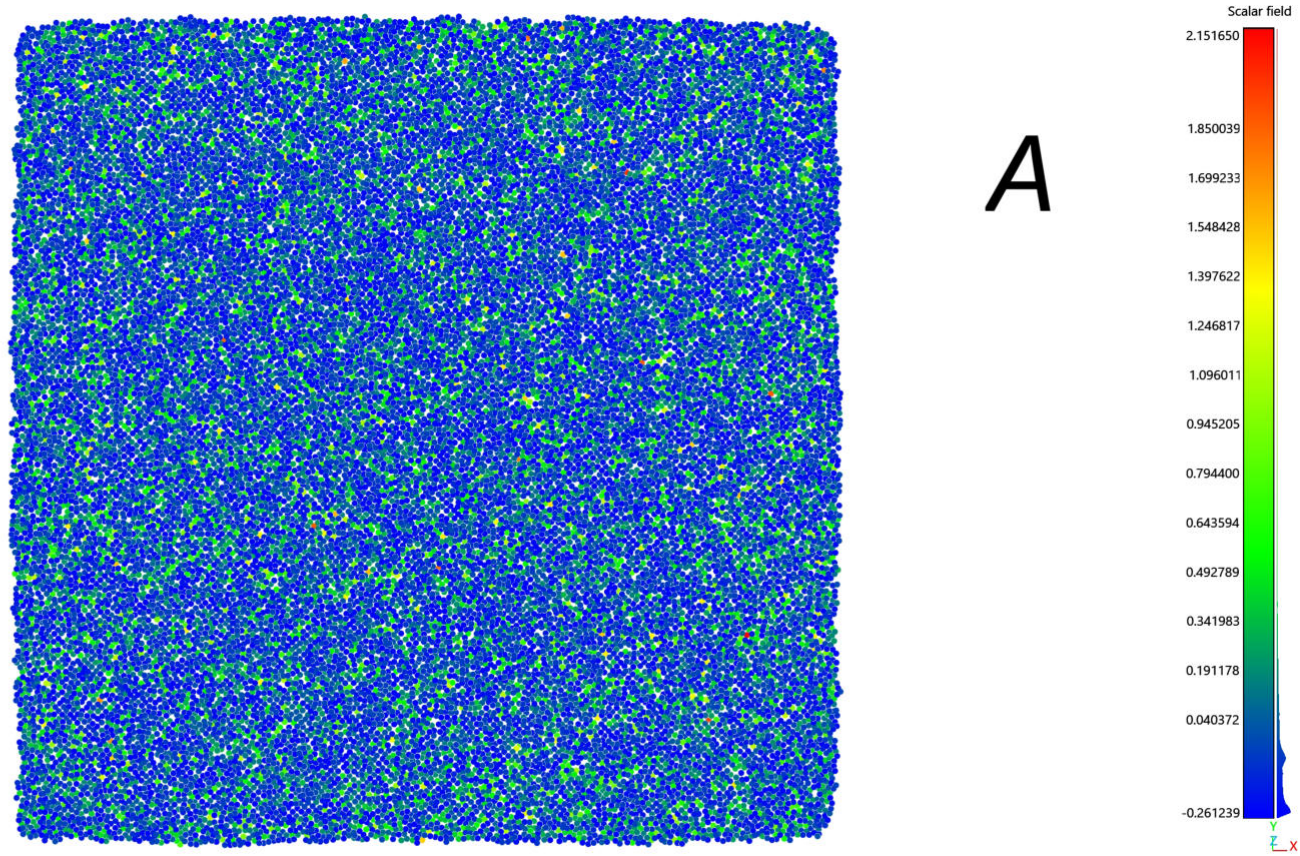}
  \includegraphics[height=5.2cm]{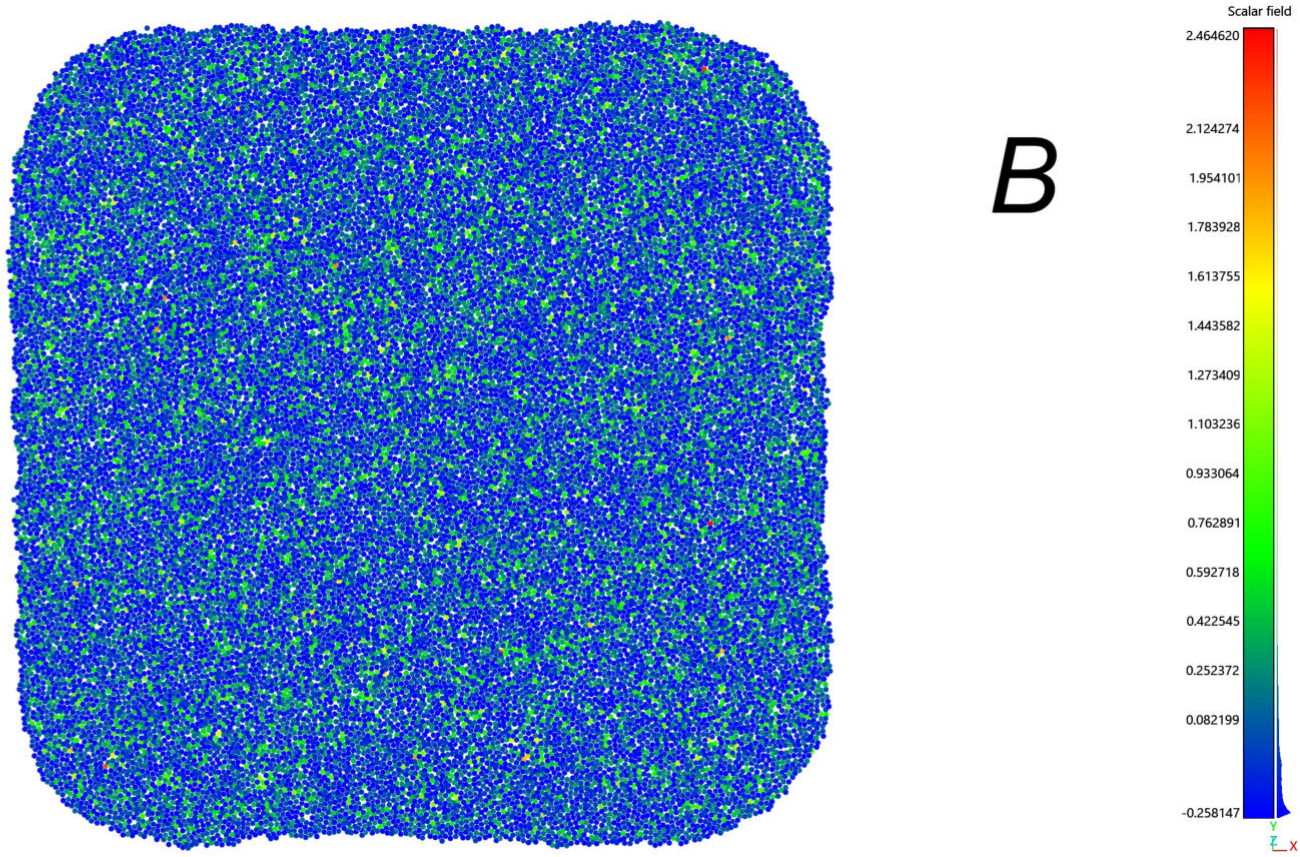}
  \includegraphics[height=5.2cm]{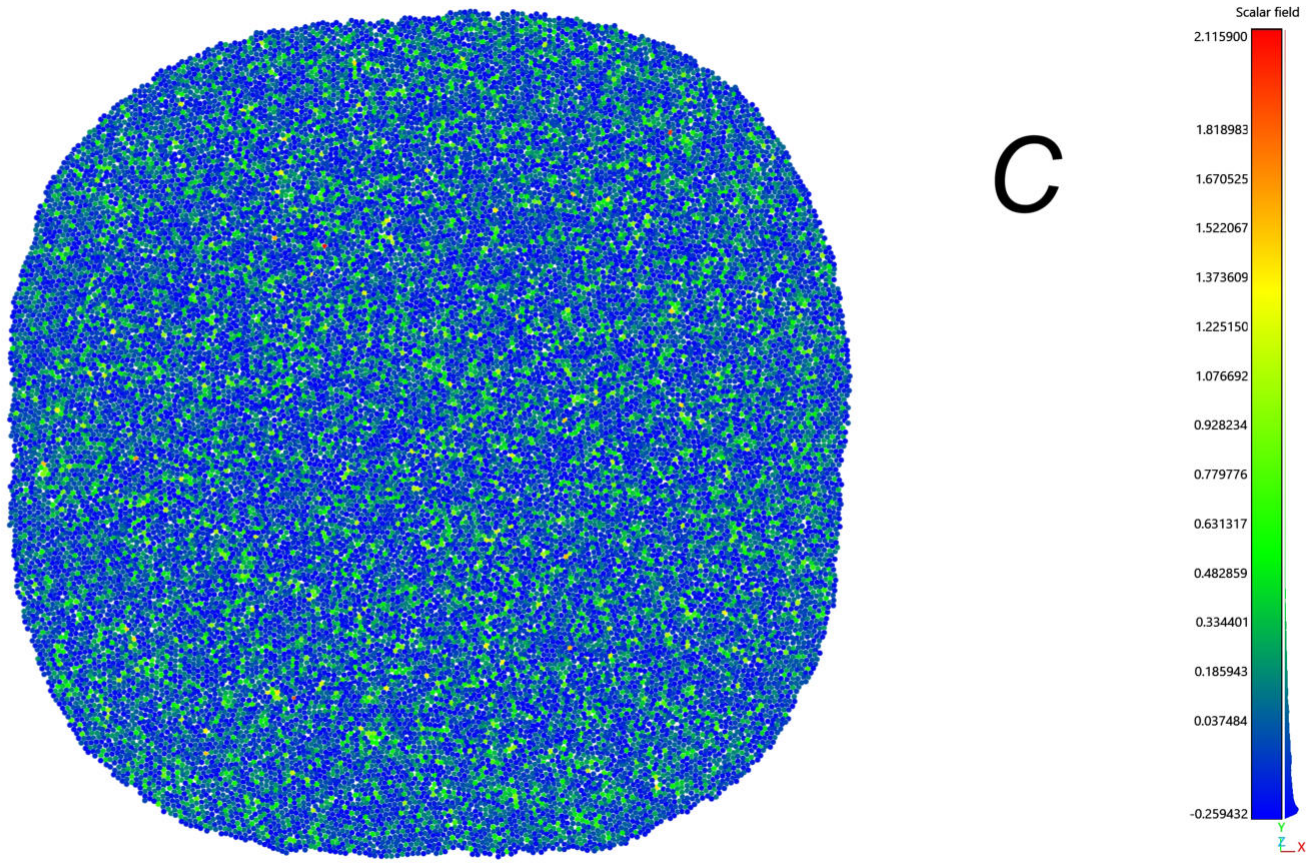}
  \includegraphics[height=5.2cm]{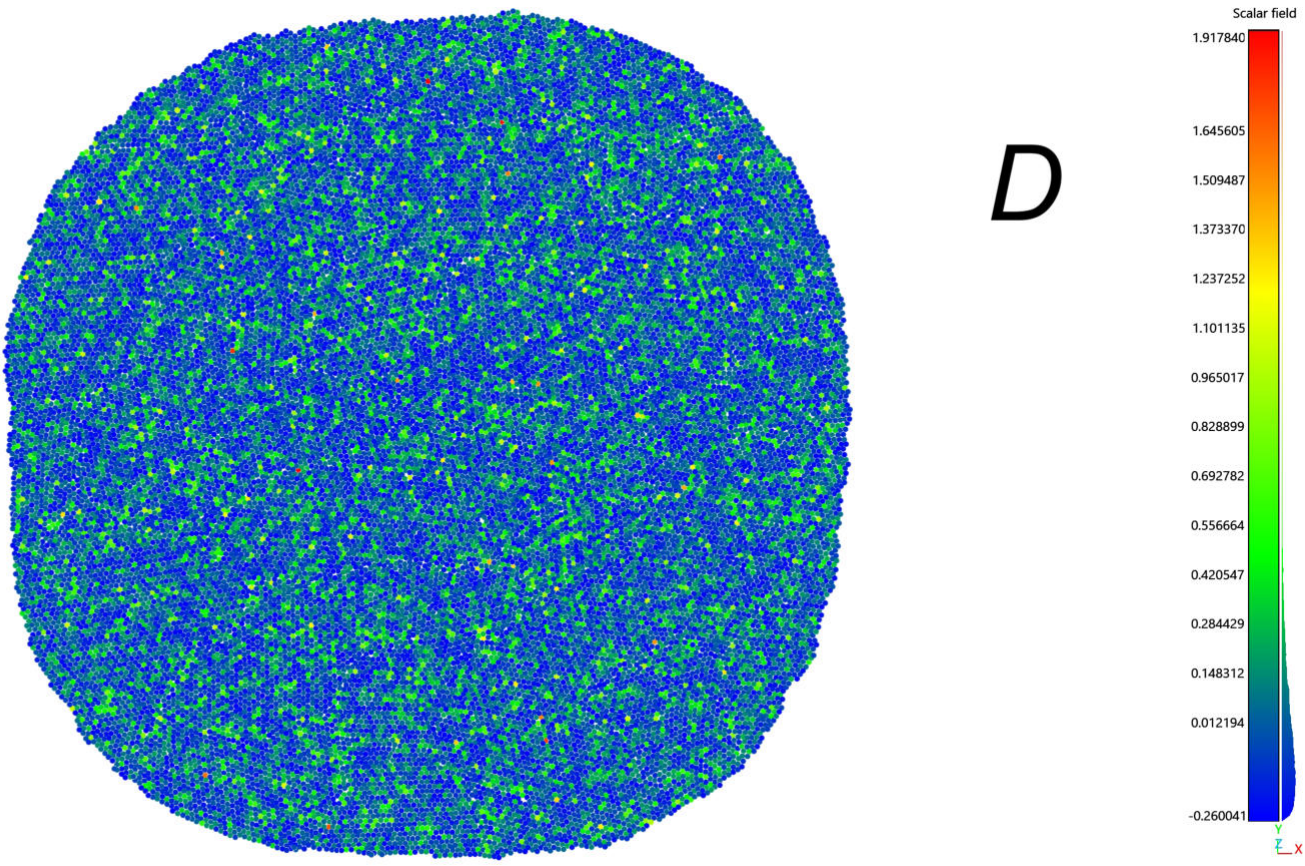}
  \caption{$F_{z}$ heatmaps of the contact atomic layer of a NP containing 1124864 atoms at time intervals $\mathit{A}$, $\mathit{B}$, $\mathit{C}$, and $\mathit{D}$ from Fig.~\ref{fig:time_deps}.}
  \label{fig:heatmaps_fcz_1M}
\end{figure*}

Fig.~\ref{fig:heatmaps_fcz_64k} and Fig.~\ref{fig:heatmaps_fcz_1M} show heatmaps of the instantaneous $F_z$ acting on the contact atomic layer of NPs containing 64000 and 1124864 atoms, respectively. Initially, when the NP is molten, attractive interactions (blue) prevail; as it approaches solidification, more atoms experience repulsive forces (green). Eventually, the contact surface of a solid NP consists of (111) domains, suggesting a polycrystalline NP, as shown by the RDF in Fig.~\ref{fig:rdfs_16_05} and by experiments~\cite{Liu2013jpcc}. Additionally, the snapshots show no clearly manifested spatial force structures. This contrasts with AFM tip adhesion measurements on the outermost surface of TiO${_2}$ NPs in aqueous solution, where adhesion forces form non-random patchy domains rather than distributing randomly across the surface~\cite{Khomami2026asusc}.

\begin{figure}[!htb]
\centering
  \includegraphics[height=3.8cm]{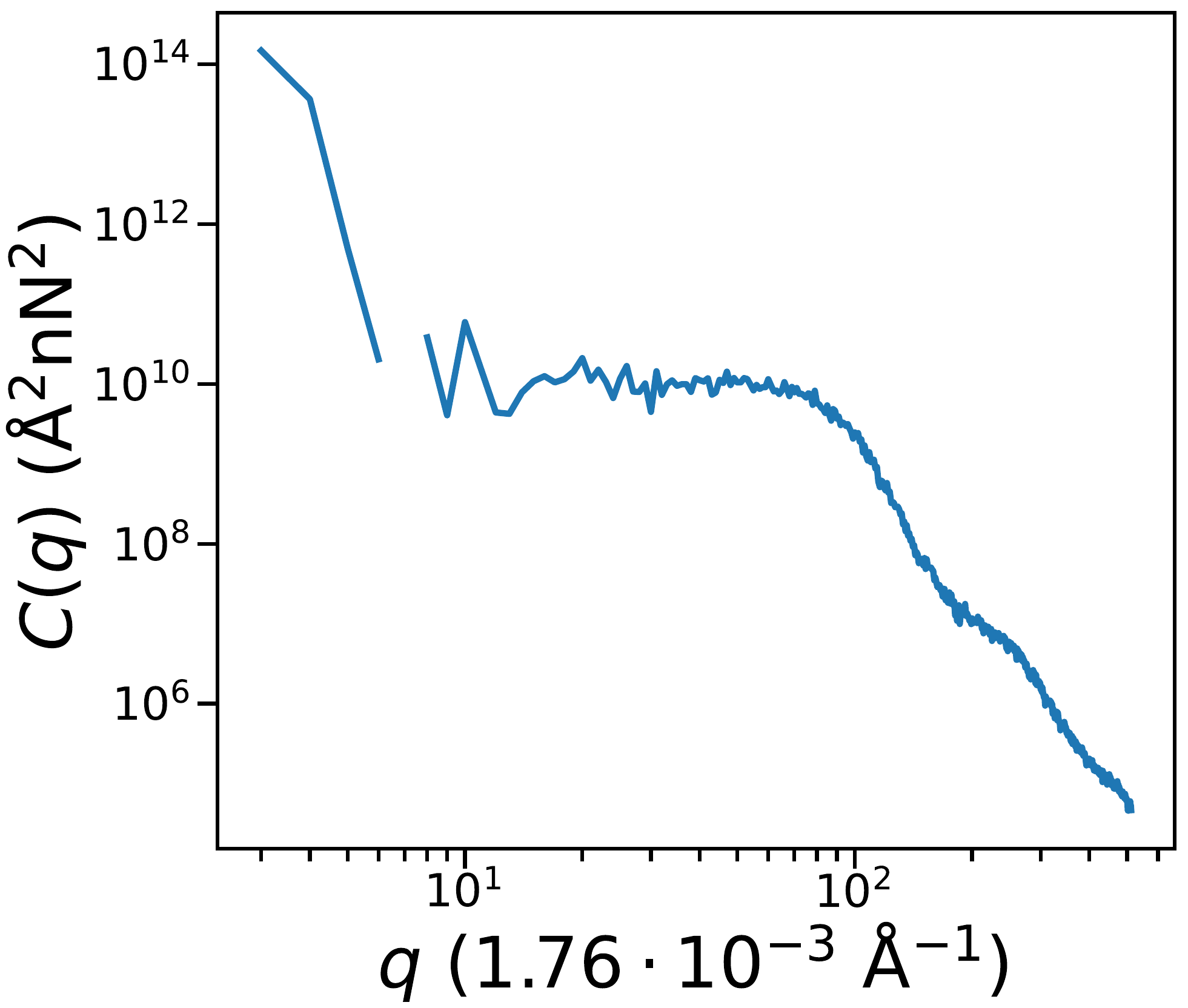}
  \includegraphics[height=3.8cm]{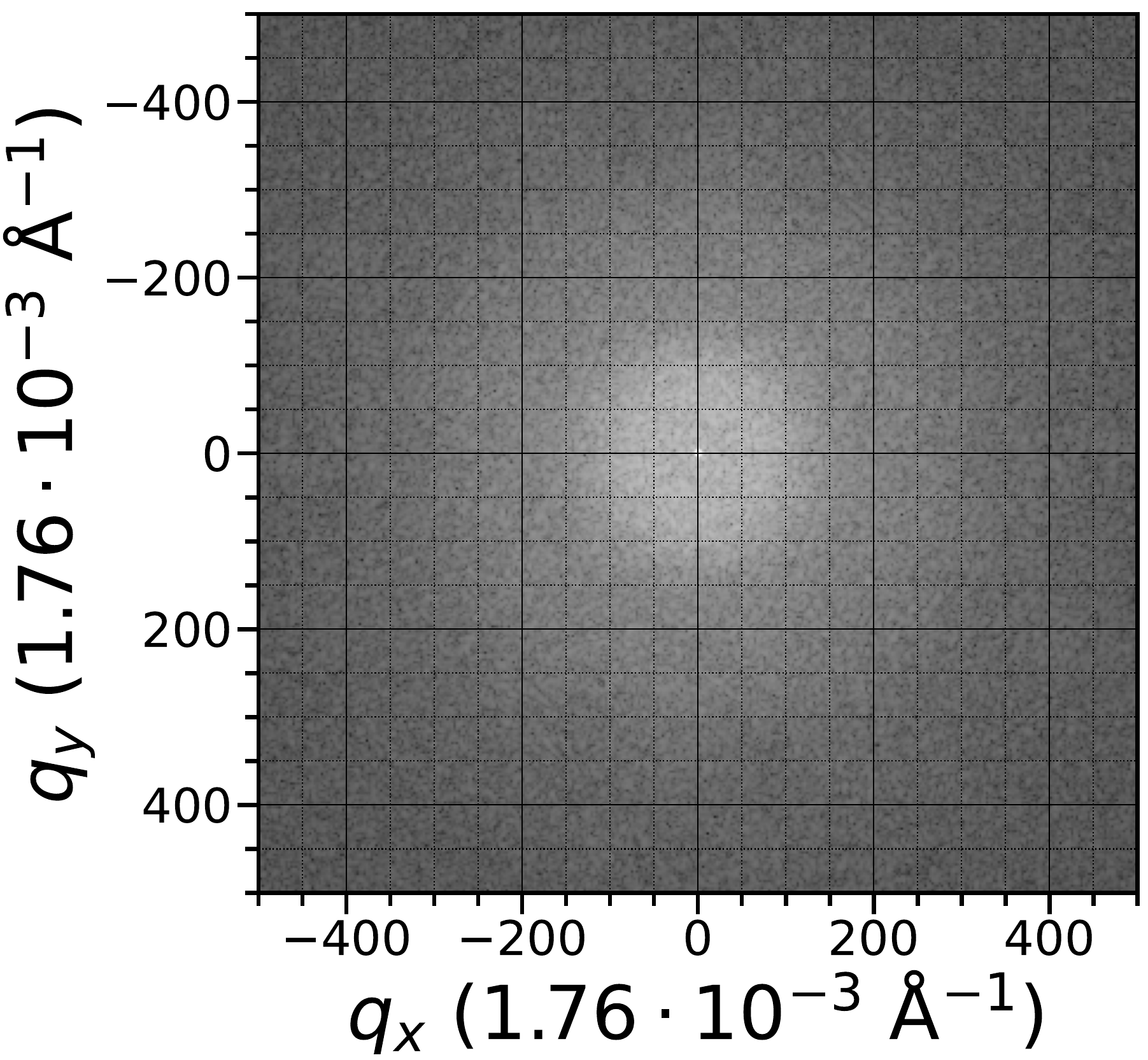}
  \includegraphics[height=3.8cm]{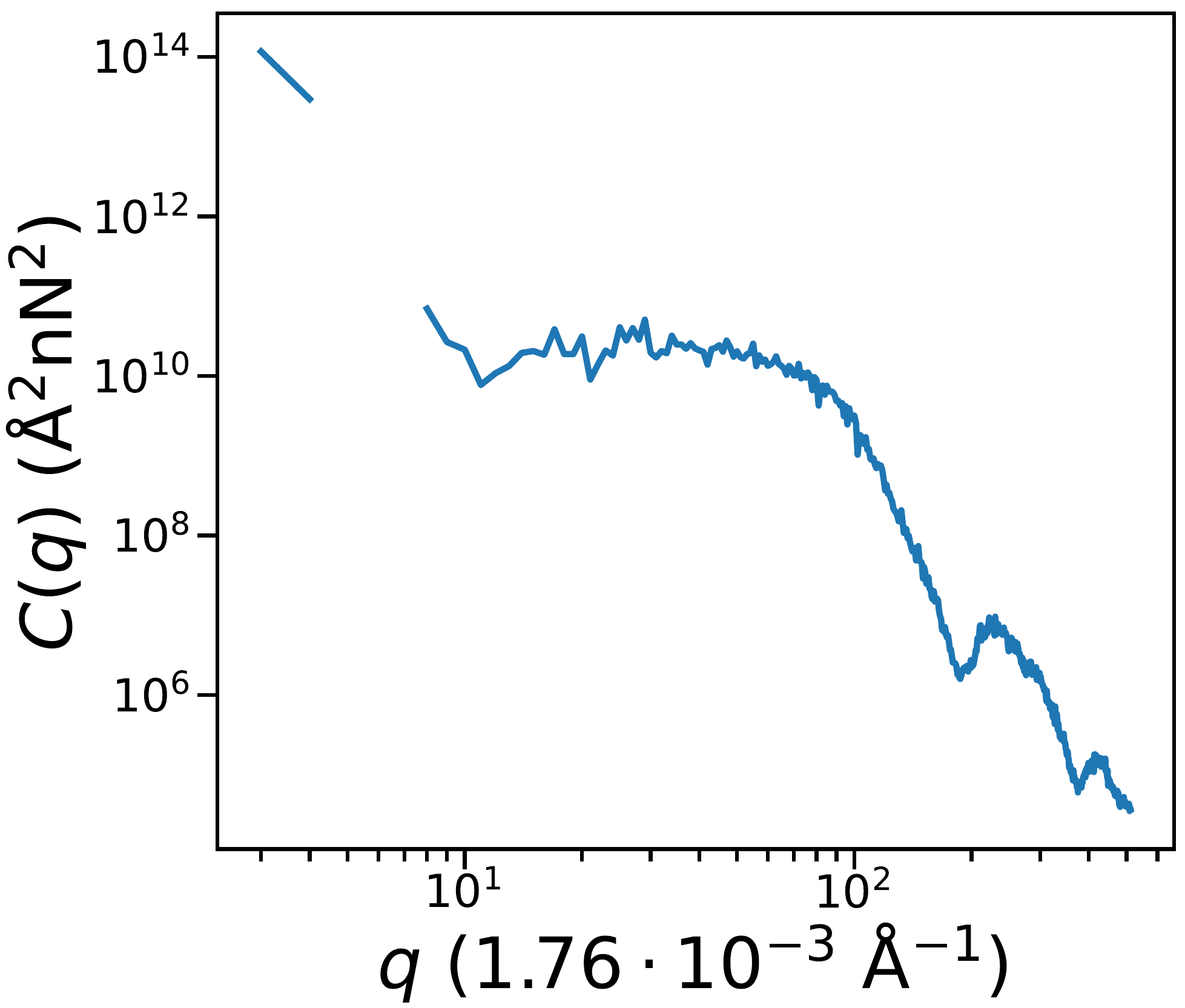}
  \includegraphics[height=3.8cm]{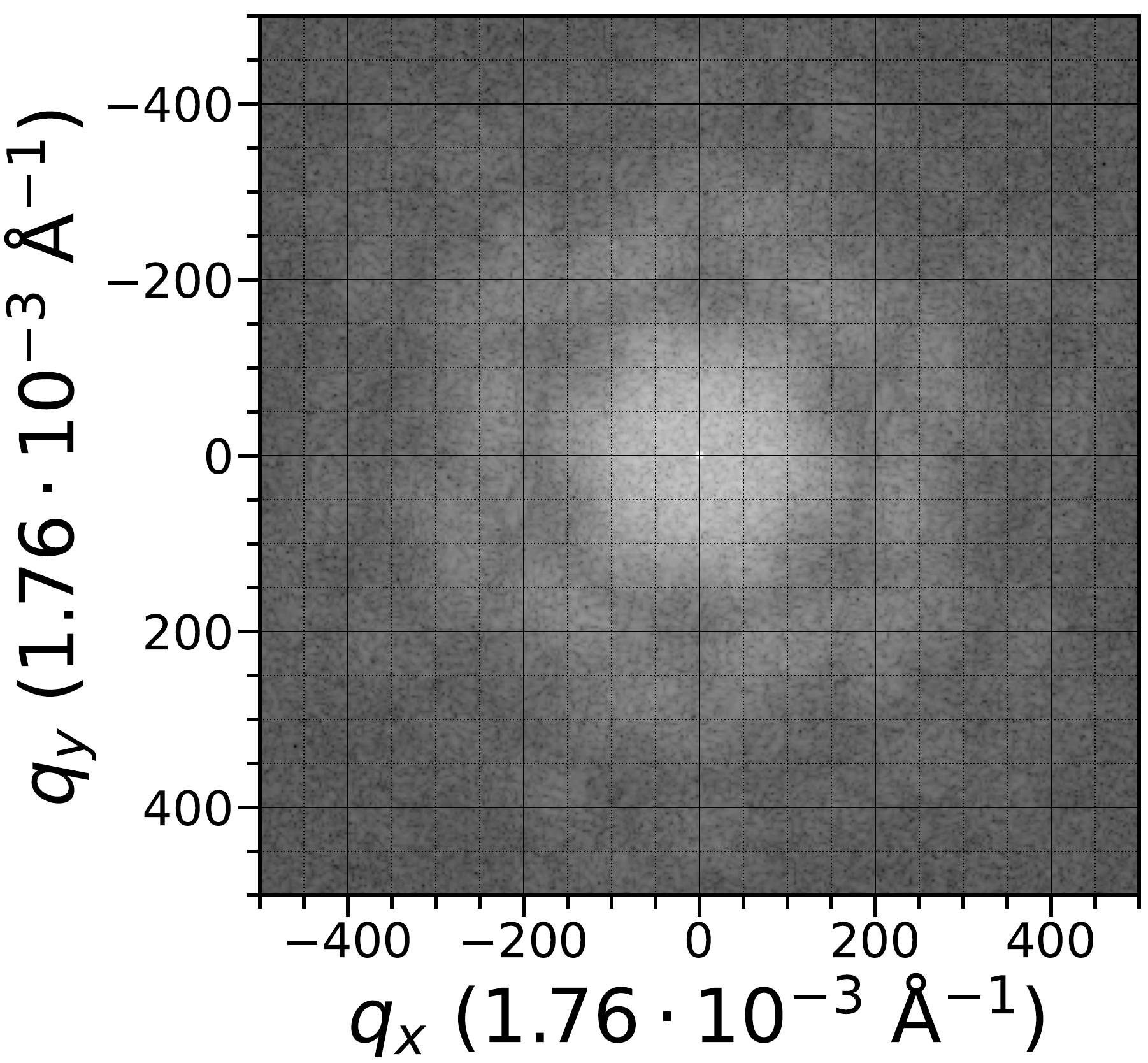}
  \caption{1D and corresponding 2D PSDs of a Pb NP containing 64000 atoms at time intervals $\mathit{B}$ (top figures, liquid NP) and $\mathit{D}$ (bottom figures, solid NP) from Fig.~\ref{fig:time_deps}.}
  \label{fig:fft_psd_12_09}
\end{figure}

\begin{figure}[!htb]
\centering
  \includegraphics[height=3.8cm]{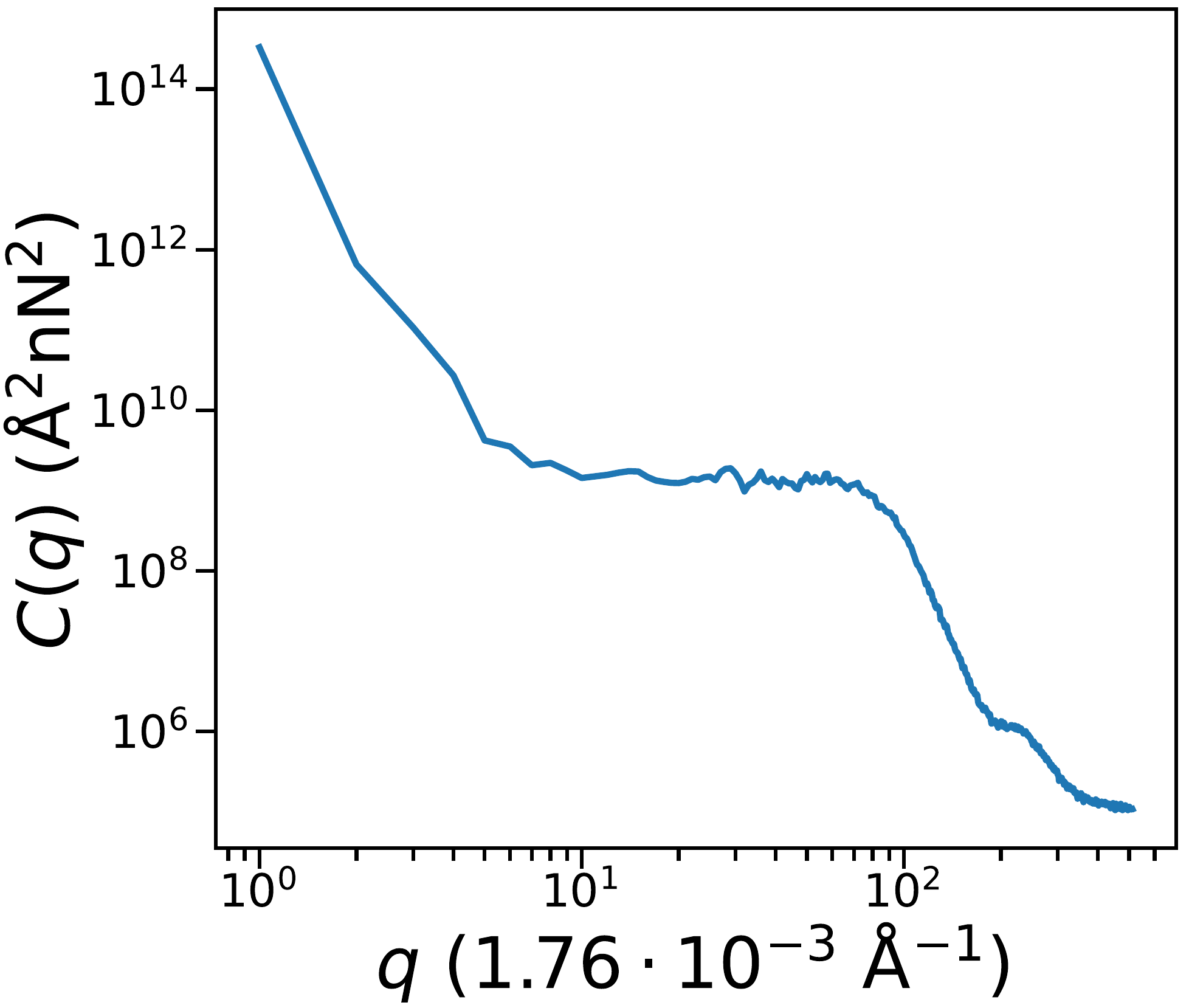}
  \includegraphics[height=3.8cm]{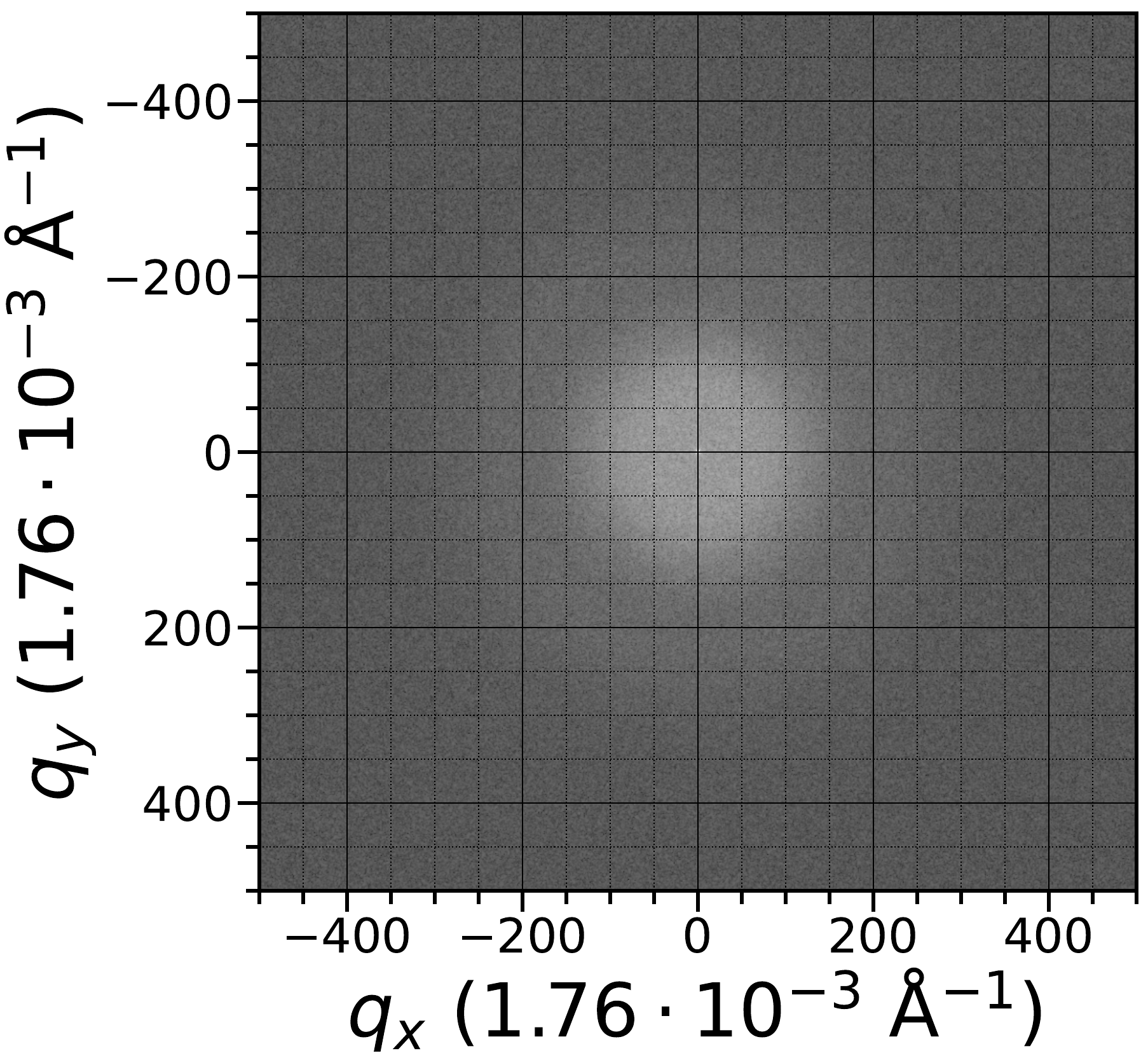}
  \includegraphics[height=3.8cm]{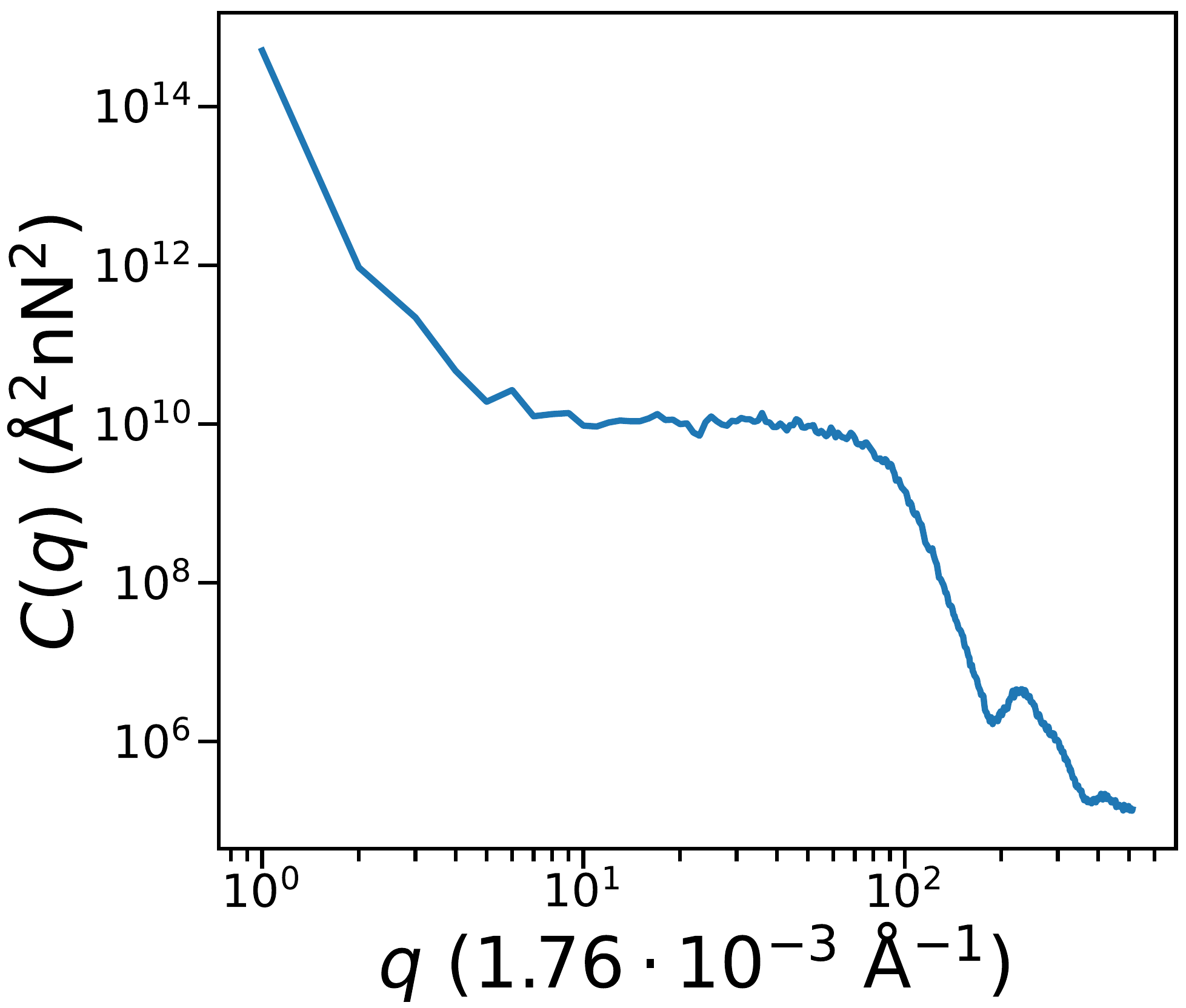}
  \includegraphics[height=3.8cm]{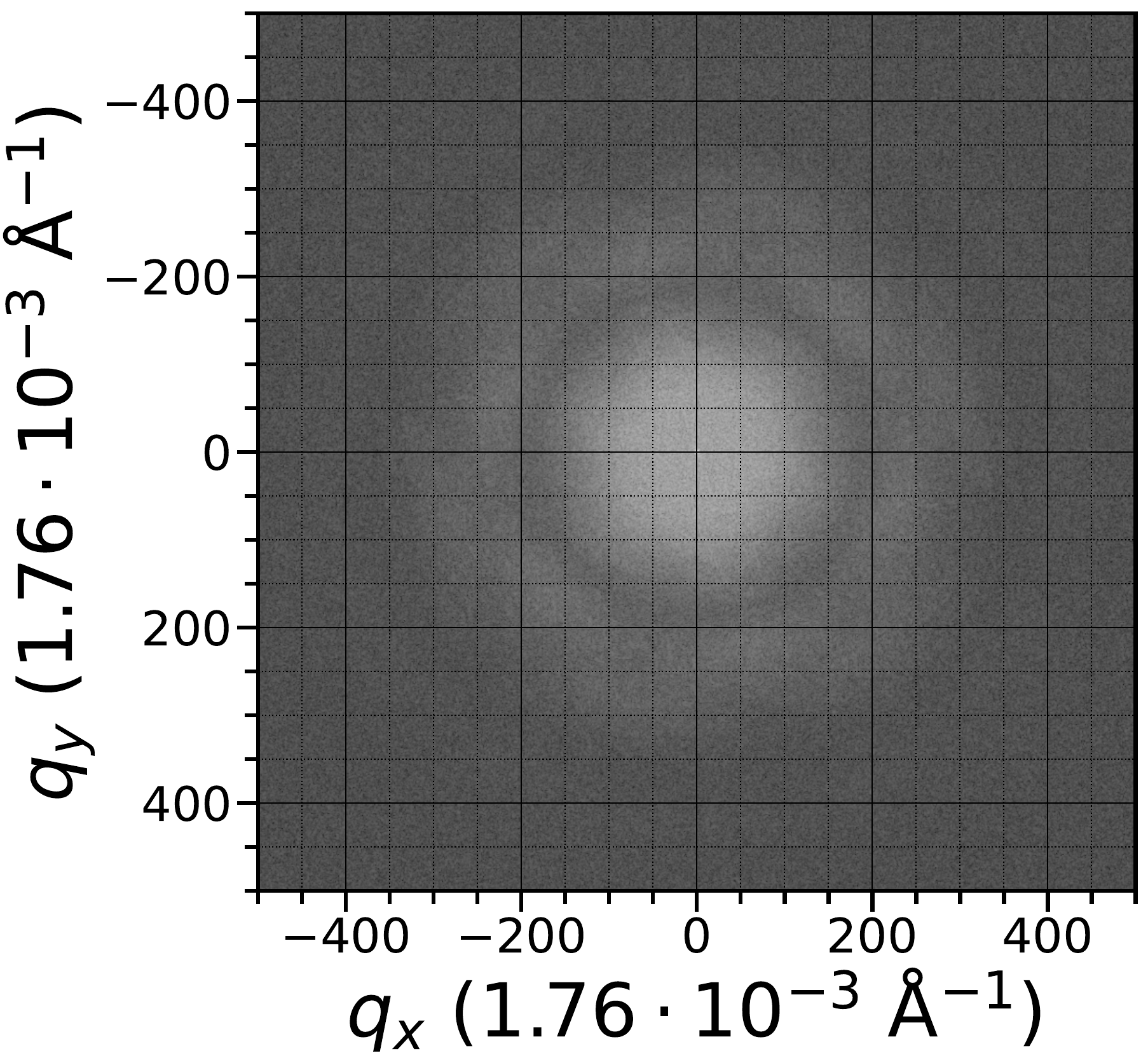}
  \caption{1D and corresponding 2D PSDs of a Pb NP containing 1124864 atoms at time intervals $\mathit{B}$ (top figures, liquid NP) and $\mathit{D}$ (bottom figures, solid NP).}
  \label{fig:fft_psd_18_05}
\end{figure}

We performed spectral analysis of the atomic forces to mathematically identify the characteristic spatial wavelength of the interfacial normal forces, cf. Fig.~\ref{fig:fft_psd_12_09} and Fig.~\ref{fig:fft_psd_18_05}. As expected, no fine structure is observed in the liquid state; the 1D PSD $C(q)$ is smeared and decays at wavevector $q_a \approx 115$ units, or 0.202~\AA$^{-1}$, which roughly corresponds to the Pb lattice constant $a = 1/q_a = 4.94$~\AA~\cite{Zhou2001}. In the solid state, some hexagonal fine structure appears in smaller NPs of up to about 64k atoms: one spike around 230 units (0.405~\AA$^{-1}$, or $\lambda_x \approx 2.47$~\AA), which is the graphene lattice constant. Finally, the smallest spike is around 0.704~\AA$^{-1}$, corresponding to the graphene covalent bond (1.42~\AA) contribution. However, for larger NPs, the highest-frequency spike is absent, and the graphene lattice-constant contribution is less pronounced than in smaller NPs. The 2D PSD of the larger NP has a circular symmetry. This behavior may be attributed to the highly incommensurate Pb-SGr lattice, which statistically smears out potentially coherent force contributions in larger NPs. The low-frequency part of the 1D and 2D PSDs does not change qualitatively with size, except that larger NPs show a stronger contribution from smaller $q$ to $C(q)$. Note that $C(q)$ has a rolloff region for $q < q_a$, where all frequencies contribute approximately equally to the power spectrum. Only the smaller wave vectors (largest wavelengths) contribute considerably more.

\subsubsection{Relative contact area kinetics}
\label{subsubsec::area_kinetics}

The contact area calculation is ambiguous at the atomic level~\cite{Yang2008euphyslett,Cheng2010triblett,
Pastewka2014pnas,Mo2010prb,Prodanov2026pccp,Almqvist2011jmps}. This is especially relevant for adhesive contacts, where not only surface-layer atoms can interact with the opposing surface, but also some atoms inside the bulk. One common approach is to consider atoms in contact if they experience a repulsive normal force from the opposing surface~\cite{Cheng2010triblett,Pastewka2014pnas}. This approach states that repulsive forces indicate the corresponding atoms are close enough to be considered in contact. In contrast, zero or attractive atomic normal forces suggest the atoms are farther apart and hence are not in contact.

In this study, we utilize the mentioned approach. The relative contact area $A_{\mathrm{r}}$ was calculated as a ratio of the number of contact atoms $N_{\mathrm{c}}$ that feel repulsion along the $z$ direction from the opposing surface, to the number of atoms $N_{\mathrm{nom}}$ in the nominal (apparent) contact area $A_0$:

\begin{equation}
A_{\mathrm{r}} = N_{\mathrm{c}}/N_{\mathrm{nom}}.
\label{eq:a_rel_via_n}
\end{equation}
In our case, $A_0$ is the area of the NP contact layer, and $N_{\mathrm{nom}}$ is the number of atoms in this layer. Eq.~(\ref{eq:a_rel_via_n}) holds under the assumption that each atom occupies the same average atomic area, $A_{\mathrm{atom}}$.

Experiments and simulations show that sufficiently large NPs (of tens of nm in linear size) can have non-ideally flat surfaces~\cite{Khomami2026asusc,Prodanov2026pccp}. Fig.~\ref{fig:np_pcd} confirms that the NP contact surface is not flat but has atomic-scale roughness, with longer wavelengths at the NP borders. The presence of the large RMS height roughness may invalidate the assumption of Eq.~(\ref{eq:a_rel_via_n}) that all the contact atoms have the same $A_{\mathrm{atom}}$, which may lead to a noticeable discrepancy between the geometrical areas of the apparent and actual NP contact surfaces.  Nevertheless, as shown in~\cite{Prodanov2026pccp}, the NP roughness RMS height is on the order of~\AA, so the discrepancy is < 1~\% for $L_x > 10$~nm and should not affect the final result for $A_{\mathrm{r}}$.

\begin{figure}[!htb]
\centering
  \includegraphics[height=3.8cm]{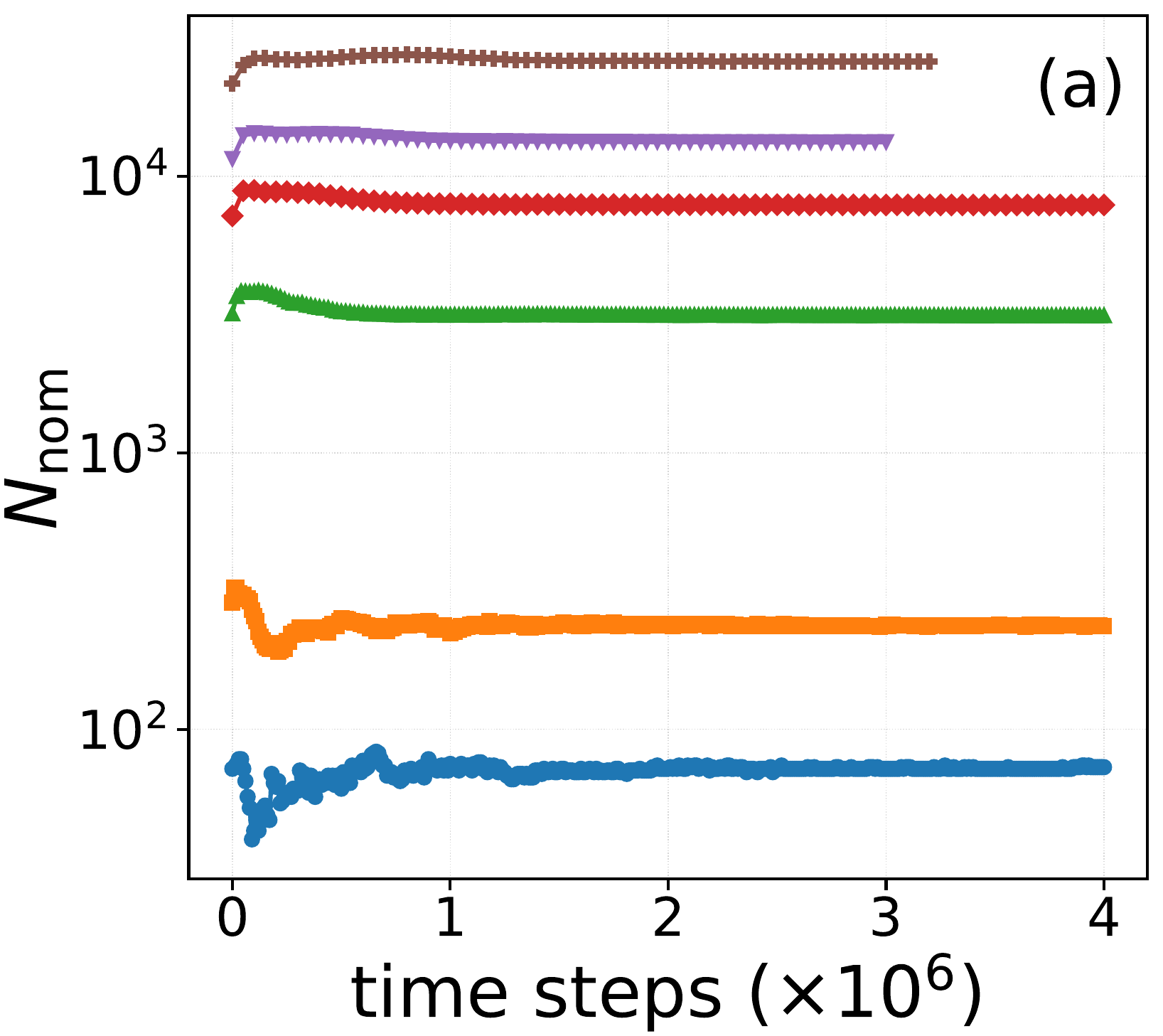}
  \includegraphics[height=3.8cm]{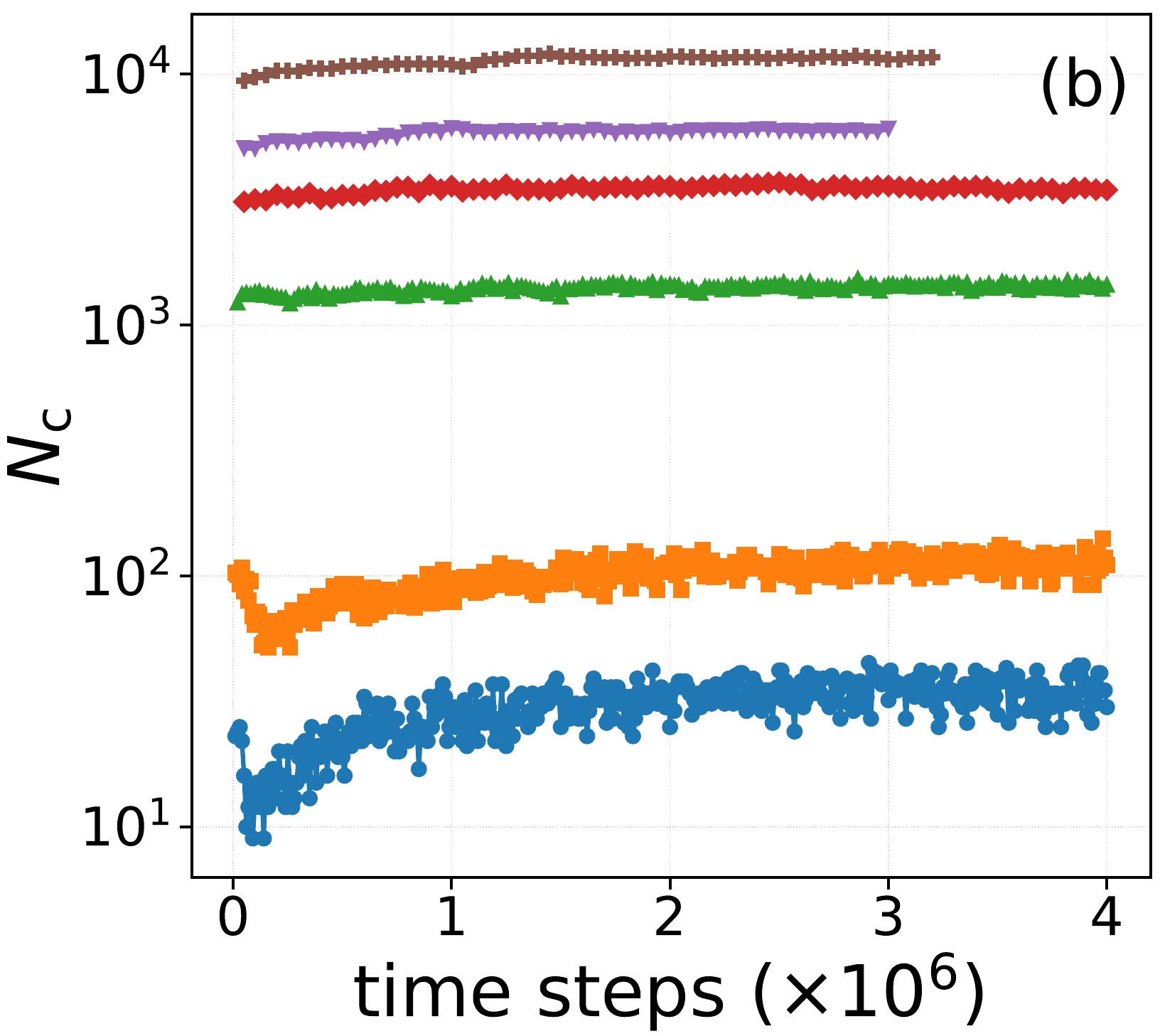}
  \caption{Time dependence of the number of atoms $N_{\mathrm{nom}}$ (a) and the number of atoms experiencing repulsion $N_{\mathrm{c}}$ (b) in the NP contact layer for NPs containing 288, 1728, 64000, 216000, 438976, and 1124864 atoms (bottom-to-top plots, respectively).}
  \label{fig:n_bot_vs_t}
\end{figure}

\begin{figure}[!htb]
\centering
  \includegraphics[height=7cm]{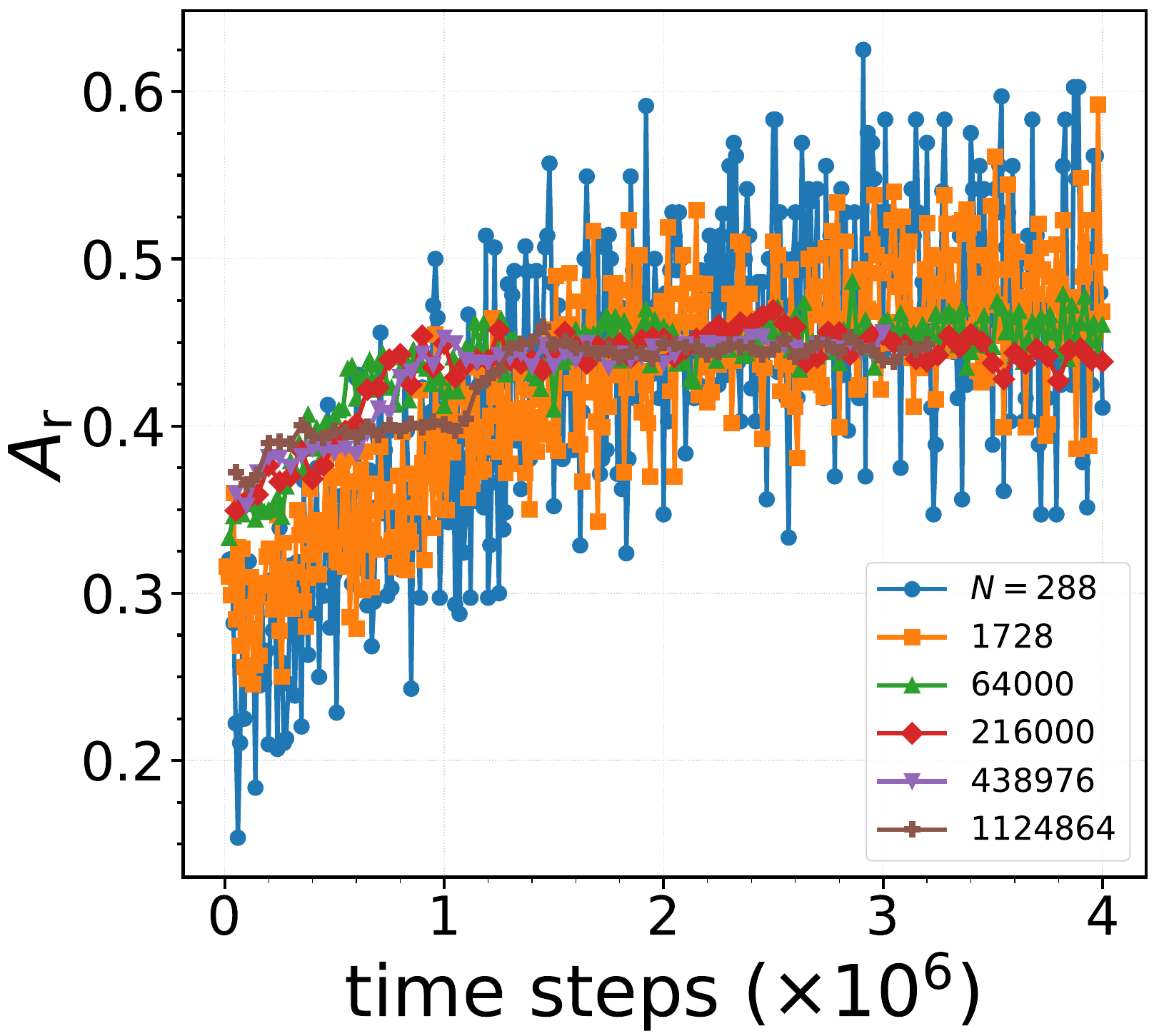}
  \caption{Time dependence of the instantaneous relative contact area $A_{\mathrm{r}}$ for NPs of various sizes.}
  \label{fig:a_rel_vs_t}
\end{figure}

The time evolution of $N_{\mathrm{nom}}$ and $N_{\mathrm{c}}$ in Fig.~\ref{fig:n_bot_vs_t} suggests that their difference between the liquid and solid states is small, especially for larger NPs. During the initial melting phase, both $N_{\mathrm{nom}}(t)$ and $N_{\mathrm{c}}(t)$ in smaller NPs initially decrease, while for larger NPs, these changes are not pronounced enough. For all the NP sizes, the quantities eventually converge to some stationary average value. Data scattering of $N_{\mathrm{c}}(t)$ in Fig.~\ref{fig:n_bot_vs_t}(b) reflects thermal fluctuations of atoms and hence the variance in the repulsive forces.

Fig.~\ref{fig:a_rel_vs_t} shows the kinetics of the instantaneous $A_{\mathrm{r}}(t)$ calculated using Eq.~(\ref{eq:a_rel_via_n}). For all the NP sizes, in the liquid state, $A_{\mathrm{r}}$ is smaller than the stationary value reached after the NP solidification. This is in accord with the kinetics of the normal force distributions in Fig.~\ref{fig:bot_pcfz_vs_t}, Fig.~\ref{fig:np_pcfx_vs_t}, and the force heatmaps in Fig.~\ref{fig:heatmaps_fcz_64k} and Fig.~\ref{fig:heatmaps_fcz_1M} that suggest that more Pb atoms experience attractions and fewer atoms feel repulsive forces in the liquid state compared to the solid one. Smaller NPs show much larger fluctuations in instantaneous values and a smaller average $A_{\mathrm{r}}$ in the liquid state, while larger ones show a smaller average $A_{\mathrm{r}}$ in the solid state. Eventually, the converged average value is around 44--48~\% depending on the NP size.

\subsection{Size scaling}
\label{subsec::scaling}

Size scaling highlights how the normal interfacial force distribution and $A_{\mathrm{r}}$ change as NP size increases and longer roughness wavelengths are added to the contact surface. It can also reveal size effects in $P(F_z)$ and $A_{\mathrm{r}}$. In particular, the geometrical criterion for calculating $A_{\mathrm{r}}$ suggests that the ensemble-averaged relative contact area of small Al and Cu NPs (a few nm in linear size) can differ by more than~15\% from that of larger ones~\cite{Prodanov2026pccp}. In this work, we check size effects using the normal-force $A_{\mathrm{r}}$ criterion for Pb NPs. For scaling, we focus on the solid NPs, which correspond to the equilibration time interval $\mathit{D}$ in Fig.~\ref{fig:time_deps}. The normal-force histograms/distributions presented below are ensemble averages of instantaneous values of the solidified, equilibrated NP over 3--7 MD runs for each system size, accounting for statistical spread and the fact that MD simulations do not produce exact atomic trajectories~\cite{Frenkel2002}. The contact area $A_{\mathrm{r}}$ calculated in the MD is an ensemble average of its time-averaged values. We time-averaged over the last 800000 time steps (160~ps) of the equilibration procedure.

\subsubsection{Forces size scaling}
\label{subsubsec::forces_scaling}

\begin{figure}[!htb]
\centering
  \includegraphics[height=7cm]{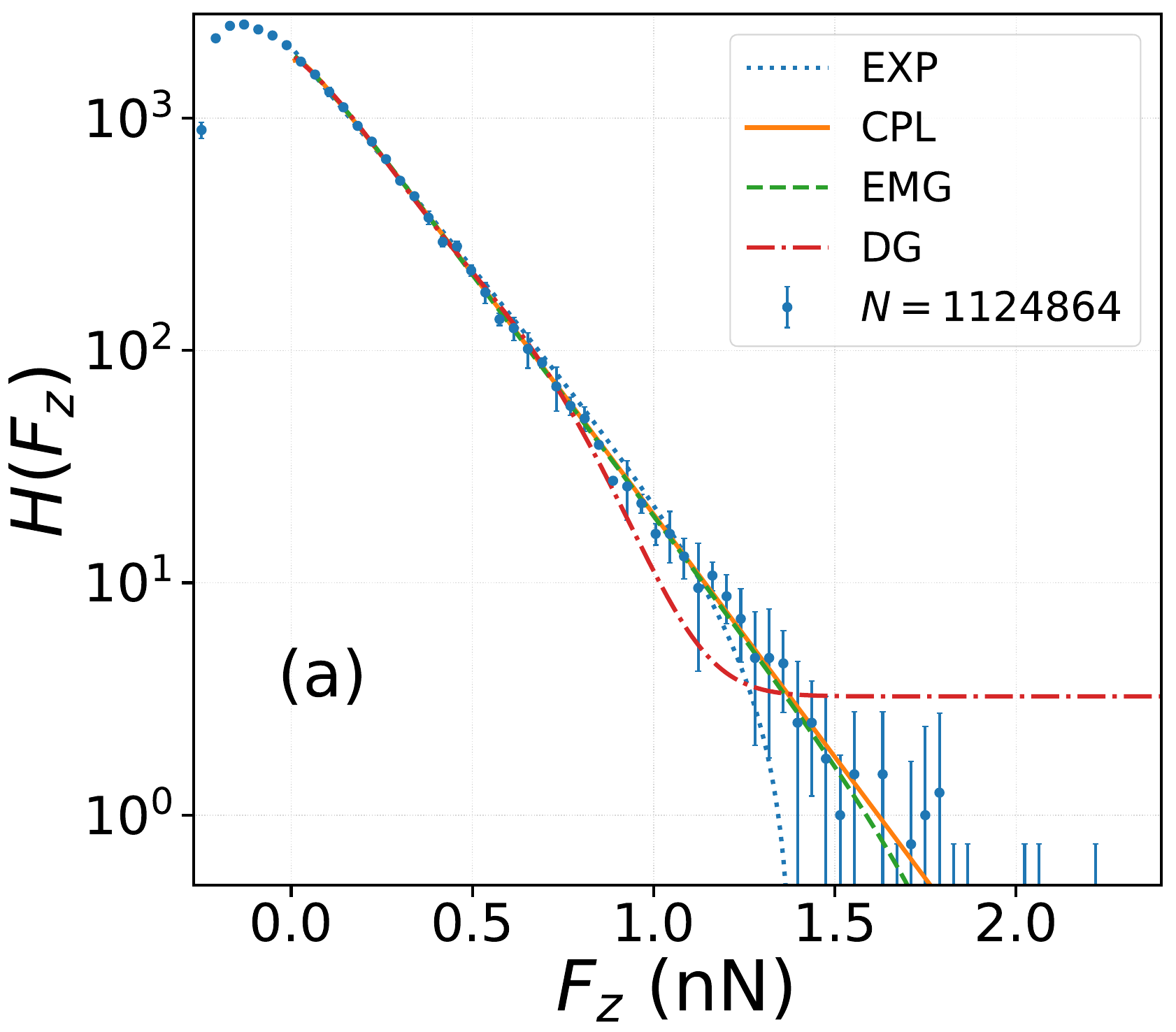}
  \includegraphics[height=7cm]{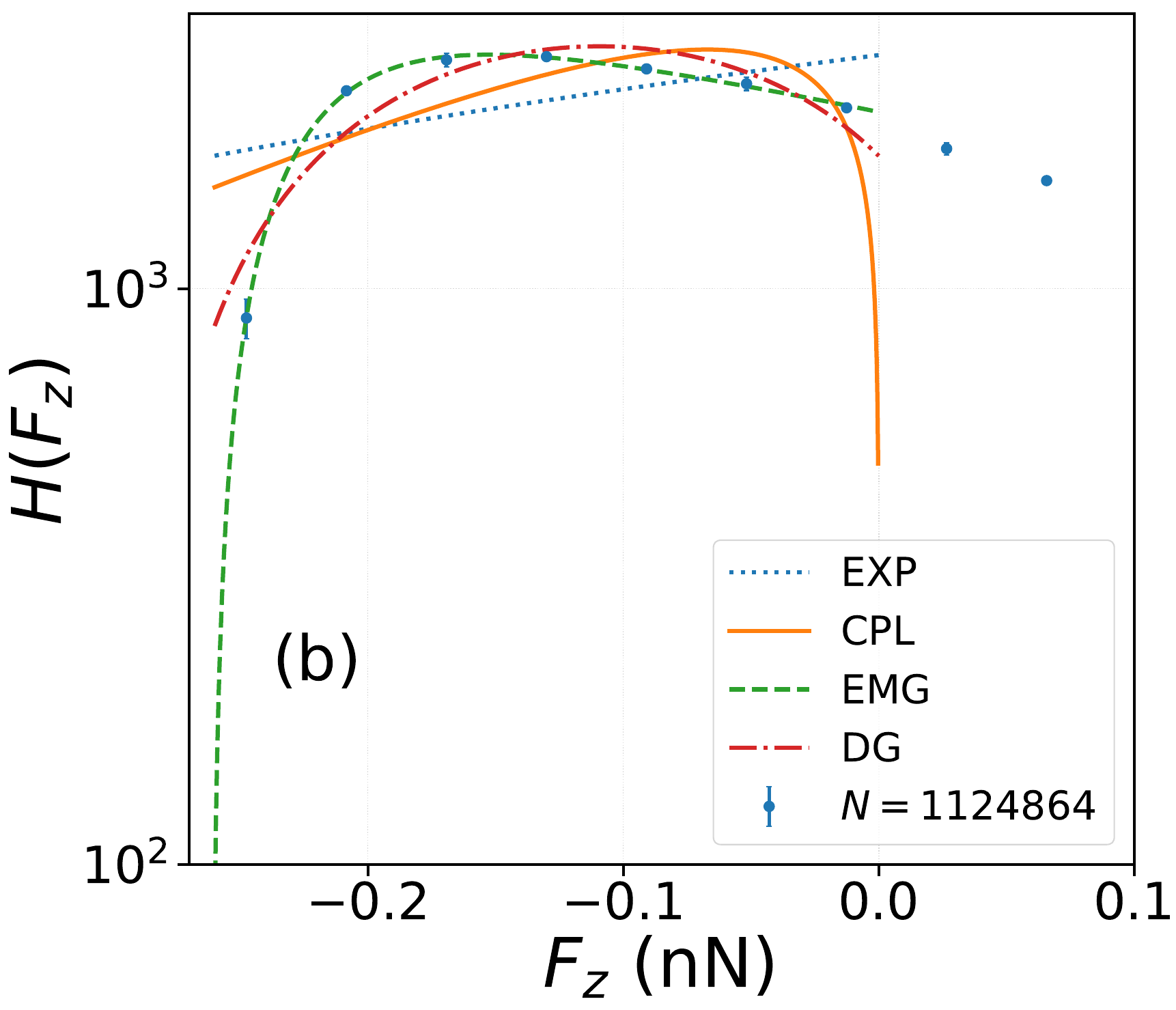}
  \caption{Example of numerical fitting of the ensemble-averaged $H(F_z)$ for the largest NP and the four fitting functions applied to positive (repulsive) (a) and negative (attractive) (b) force values.}
  \label{fig:bot_ens_fitting_18}
\end{figure}

Along with the scaling study, we fitted the normal intersurface force histograms/distribution plots using the following functions with adjustable parameters.

\begin{enumerate}
\item Exponential (EXP) dependency to compare with the results of Cheng et al.~\citep{Cheng2010triblett}:
\begin{equation}
P_{\mathrm{EXP}}(F) = P_0e^{\lambda F} + C_0,
\label{eq:exp}
\end{equation}
where $P_0$ is the amplitude, $\lambda$ is the exponential rate, $C_0$ is the constant shift.

\item Cutoff power law (CPL) combines a power law part with an exponentially decaying tail:
\begin{equation}
P_{\mathrm{CPL}}(F) = P_0F^\alpha e^{-F/F_c},
\label{eq:cpl}
\end{equation}
where $P_0$ is the amplitude, $\alpha$ is the power-law exponent, $F_c>0$---exponential cutoff scale. The power law dominates at $F \ll F_c$, while for $F\gtrsim F_c$ the exponential decay prevails, causing the tail to decay faster than a pure power law.

\item
Exponentially modified Gaussian (EMG), applied separately to $F=|F_z|$:
\begin{multline}
P_{\mathrm{EMG}}(F) =\\
P_0 \frac{\lambda}{2} \exp\!\left[\lambda(\mu-F) +\frac{\lambda^2\sigma^2}{2} \right] \operatorname{erfc} \!\left[ \frac{\mu+\lambda\sigma^2 - F} {\sqrt{2}\sigma} \right]\\ + C_0,
\label{eq:emg}
\end{multline}
where $\operatorname{erfc}$ is the complementary error function
$\operatorname{erfc}(F)=1-\operatorname{erf}(F)$, $P_0$, $\mu$, $\sigma$, $\lambda$, $C_0$ are the adjustable parameters. The EMG is the convolution of a Gaussian with an exponential distribution, making it useful for asymmetric distributions with a Gaussian-like core and an exponential tail, which may hold for the normal-force distributions under consideration.

\item The double Gaussian (DG) as a difference of two Gaussians with opposite-sign centers as predicted by Persson theory, Eq.~(\ref{eq:persson_double_gaussian}):
\begin{equation}
P_{\mathrm{DG}}(F) = P_1 e^{-(F-s)^2/(2\sigma_1^2)} - P_2 e^{-(F+s)^2/(2\sigma_2^2)} + C_0.
\label{eq:double_gaussian}
\end{equation}
with fitting parameters $P_1$, $P_2$, $s$, $\sigma_1$, $\sigma_2$, $C_0$.
\end{enumerate}

\begin{figure}[!htb]
\centering
  \includegraphics[height=7cm]{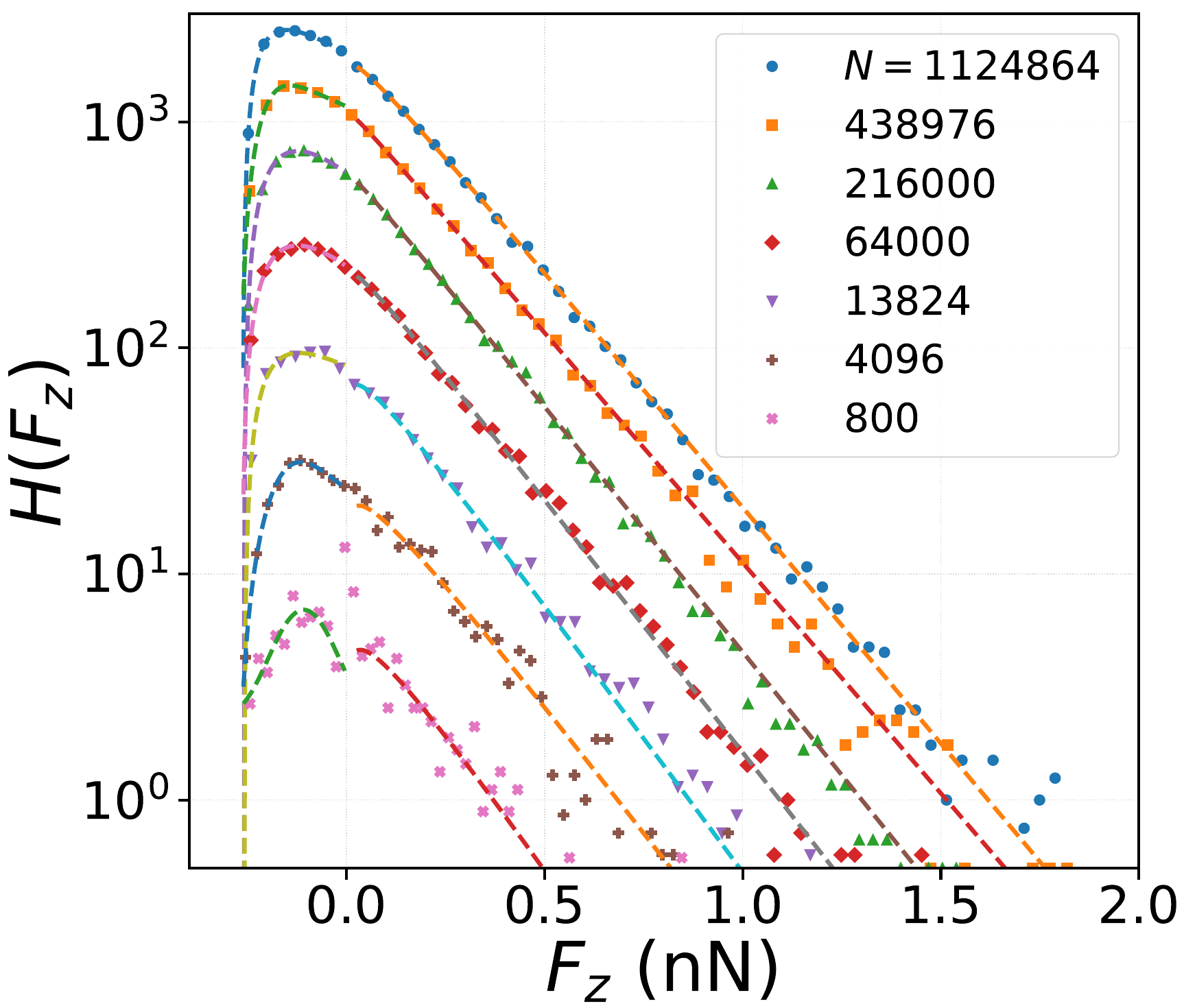}
  \caption{Numerical fitting (the dashed lines) of the histogram $H(F_z)$ data (symbols). We fit the negative and positive forces separately using EMG and CPL, respectively.}
  \label{fig:bot_ens_scaling_emg_cpl}
\end{figure}

\begin{figure}[!htb]
\centering
  \includegraphics[height=7cm]{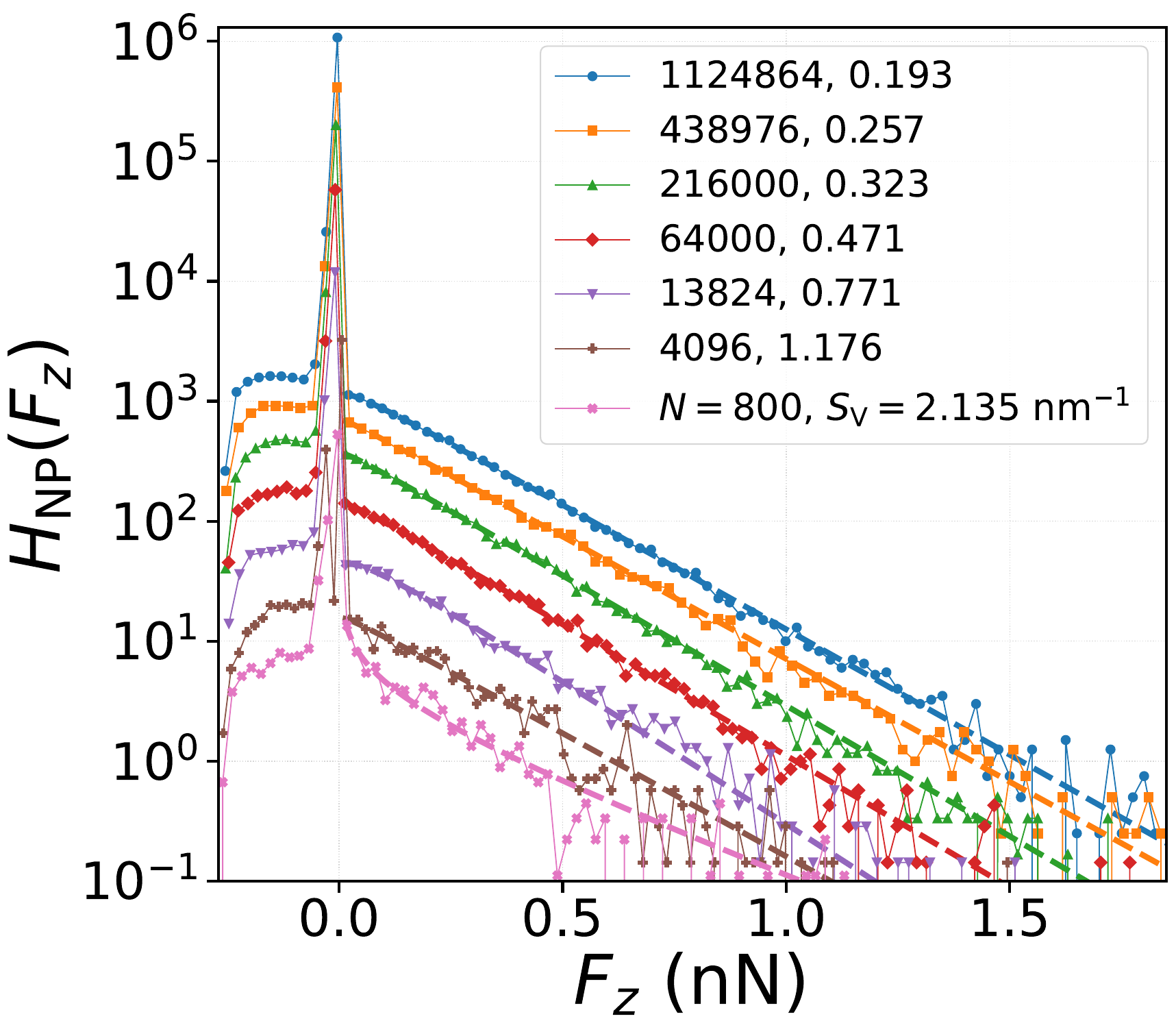}
  \caption{$H_{\mathrm{NP}}(F_z)$ of normal forces from C atoms on all Pb NP atoms. The dashed lines are CPL fittings of the repulsive tails. The solid lines are shown to guide the eye. The legend also depicts the surface area-to-volume ratio $S_{\mathrm{V}}$.}
  \label{fig:np_ens_scaling}
\end{figure}

\begin{table*}[htb]
\centering
\small
\caption{Fitting quality coefficients for $H(F_z)$ plots in Fig.~\ref{fig:bot_ens_scaling_emg_cpl} (positive $F_z$). Best quality coefficients are highlighted in bold.}
\begin{tabular}{|c|l|r|r|r|r|}
\hline
$N$ & Method & RMS & log RMSE & $R^2_{\log}$ & log AIC \\
\hline
800 & EXP & $0.46$ & \textbf{\boldmath$0.49$} & \textbf{\boldmath$0.86$} & \textbf{\boldmath$-50.69$} \\
 & CPL & \textbf{\boldmath$0.41$} & $0.56$ & $0.82$ & $-39.72$ \\
 & EMG & $0.46$ & \textbf{\boldmath$0.49$} & \textbf{\boldmath$0.86$} & $-46.69$ \\
 & DG & $0.60$ & $0.55$ & $0.83$ & $-36.41$ \\
\hline
4096 & EXP & \textbf{\boldmath$0.78$} & $0.87$ & $0.71$ & $-4.50$ \\
 & CPL & $0.79$ & \textbf{\boldmath$0.52$} & \textbf{\boldmath$0.90$} & \textbf{\boldmath$-45.25$} \\
 & EMG & \textbf{\boldmath$0.78$} & $0.87$ & $0.71$ & $-0.51$ \\
 & DG & $1.11$ & $0.58$ & $0.87$ & $-29.92$ \\
\hline
13824 & EXP & $1.31$ & $1.24$ & $0.60$ & $21.03$ \\
 & CPL & \textbf{\boldmath$0.77$} & \textbf{\boldmath$1.06$} & \textbf{\boldmath$0.75$} & \textbf{\boldmath$10.80$} \\
 & EMG & $1.31$ & $1.24$ & $0.60$ & $25.24$ \\
 & DG & $2.13$ & $1.14$ & $0.70$ & $22.81$ \\
\hline
64000 & EXP & $1.92$ & \textbf{\boldmath$0.34$} & \textbf{\boldmath$0.97$} & \textbf{\boldmath$-78.95$} \\
 & CPL & \textbf{\boldmath$1.50$} & $0.54$ & $0.94$ & $-48.91$ \\
 & EMG & $1.92$ & \textbf{\boldmath$0.34$} & \textbf{\boldmath$0.97$} & $-74.94$ \\
 & DG & $7.12$ & $1.51$ & $0.57$ & $49.09$ \\
\hline
216000 & EXP & $2.57$ & $0.51$ & $0.94$ & $-47.22$ \\
 & CPL & $2.20$ & \textbf{\boldmath$0.40$} & \textbf{\boldmath$0.97$} & \textbf{\boldmath$-82.80$} \\
 & EMG & \textbf{\boldmath$2.18$} & $0.62$ & $0.92$ & $-29.69$ \\
 & DG & $18.34$ & $1.69$ & $0.51$ & $62.33$ \\
\hline
438976 & EXP & $4.46$ & \textbf{\boldmath$0.36$} & \textbf{\boldmath$0.97$} & \textbf{\boldmath$-65.26$} \\
 & CPL & \textbf{\boldmath$3.66$} & $0.75$ & $0.92$ & $-21.28$ \\
 & EMG & $4.46$ & \textbf{\boldmath$0.36$} & \textbf{\boldmath$0.97$} & $-61.26$ \\
 & DG & $38.17$ & $2.06$ & $0.41$ & $80.03$ \\
\hline
1124864 & EXP & $12.73$ & $0.41$ & $0.95$ & $-56.30$ \\
 & CPL & \textbf{\boldmath$5.74$} & $0.37$ & \textbf{\boldmath$0.98$} & \textbf{\boldmath$-92.94$} \\
 & EMG & $5.76$ & \textbf{\boldmath$0.36$} & \textbf{\boldmath$0.98$} & $-85.42$ \\
 & DG & $56.22$ & $2.10$ & $0.41$ & $86.19$ \\
\hline
\end{tabular}
\label{tab:unnorm_fitting_pos}
\end{table*}

Fig.~\ref{fig:bot_ens_fitting_18} visually illustrates the fitting results of the repulsive (a) and attractive (b) parts of the histogram $H(F_z)$ for the contact atomic layer of the largest NP. For repulsion, CPL Eq.~(\ref{eq:cpl}) and EMG Eq.~(\ref{eq:emg}) give the best fit, with slight deviations at the largest forces, while DG Eq.~(\ref{eq:double_gaussian}) gives the poorest fit. For negative attractive forces, EMG produces the best fit.

\begin{table*}[!htb]
\centering
\small
\caption{Fitting quality coefficients for the $H(F_z)$ plots in Fig.~\ref{fig:bot_ens_scaling_emg_cpl} (negative $F_z$). Best quality coefficients are highlighted in bold.}
\begin{tabular}{|c|l|r|r|r|r|}
\hline
$N$ & Method & RMS & log RMSE & $R^2_{\log}$ & log AIC \\
\hline
800 & EXP & $1.31$ & $0.26$ & $0.30$ & $-24.03$ \\
 & CPL & \textbf{\boldmath$0.70$} & \textbf{\boldmath$0.12$} & \textbf{\boldmath$0.84$} & \textbf{\boldmath$-39.90$} \\
 & EMG & $0.72$ & $0.13$ & $0.81$ & $-34.21$ \\
 & DG & $1.02$ & $0.19$ & $0.61$ & $-24.35$ \\
\hline
4096 & EXP & $6.03$ & $0.43$ & $0.46$ & $-10.84$ \\
 & CPL & $4.94$ & $0.40$ & $0.53$ & $-12.26$ \\
 & EMG & \textbf{\boldmath$0.65$} & \textbf{\boldmath$0.03$} & \textbf{\boldmath$1.00$} & \textbf{\boldmath$-60.24$} \\
 & DG & $2.17$ & $0.08$ & $0.98$ & $-38.82$ \\
\hline
13824 & EXP & $15.33$ & $0.27$ & $0.45$ & $-12.37$ \\
 & CPL & $10.51$ & $0.21$ & $0.67$ & $-15.88$ \\
 & EMG & \textbf{\boldmath$3.69$} & \textbf{\boldmath$0.04$} & \textbf{\boldmath$0.99$} & \textbf{\boldmath$-33.94$} \\
 & DG & $4.52$ & $0.06$ & $0.97$ & $-27.81$ \\
\hline
64000 & EXP & $44.73$ & $0.24$ & $0.33$ & $-16.60$ \\
 & CPL & $34.04$ & $0.20$ & $0.55$ & $-19.80$ \\
 & EMG & \textbf{\boldmath$4.92$} & \textbf{\boldmath$0.02$} & \textbf{\boldmath$1.00$} & \textbf{\boldmath$-51.98$} \\
 & DG & $53.59$ & $0.30$ & $-0.02$ & $-7.29$ \\
\hline
216000 & EXP & $132.89$ & $0.38$ & $0.49$ & $-7.70$ \\
 & CPL & $94.41$ & $0.31$ & $0.65$ & $-10.40$ \\
 & EMG & \textbf{\boldmath$4.14$} & \textbf{\boldmath$0.01$} & \textbf{\boldmath$1.00$} & \textbf{\boldmath$-60.64$} \\
\hline
438976 & EXP & $250.85$ & $0.28$ & $0.42$ & $-9.15$ \\
 & CPL & $178.92$ & $0.22$ & $0.66$ & $-12.41$ \\
 & EMG & \textbf{\boldmath$8.28$} & \textbf{\boldmath$0.01$} & \textbf{\boldmath$1.00$} & \textbf{\boldmath$-51.19$} \\
 & DG & $203.87$ & $0.22$ & $0.65$ & $-6.11$ \\
\hline
1124864 & EXP & $461.04$ & $0.29$ & $0.26$ & $-11.14$ \\
 & CPL & $366.85$ & $0.25$ & $0.48$ & $-13.63$ \\
 & EMG & \textbf{\boldmath$18.54$} & \textbf{\boldmath$0.01$} & \textbf{\boldmath$1.00$} & \textbf{\boldmath$-57.46$} \\
 & DG & $392.94$ & $0.24$ & $0.50$ & $-7.80$ \\
\hline
\end{tabular}
\label{tab:unnorm_fitting_neg}
\end{table*}

Fig.~\ref{fig:bot_ens_scaling_emg_cpl} shows the size scaling of $H(F_z)$ of the contact NP atomic layer. The plots also show CPL for positive $F_z$ and EMG fitting for negative $F_z$. The histogram's force range widens with increasing NP size, meaning larger NPs are more likely to experience larger interfacial normal forces. As expected, the statistical spread is larger for smaller NPs, especially near the largest $F_z$. According to Table~\ref{tab:unnorm_fitting_pos} (cf. subsection~\ref{subsec::analysis} for the description of the fitting quality coefficients), CPL provides the best fit for repulsive forces for most NP sizes, giving the smallest RMS and reflecting the best fit in the high-probability (close to zero) $F_z$ interval. Additionally, CPL is the only fitting function with 3 or more best-fitting quality coefficients for several system sizes, meaning it also describes the larger $F_z$ region reasonably well. Nevertheless, the fitting quality of low-probability $H(F_z)$ at the larger $F_z$ (assessed using logarithmic residuals, e.g., $R^2_{\log}$, log AIC) is comparable for EXP, CPL, and EMG, especially for the larger NPs, which have better data statistics, suggesting that the exponential decay included in all the mentioned functions prevails for larger $F_z$. As for attractive forces, EMG is the undisputed best-fitting function for most of the NP sizes except for the smallest ones, where the statistical spread still may not allow determining fitting results reliably, cf. Table~\ref{tab:unnorm_fitting_neg}.

The scaling of the histograms $H_{\mathrm{NP}}(F_z)$ of forces acting on each NP atom in Fig.~\ref{fig:np_ens_scaling} is similar to the contact layer's $H(F_z)$. The repulsive part of $H_{\mathrm{NP}}(F_z)$ follows the same trends as for the contact layer. As shown in subsection~\ref{subsubsec::kinetics_forces_np}, $H_{\mathrm{NP}}(F_z)$ differs from $H(F_z)$ by a tall spike around zero force, reflecting the non-interacting Pb atoms and bulk atoms that feel a small attraction from the substrate.

Both Fig.~\ref{fig:bot_ens_scaling_emg_cpl} and Fig.~\ref{fig:np_ens_scaling} confirm the conclusion of subsection~\ref{subsubsec::forces_scaling} that the functional form of the histograms does not change qualitatively with NP size, cf. also Fig.~\ref{fig:bot_pcfz_vs_t}. Similar to the model involving atomically smooth surfaces~\cite{Cheng2010triblett}, the best fits for the ensemble-averaged instantaneous forces on atoms provide functional forms that include an exponentially decaying tail, in contrast to the pure double Gaussian observed for randomly rough surfaces~\cite{Almqvist2026pre,Prodanov2014triblett,Yang2008prl}. A mean-field statistical model can explain the exponential relation by using the Boltzmann equation for the probability of atomic height and force~\cite{Cheng2010triblett}.

\begin{table}[!htbp]
\centering
\small
\caption{CPL fitting parameters for $P(F_z)$ for positive $F_z$.}
\label{tab:cpl_params_positive_norm}
\setlength{\tabcolsep}{9pt}
\begin{tabular}{|l|l|l|l|}
\hline
$N$ & $P_0$ & $\alpha$ & $F_c$ \\ \hline
288 & $0.549251$ & $-0.387122$ & $0.467$ \\  \hline
392 & $2.43478$ & $0.0465822$ & $0.202918$ \\ \hline
512 & $4.12337$ & $0.149156$ & $0.159468$ \\ \hline
648 & $2.41241$ & $0.0381268$ & $0.222683$ \\ \hline
800 & $0.543008$ & $-0.454203$ & $0.394218$ \\ \hline
1728 & $2.98506$ & $0.0978769$ & $0.188484$ \\ \hline
4096 & $2.49504$ & $0.0380278$ & $0.210203$ \\ \hline
8000 & $3.11572$ & $0.0986338$ & $0.187093$ \\ \hline
13824 & $3.36799$ & $0.132095$ & $0.179476$ \\ \hline
32768 & $2.64982$ & $0.0636418$ & $0.197438$ \\ \hline
64000 & $2.12638$ & $-0.00533541$ & $0.211871$ \\ \hline
110592 & $2.47185$ & $0.0353064$ & $0.197796$ \\ \hline
216000 & $2.06768$ & $0.0100845$ & $0.216991$ \\ \hline
438976 & $2.18551$ & $0.0186424$ & $0.212785$ \\ \hline
1124864 & $2.49056$ & $0.0683554$ & $0.205644$ \\ \hline
\end{tabular}
\end{table}

\begin{figure}[!htb]
\centering
  \includegraphics[height=7.5cm]{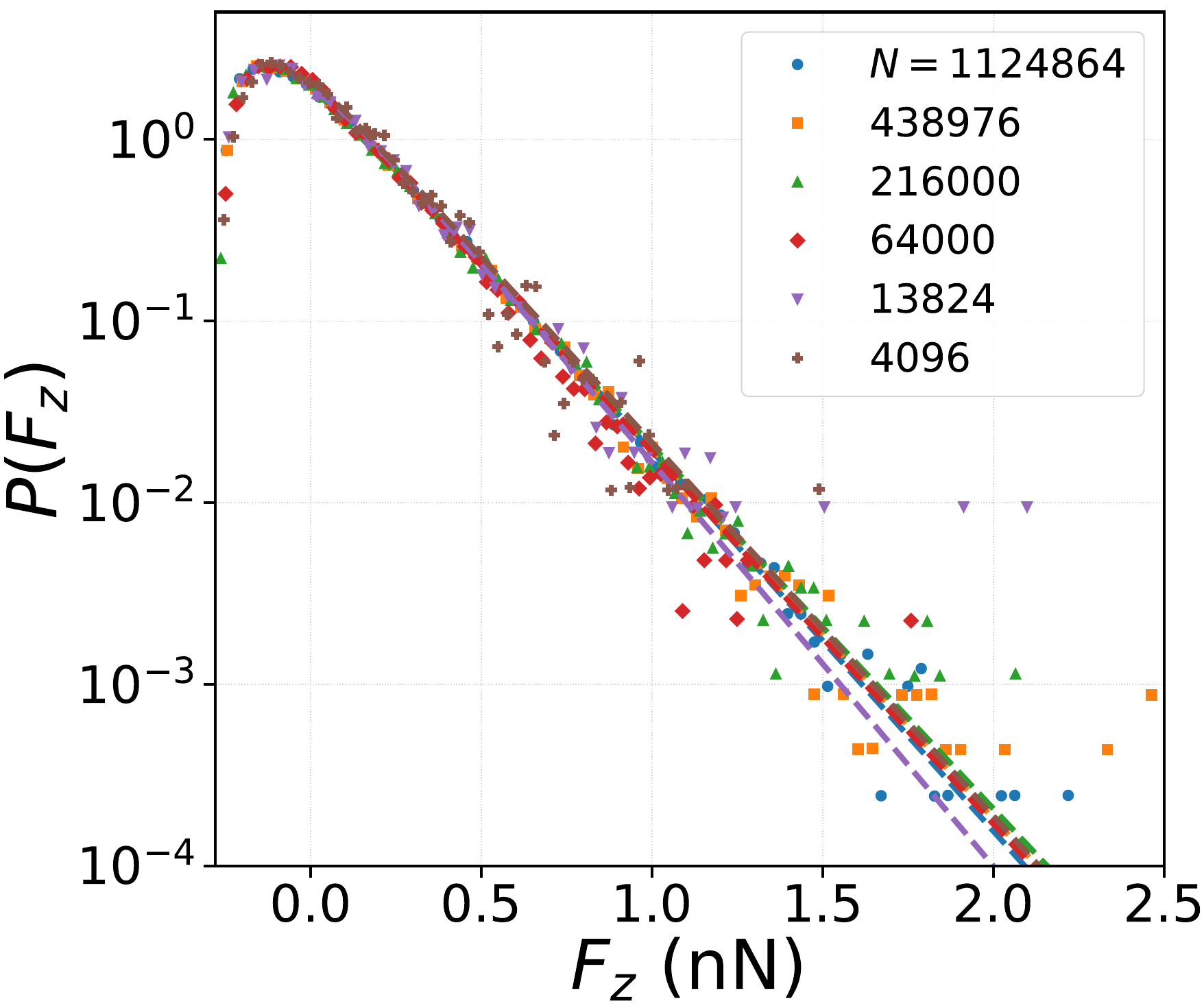}
  \caption{Normalized, ensemble-averaged $P(F_z)$ for contact-layer NP atoms. The dashed lines are the corresponding CPL fits.}
  \label{fig:bot_ens_norm}
\end{figure}

To find the universal CPL and EMG fitting parameters that best reproduce the data, we calculated a normalized, ensemble-averaged probability distribution $P(F_z)$ of the normal intersurface forces $F_z$ acting on the contact atomic NP layer shown in Fig.~\ref{fig:bot_ens_norm}. The standard normalization condition is:
\begin{equation}
\int_{F_{\mathrm{min}}}^\infty P(F_z)\,dF_z = 1,
\label{eq:pfz_norm}
\end{equation}
with $F_{\mathrm{min}}$ being the largest (by absolute value) attractive force. As expected, the smallest NPs in Fig.~\ref{fig:bot_ens_norm} show considerable data spread even at high probabilities, while for larger NPs it remains noticeable mostly in the repulsive-tail outliers. The corresponding CPL fits are very close for most NP sizes.

Table~\ref{tab:cpl_params_positive_norm} contains the CPL fitting parameters obtained for different NP sizes $N$. At around $N = 13824$, the parameter values start to converge to $P_0 \approx 2.28 \pm 0.21$, and $F_c \approx 0.2 \pm 0.01$~nN. However, $\alpha$ shows a noticeable scatter between $-0.005$ and $0.07$. According to CPL Eq.~(\ref{eq:cpl}), the power law contributes mainly over a narrow interval of high-probability forces, $F_z \ll F_c$. For most values of $F_z \gtrsim F_c$, $P(F_z)$ is close to the exponential function with the exponent of $\lambda = -1/F_c \approx -5$ nN$^{-1}$. Fitting parameters of EXP function Eq.~(\ref{eq:exp}) in Table~\ref{tab:exp_params_positive_norm} confirm that the exponent $\lambda$ is close to -5~nN$^{-1}$.

Regarding the attractive-force fitting: although the EMG clearly gives the best fit, identifying a unique combination of fitting parameters is not trivial because the five fitting parameters can converge to a local minimum. This means the five-parameter EMG model is poorly constrained by our data, so the optimizer can find very different parameter combinations that reproduce nearly identical curves. Because attractive forces are not the main focus of the current work, we postponed identifying EMG fitting parameters to a future study.

\begin{table}[!htb]
\centering\small
\caption{Exponential fitting parameters for $P(F_z)$ for positive $F_z$.}
\begin{tabular}{|c|r|r|r|}
\hline
$N$ & $P_0$ & $\lambda$ & $C_0$ \\ \hline
288 & $2.04119$ & $-5.40143$ & $0.0500746$ \\ \hline
392 & $2.04904$ & $-4.22437$ & $-0.0363615$ \\ \hline
512 & $2.35963$ & $-4.62116$ & $-0.0317901$ \\ \hline
648 & $2.14623$ & $-3.50789$ & $-0.122166$ \\ \hline
800 & $2.86909$ & $-7.39896$ & $0.0623652$ \\ \hline
1728 & $2.14122$ & $-4.31769$ & $-0.0355119$ \\ \hline
4096 & $2.19691$ & $-4.34896$ & $-0.0182141$ \\ \hline
8000 & $2.18642$ & $-4.32497$ & $-0.0279414$ \\ \hline
13824 & $2.11825$ & $-4.4609$ & $-0.00682234$ \\ \hline
32768 & $2.11993$ & $-4.48763$ & $-0.00936607$ \\ \hline
64000 & $2.17447$ & $-4.73546$ & $-0.00502457$ \\ \hline
110592 & $2.19855$ & $-4.77771$ & $-0.0029491$ \\ \hline
216000 & $2.0007$ & $-4.52339$ & $-0.00186512$ \\ \hline
438976 & $2.03769$ & $-4.51345$ & $-0.00227087$ \\ \hline
1124864 & $1.9821$ & $-4.34057$ & $-0.00483114$ \\ \hline
\end{tabular}
\label{tab:exp_params_positive_norm}
\end{table}

Notice that even though longer wavelengths are added to the contact of larger NPs, our simulations in Fig.~\ref{fig:bot_ens_norm} indicate that the asymmetric distribution $P(F_z)$ does not change. This may seem to contrast with Persson theory, where resolving shorter roughness wavelengths broadens $P(F_z)$. However, we always resolve roughness down to the shortest, atomic wavelengths. Increasing the NP adds longer wavelengths, cf. the PSD in Fig.~\ref{fig:fft_psd_12_09} and Fig.~\ref{fig:fft_psd_18_05}, but not the short ones. Therefore, $P(F_z)$ behavior does not contradict Persson's theory.

\subsubsection{Contact area scaling}
\label{subsubsec::area_scaling}

\begin{figure}[!htb]
\centering
  \includegraphics[height=3.85cm]{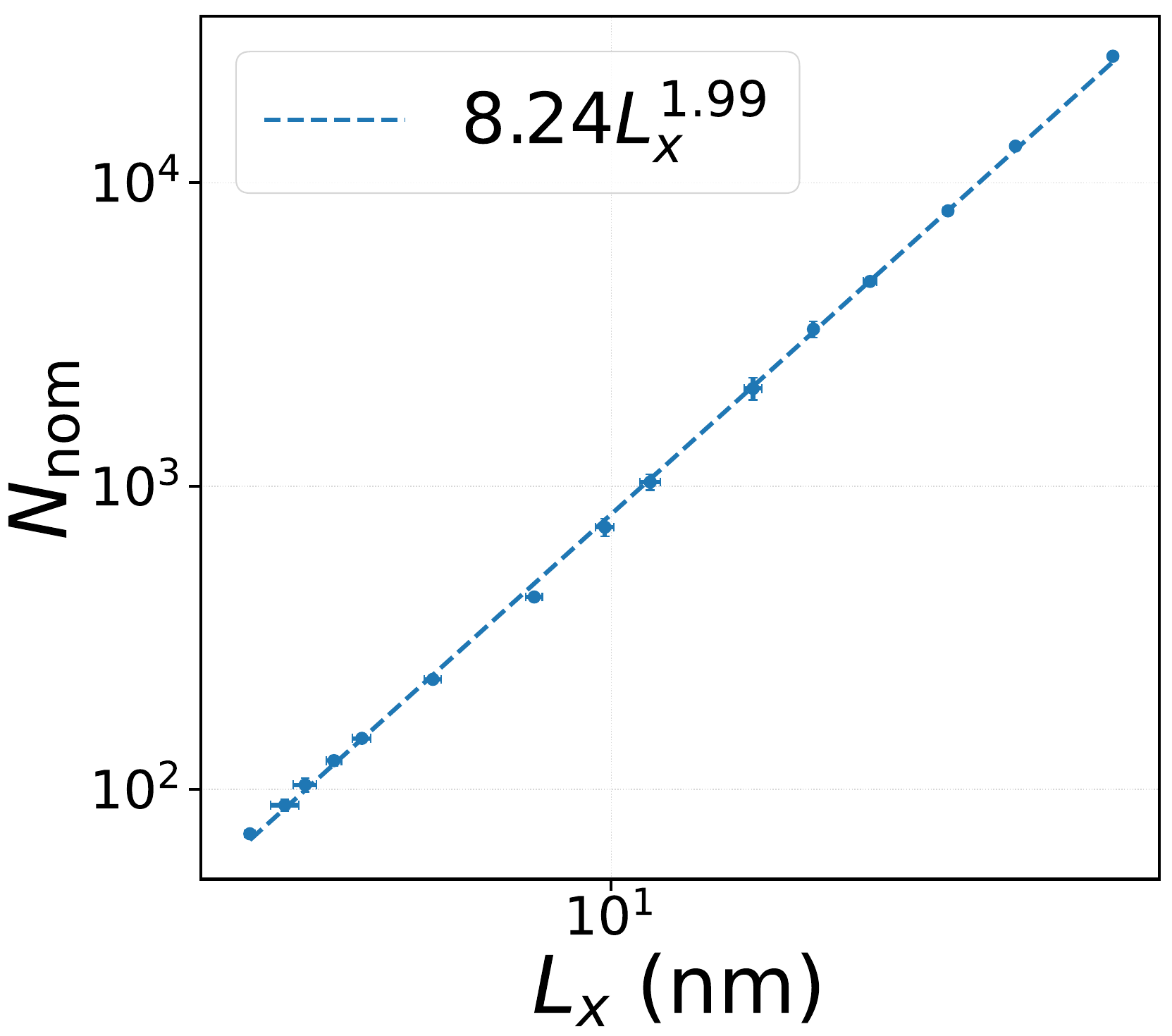}
  \includegraphics[height=3.85cm]{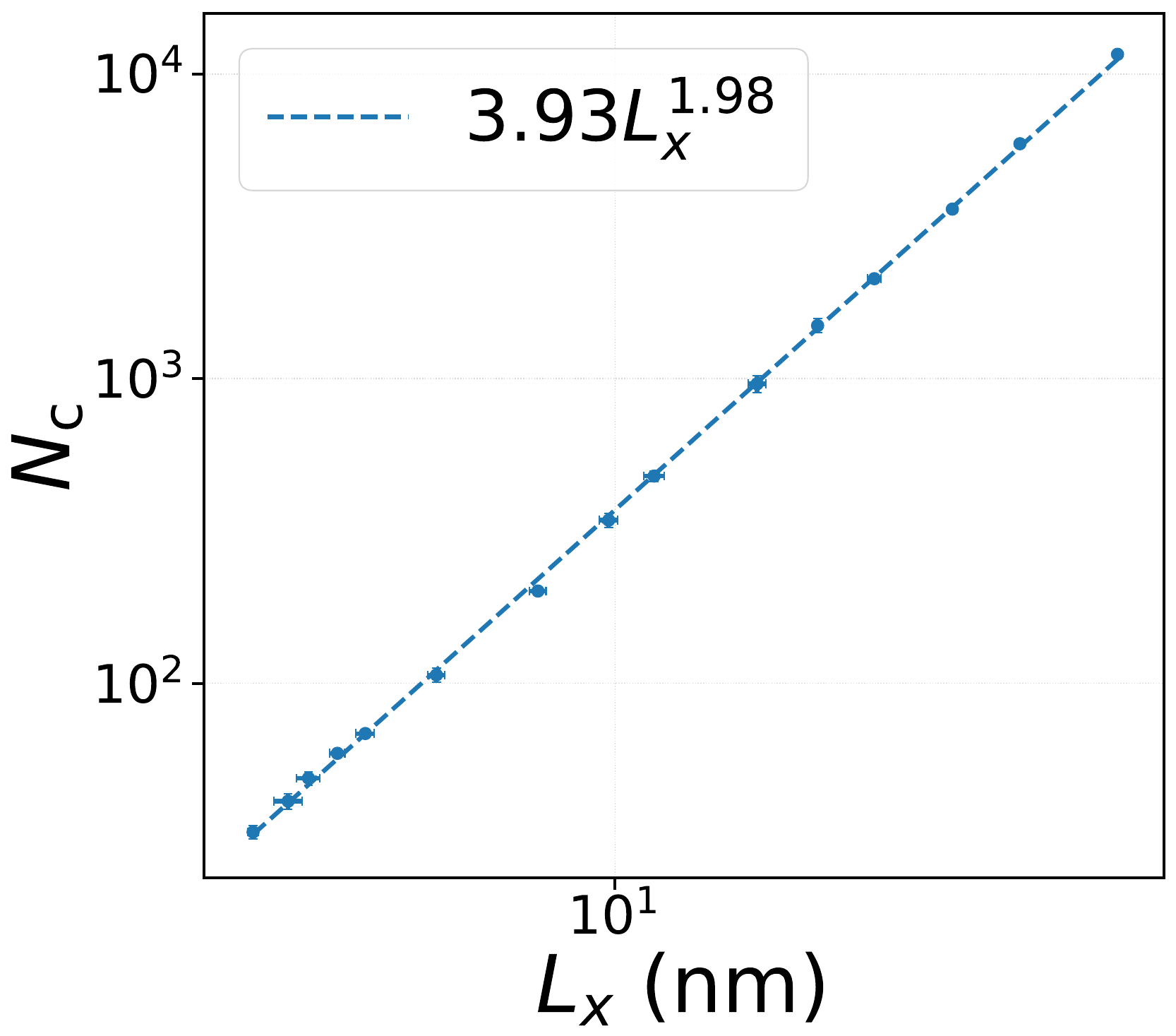}
  \caption{Scaling of $N_{\mathrm{nom}}$ (left) and $N_{\mathrm{c}}$ (right) with the NP linear size $L_{x}$.}
  \label{fig:np_surface_repuls_atoms_vs_lx}
\end{figure}

Let's consider size scaling of the time-averaged values of $N_{\mathrm{nom}}(t)$, $N_{\mathrm{c}}(t)$ and $A_{\mathrm{r}}(t)$. Fig.~\ref{fig:np_surface_repuls_atoms_vs_lx} shows the time-averaged numbers of atoms $N_{\mathrm{nom}}$ and $N_{\mathrm{c}}$ as functions of the lateral size $L_{x}$ of the NP. As expected, they follow a power law with an exponent close to 2.

\begin{figure}[!htb]
\centering
  \includegraphics[height=7cm]{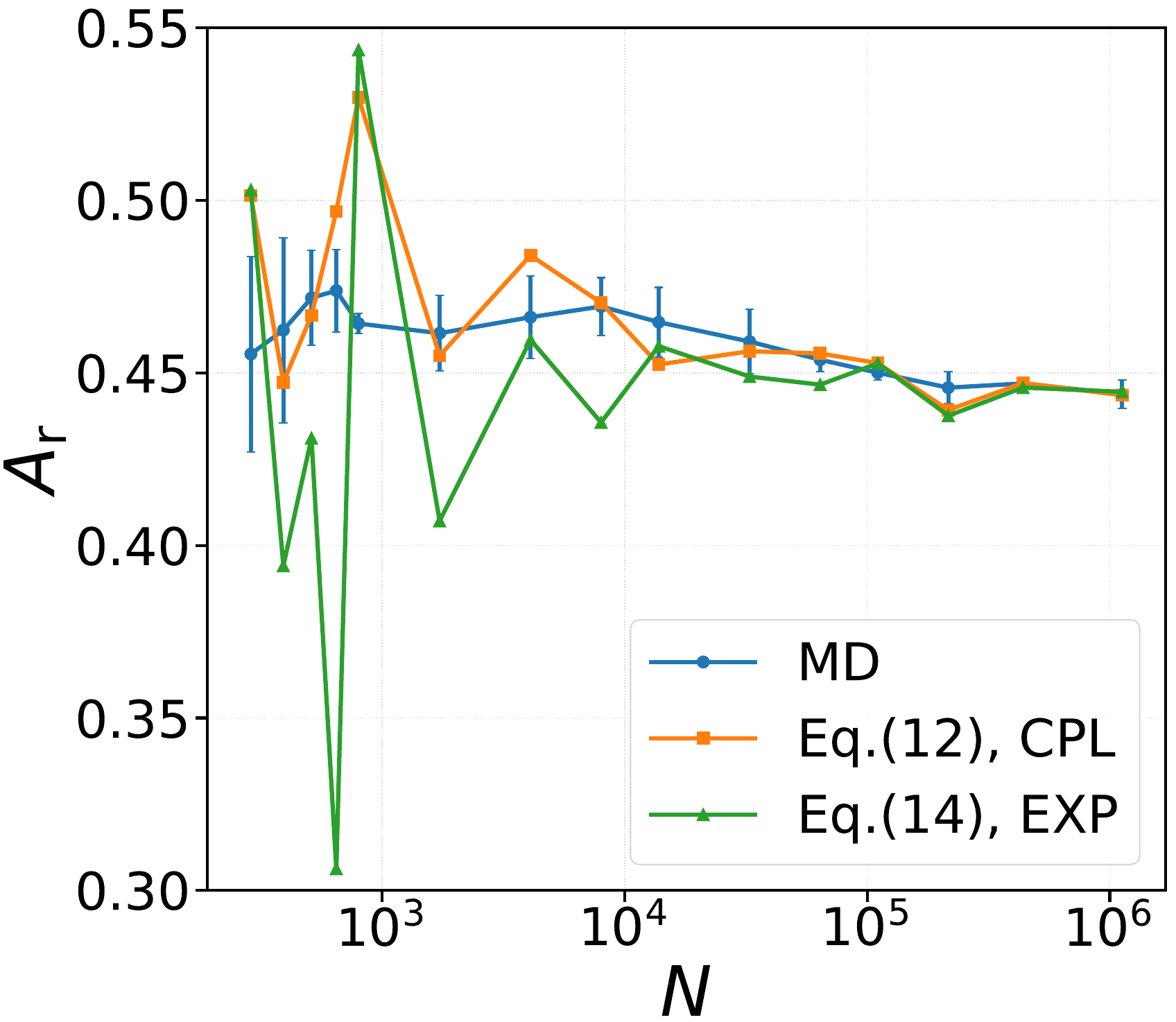}
  \caption{Time-averaged $A_{\mathrm{r}}$ vs $N$ obtained in MD and calculated using the CPL Eq.~(\ref{eq:ar_cpl}) and EXP Eq.~(\ref{eq:ar_exp}) fitting functions with the parameters from Tables~\ref{tab:cpl_params_positive_norm} and~\ref{tab:exp_params_positive_norm}, respectively.}
  \label{fig:a_rel_vs_n}
\end{figure}

\begin{figure}[!htb]
\centering
  \includegraphics[height=7cm]{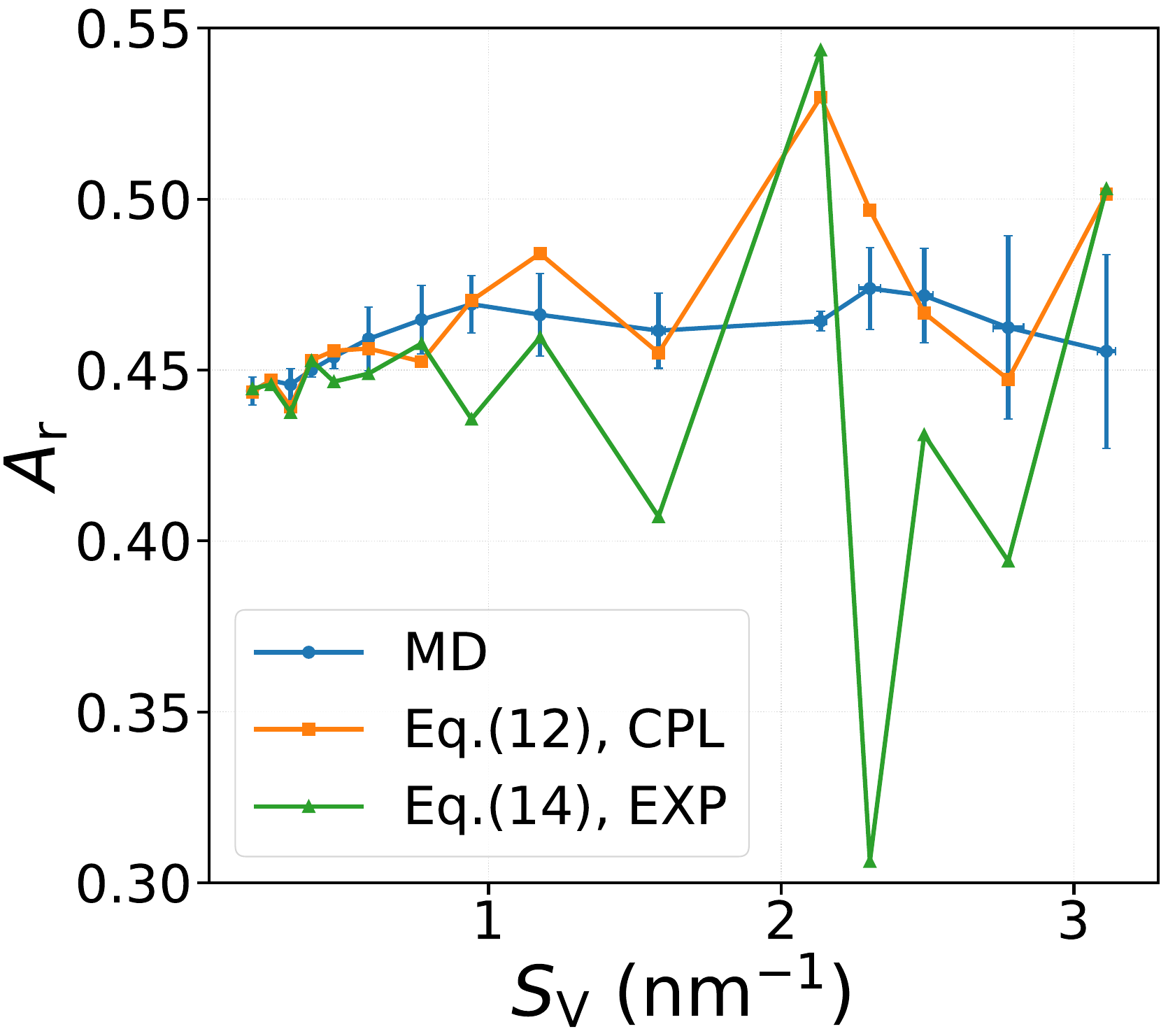}
  \caption{$A_{\mathrm{r}}$ vs $S_{\mathrm{V}}$ obtained in MD and calculated using the CPL and EXP fitting functions.}
  \label{fig:a_rel_vs_sv}
\end{figure}

Fig.~\ref{fig:a_rel_vs_n} and Fig.~\ref{fig:a_rel_vs_sv} show how $A_{\mathrm{r}}$ scales with the number of NP atoms $N$ and the surface area-to-volume ratio $S_{\mathrm{V}}$, respectively. The blue line (circles) represents the data obtained in MD. Other plots are obtained by calculating $A_{\mathrm{r}}$ using the corresponding fitting functions of $P(F_z)$ and Eq.~(\ref{eq:a_r}). Substituting the CPL fitting Eq.~(\ref{eq:cpl}) into Eq.~(\ref{eq:a_r}) gives an analytical expression for $A_{\mathrm{r}}$:
\begin{multline}
A_{\mathrm{r}} = \int_{F_{\mathrm{min}}}^{F_{\mathrm{max}}}
P_0F^\alpha e^{-F/F_c}\,dF\\
=P_0F_c^{\alpha+1}\Biggl[
\Gamma\left(\alpha+1,\frac{F_{\mathrm{min}}}{F_c}\right)\\
-\Gamma\left(\alpha+1,\frac{F_{\mathrm{max}}}{F_c}\right)
\Biggr],
\label{eq:ar_cpl}
\end{multline}
where 
\begin{equation}
\Gamma(s,x)=\int_x^\infty u^{s-1}e^{-u}\,du
\label{eq:gamma}
\end{equation}
is the upper incomplete gamma function.

The MD data in Fig.~\ref{fig:a_rel_vs_n} and Fig.~\ref{fig:a_rel_vs_sv} show more than 10\% spread for the smallest NPs, with slightly larger average $A_{\mathrm{r}}$ values than for the larger NPs. The larger NPs have a smaller statistical spread in $A_{\mathrm{r}}$ and converge to $A_{\mathrm{r}} \approx 0.44$. In general, the ensemble-averaged $A_{\mathrm{r}}$ is mostly size-independent: the spread of the average values is no more than 5\%. At $S_{\mathrm{V}} \lesssim 1$~nm$^{-1}$ or $L_x \gtrsim 1$~nm or $N \gtrsim 10^4$ atoms, $A_{\mathrm{r}}$ converges to an asymptotic value of around 0.44 with decreasing fluctuations.

$A_{\mathrm{r}}$ calculated by numerically integrating Eq.~(\ref{eq:ar_cpl}) using the fitting parameters from Table~\ref{tab:cpl_params_positive_norm} with $F_{\mathrm{min}} = 10^{-5}$~nN (to reflect the $0^+$ lower integration bound that excludes the non-contact atoms) in Fig.~\ref{fig:a_rel_vs_n} and Fig.~\ref{fig:a_rel_vs_sv} follows the MD result within the statistical error for NPs with $N \gtrsim 10^4$, coinciding for the largest NP size. For smaller NPs, especially the smallest ones with a few hundred atoms, Eq.~(\ref{eq:ar_cpl}) can manifest larger deviations of more than 10~\% compared to the MD results.

Fig.~\ref{fig:a_rel_vs_n} and Fig.~\ref{fig:a_rel_vs_sv} also show $A_{\mathrm{r}}$ calculated using the EXP function Eq.~(\ref{eq:exp}):
\begin{multline}
A_{\mathrm{r}} = \int_{F_{\mathrm{min}}}^{F_{\mathrm{max}}}
\left(P_0e^{\lambda F}+C_0\right)\,dF\\
= \frac{P_0}{\lambda}
\left(e^{\lambda F_{\mathrm{max}}}
-e^{\lambda F_{\mathrm{min}}}\right)\\
+ C_0(F_{\mathrm{max}}-F_{\mathrm{min}}),
\label{eq:ar_exp}
\end{multline}
with fitting parameters taken from Table~\ref{tab:exp_params_positive_norm}. Again, for $N \gtrsim 10^4$, Eq.~(\ref{eq:ar_exp}) gives results within statistical error of MD and very close to CPL, coinciding for the largest NP at $A_{\mathrm{r}} \approx 0.44$. For smaller NPs, the exponential function gives even larger data spread than CPL. These results indicate that, in general, CPL gives more precise results than EXP, even though they are very close for the largest NP sizes. Although the distributions $P(z)$ for NPs of different sizes are close, they are not identical in the covered $F_z$ interval and data spread, especially for smaller NPs, which explains the observed difference in relative contact area.

The simulations suggest that even at zero external load, $A_{\mathrm{r}} \approx 44$~\% for Pb NP adsorbed on SGr, which is much higher than $A_{\mathrm{r}}$ of randomly rough surfaces under finite load, e.g., a tire and the road surface, where $A_{\mathrm{r}}$ can be just a few~\%~\citep{Persson2005jpcm}. $A_{\mathrm{r}}$ values in Fig.~\ref{fig:a_rel_vs_n} are close to the estimate obtained for Al and Cu NP using the geometrical water leakage criterion in ~\cite{Prodanov2026pccp}. Given that isotropic randomly rough surfaces percolate at $A_{\mathrm{r}} \approx 0.48$ (though at non-zero load)~\cite{Persson2012}, the NP interface should be close to percolation, as also shown by the heatmaps in Fig.~\ref{fig:heatmaps_fcz_64k}, Fig.~\ref{fig:heatmaps_fcz_1M}. The observed large $A_{\mathrm{r}}$ of NPs at zero load may be due to the much smaller contact surface roughness of the NP-SGr interface, on the order of \AA~\cite{Prodanov2026pccp}, compared with that of randomly rough surfaces.

\section{Conclusions}
\label{sec::conclusions}

We utilized atomistic simulations to investigate the histograms $H(F_z)$ and probability distributions $P(F_z)$ of interfacial force $F_z$ and relative contact area $A_{\mathrm{r}}$ of Pb NPs physisorbed on SGr at zero external load. We considered two scenarios: their time evolution during NP formation and size scaling for solidified NPs.

The results suggest that histograms $H(F_z)$ differ between the liquid and solid states of the NP. In the liquid state, more metal atoms experience stronger attraction and less repulsion, likely because larger thermal fluctuations at higher temperatures reduce the effective repulsion. The repulsive tail of $H(F_z)$ decays faster and covers a smaller force range in the solid state. The lateral contact force distributions are symmetric around 0. The force power spectra are smeared in the liquid state, can show hexagonal symmetry for relatively small NPs, and do not exhibit hexagonal symmetry for the largest NPs. The characteristic wavelengths of the PSD correspond to the lattice structures of Pb and graphene.

Ensemble-averaged histograms of the instantaneous normal forces exhibit the size-independent functional forms for $N \gtrsim 10^4$. For smaller sizes, the data spread does not allow reliable fitting. To first order, the repulsive tail decays exponentially as $P_0e^{\lambda F_z}+C_0$ in accord with the previous MD simulations of atomically-smooth surfaces~\cite{Cheng2010triblett}, while the cutoff power law $P_0F_z^\alpha e^{\lambda F_z}$ fits best, especially for high-probability $F_z$ values. The exponential decay exponent is close to $\lambda \approx -5$~nN$^{-1}$. The exponentially modified Gaussian is the best-fitting function for the attractive part for most NP sizes.

The time-averaged relative contact area $A_{\mathrm{r}}$, calculated directly in the simulations, shows considerable fluctuations of more than 10\% for the smallest NPs, a few nm in diameter. Data scatter decreases with NP size, with $A_{\mathrm{r}}$ converging to about 44~\%. We derived analytical expressions for $A_{\mathrm{r}}$ using the corresponding CPL and EXP fits of $P(F_z)$. The analytical results agree with the MD data within statistical error for NPs with $N \gtrsim 10^4$.

Our findings may serve as reference data for potential analytical theories and for future experiments targeting interfacial force values and distributions. Additionally, the derived analytical expressions provide a simple practical way to estimate $A_{\mathrm{r}}$.

Further studies should verify the generality of the results by testing other metals and 2D materials. Identifying a unique set of EMG fitting parameters and relating the analytical fit coefficients to system parameters (e.g., lattice constant, interaction energy, etc.) are additional tasks. Another step is to check how $A_{\mathrm{r}}$ depends on load in both elastic and plastic deformation regimes when the NP is subjected to external pressure, as in indentation and ``tip-on-top'' tribological experiments, and compare with existing experimental and theoretical results~\cite{Almqvist2026pre,Dietzel2009apl,Lyashenko2013,Troshchenko_CMP2015,TPL_12,lyashenko_thermod}.

\section{CRediT authorship contribution statement}
\label{sec::credit}
Mykola Prodanov: conceptualization, methodology, software, running the simulations, data processing and analysis, writing---original draft, review and editing. Anton Samilyk: data analysis, writing---review and editing. Oleksii Khomenko: project administration, resources, conceptualization, supervision, writing---review and editing. Antonina Biesiedina: resources, conceptualization, writing---review and editing.

\section{Declaration of generative AI and AI-assisted technologies in the
writing process}
We used ChatGPT to help develop the fitting, data visualization, and data-processing scripts, but we executed the scripts and manually verified the results. Additionally, at the final stage of manuscript preparation, the corresponding author used Grammarly (www.grammarly.com) to check the text for grammar and other issues. The authors manually verified each potential issue identified by the program and accepted or rejected it. The authors again verified and approved the final manuscript.

\section{Declaration of competing interest}
\label{sec::interest}
The authors have no conflicts of interest to declare.

\section{Acknowledgments}
\label{sec::ack}

This work was supported by the Ministry of Education and Science of Ukraine, grant number \foreignlanguage{ukrainian}{БФ/26-2026.1}.

\section{Data availability}
\label{sec::data}
Raw MD data is publicly available in the Harvard Dataverse repository at https://doi.org/10.7910/DVN/NG0LXA.

MD source code is publicly available at https://github.com/prodk/SurfaceGrowthCrossPlatform.

\bibliographystyle{elsarticle-num}
\bibliography{pb_np_forces}





%
\end{document}